\documentclass[aps,reprint,longbibliography,floatfix]{revtex4-2}
\usepackage{graphicx}
\usepackage{amssymb}
\usepackage[font=small,skip=3pt]{caption}
\usepackage{subcaption}
\usepackage[fleqn]{amsmath}
\usepackage{siunitx}
\usepackage[version=4]{mhchem}
\usepackage{booktabs}
\usepackage{hyperref}
\hypersetup{
  pdftitle={Scalable Production of Lead-212 and Actinium-225 Generators with Fusion Neutrons},
  pdfauthor={J. F. Parisi and A. Rutkowski}
}
\usepackage{cleveref}
\usepackage{url}
\usepackage{bm}
\let\mrm\mathrm

\newcommand{\thalf}{t_{1/2}}
\newcommand{\ntn}{\ensuremath{(\mrm{n},2\mrm{n})}}
\newcommand{\ngamma}{\ensuremath{(\mrm{n},\gamma)}}
\newcommand{\betam}{\beta^{-}}



\usepackage{placeins}

\begin{document}

\title{Scalable Production of Lead-212 and Actinium-225 Generators with Fusion Neutrons}

\author{J. F. Parisi}
\email{jason@marathonfusion.com}
\author{A. Rutkowski}
\affiliation{Marathon Fusion, 150 Mississippi Street, San Francisco, CA 94107, USA}

\begin{abstract}
Targeted alpha therapy will require a large expansion of ${}^{212}$Pb and ${}^{225}$Ac production. We propose neutron- and photon-driven routes that convert ${}^{230}$Th, ${}^{231}$Pa, ${}^{232}$Th, and ${}^{237}$Np into generator parents ${}^{228}$Th and ${}^{229}$Th. The most direct ${}^{225}$Ac route is ${}^{230}$Th(n,2n)${}^{229}$Th. A 10 MW deuterium-tritium (D-T) neutron source irradiating thorium with a 27\% ${}^{230}$Th isotopic fraction accumulates about 250 Ci of ${}^{229}$Th in six months, initially enough for five million ${}^{225}$Ac dose-equivalents per year, and its 7916 yr half-life makes it useful for millennia. This requires only $\sim$2 gigawatt-days of D-T fusion operations. Thermal-neutron irradiation of tens of grams of ${}^{230}$Th produces gram quantities of ${}^{232}$U, a decades-long source of ${}^{228}$Th for millions of ${}^{212}$Pb dose-equivalents per year. Alternate routes proceed through ${}^{231}$Pa: fusion neutrons produce 1 to 2 tonnes of ${}^{231}$Pa per gigawatt-year from ${}^{232}$Th, and a thermal reactor converts up to 0.4 g of ${}^{232}$U per gram of ${}^{231}$Pa. Each gram of stored ${}^{232}$U produces 10 mg (8.2 Ci) of ${}^{228}$Th per year. Subsequent ${}^{228}$Th(n,$\gamma$)${}^{229}$Th converts 0.07 to 0.34 g of ${}^{229}$Th per gram of ${}^{228}$Th. We also analyze the ${}^{237}$Np(n,2n)${}^{236\mathrm{m}}$Np route to ${}^{228}$Th. These pathways scale from kilowatt to megawatt D-T sources and could secure millions of ${}^{212}$Pb and ${}^{225}$Ac dose-equivalents per year.
\end{abstract}

\maketitle

\section{Introduction}\label{sec:intro}

Targeted alpha therapy (TAT) uses short-lived alpha emitters bound to tumor-seeking ligands for radioligand therapy (RLT)~\cite{Yong2015}. At present, the radionuclides in clinical trials for alpha-emitting RLT include $^{225}$Ac, $^{211}$At, and $^{212}$Pb. As of September 2026, ${}^{225}$Ac has reached phase III~\cite{Pedersen2026TAT,CTGovAcNET3,CTGovAcProstate3}, ${}^{212}$Pb phase II~\cite{Pedersen2026TAT,CTGovPbNET2}, and ${}^{211}$At phase I/II~\cite{Pedersen2026TAT,CTGovAtLeukemia,CTGovAtPSMA}. Clinical trials for all three address a range of cancers. Although ${}^{212}$Pb is a beta emitter, it acts in vivo as a generator of the alpha emitter ${}^{212}$Bi, whose half-life is only 61 minutes.  The only FDA-approved alpha therapeutic, ${}^{223}$RaCl${}_2$ for bone metastases, is bone-seeking rather than ligand-targeted, and is currently produced by $^{226}$Ra reactor neutron capture~\cite{Radchenko2021Supply}. In this work we focus on new transmutation pathways for producing the generators for ${}^{212}$Pb and ${}^{225}$Ac: $^{228}$Th and $^{229}$Th, enabled by the growing availability of 14 MeV deuterium-tritium (D-T) fusion neutrons.

Supply of both ${}^{212}$Pb and ${}^{225}$Ac is currently limited \cite{Kokov2022}, although there are ongoing efforts to expand supply significantly. There are several pathways for $^{225}$Ac production: $^{233}$U$\to$$^{229}$Th, which is milked for $^{225}$Ac, mainly held at three sites~\cite{robertson2018ac225,morgenstern2018}. Proton spallation on thorium produces significant $^{225}$Ac yields, although there are ongoing concerns around $^{227}$Ac contamination. Proton~\cite{Nagatsu2022} and photon~\cite{diamond2021actinium} irradiation of $^{226}$Ra via $^{226}$Ra(p,2n) and $^{226}$Ra($\gamma$,n) is also being developed~\cite{ANS2026radium}. Finally, fast-neutron irradiation via $^{226}$Ra(n,2n) is also under evaluation~\cite{morrell2021next}.

Unlike $^{225}$Ac, which has no naturally occurring generator ($^{229}$Th), the $^{228}$Th generator for $^{212}$Pb is in the $^{232}$Th decay chain. Therefore $^{228}$Th recovery from $^{232}$Th decay (via the chemical separation of its precursor ${}^{228}$Ra) is one of the main pathways for $^{212}$Pb supply, although given the 14 billion year $^{232}$Th half-life, kilotonne ${}^{232}$Th stockpiles are required to give useful ${}^{228}$Th activity~\cite{McAlister2018}. Hospitals draw ${}^{212}$Pb on site from ${}^{228}$Th/${}^{224}$Ra generators. $^{228}$Th can also be milked from ${}^{232}$U~\cite{pruszynski2021radiochemical} or bred by double neutron capture on ${}^{226}$Ra~\cite{Kuznetsov2012,melville2013theoretical}. All of this production occurs at a handful of facilities, and more than 15 companies are building ${}^{212}$Pb supply on them, at a scale of $>10^{4}$ doses per year in 2025 and $>10^{5}$ by 2030~\cite{Zimmermann2024}. Earlier cyclotron routes to ${}^{236}$Pu via ${}^{238}$U(p,3n) and ${}^{237}$Np(p,2n)~\cite{Morgenstern2003,Aaltonen1993} were not pursued at scale.

\begin{table}[b]
\centering
\caption{Annual dose-equivalents supported by one curie of generator parent (one dose is one administration, and a course of therapy is usually four to six doses). Ideal values assume secular equilibrium, prompt milking, and no losses. The reference ${}^{212}$Pb value applies a 67\% recovery benchmark reported for generator systems~\cite{Li2023generator,Radchenko2021Supply}. The ${}^{225}$Ac value uses the annual output of the ORNL ${}^{229}$Th cow~\cite{robertson2018ac225,morgenstern2018}. Actual administered activity also depends on the collection schedule and delivery time, as we show in Appendix~\ref{app:pb212_doses}.}
\label{tab:doses_per_ci}
\small
\setlength{\tabcolsep}{3pt}
\begin{tabular}{lccrr}
\toprule
Parent & Ci/g & Dose & \multicolumn{2}{c}{Dose-eq./yr per Ci} \\
 & & & ideal & reference \\
\midrule
${}^{228}$Th (1.91 yr) & 820 & 2.7 mCi ${}^{212}$Pb & $2.1\cdot10^{5}$ & $1.4\cdot10^{5}$ \\
${}^{229}$Th (7916 yr) & 0.20 & 0.2 mCi ${}^{225}$Ac & $1.3\cdot10^{5}$ & $2.4\cdot10^{4}$ \\
\bottomrule
\end{tabular}
\end{table}

\begin{figure*}[t]
\centering
\begin{subfigure}[t]{0.48\textwidth}
\centering
\includegraphics[width=0.99\linewidth]{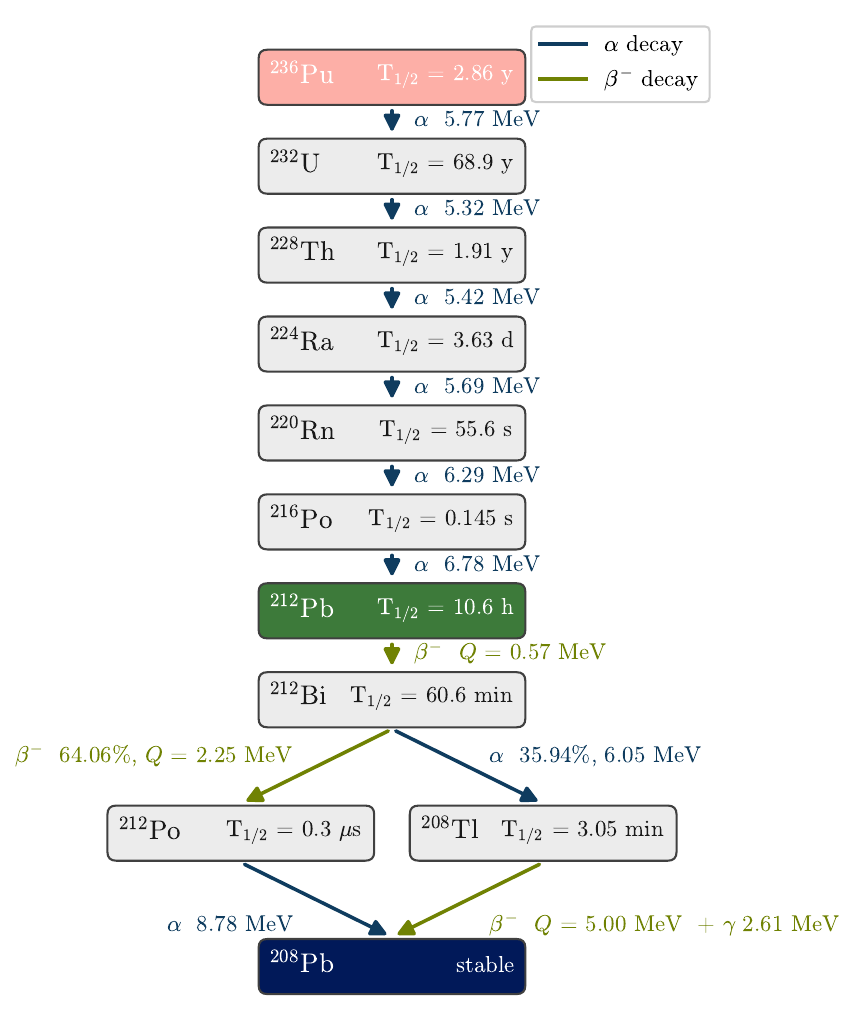}
\caption{${}^{236}$Pu decay chain to stable ${}^{208}$Pb.}
\label{fig:decay_pu236}
\end{subfigure}\hfill
\begin{subfigure}[t]{0.48\textwidth}
\centering
\includegraphics[width=0.99\linewidth]{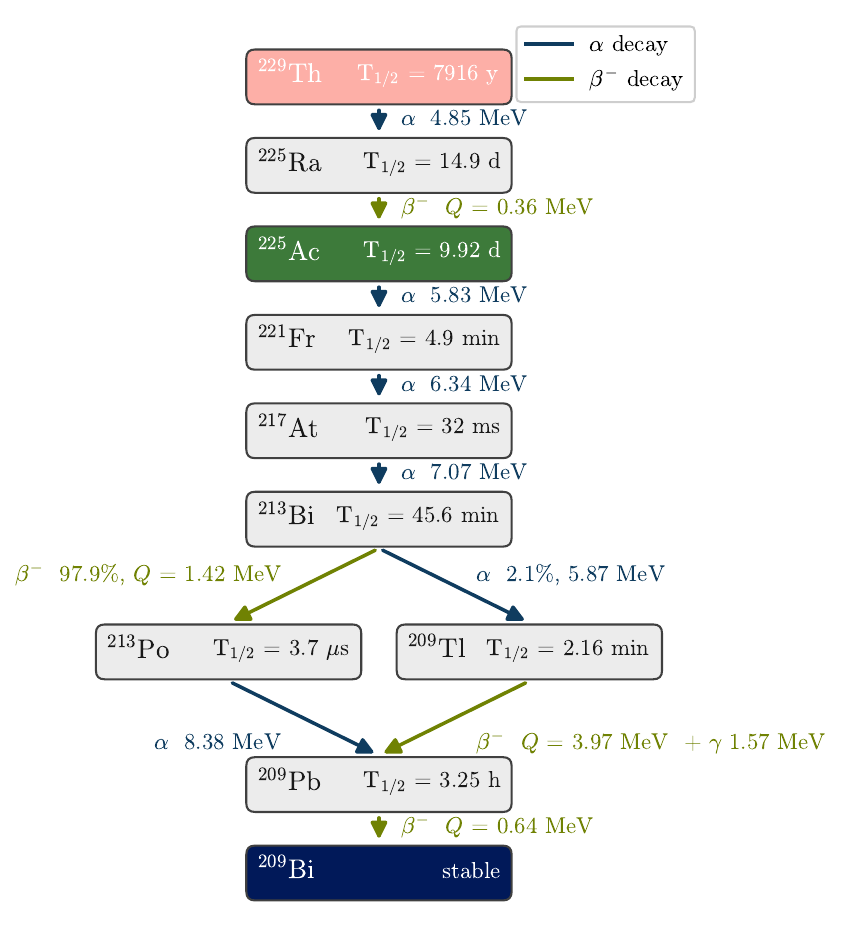}
\caption{${}^{229}$Th decay chain to stable ${}^{209}$Bi.}
\label{fig:decay_th229}
\end{subfigure}
\caption{The two main decay schemes considered in this work.}
\label{fig:decay}
\end{figure*}

In this work we consider only neutron- and photon-driven routes to ${}^{212}$Pb and ${}^{225}$Ac generators. Because the generator parents live much longer than their therapeutic daughters, they must be produced in substantial quantities, especially ${}^{229}$Th with its 7916 yr half-life. Fast neutrons can irradiate thick targets because they do not lose energy through Coulomb interactions. One megawatt of D-T fusion power corresponds to $\sim3.6\cdot10^{17}$ neutrons per second, compared with $\sim6\cdot10^{15}$ protons per second in a 1 mA cyclotron beam (considered a very large cyclotron facility). In the routes analyzed here, the useful reaction yield per source particle is 10 to 1000 times larger for neutrons than for protons. This combination of high neutron rate and high yield per source neutron enables gram-to-kilogram-scale ${}^{229}$Th production with kilowatt-to-megawatt years of D-T fusion power. We also include photonuclear routes because existing linear accelerators produce much larger photon fluxes than currently available D-T fusion-neutron facilities and drive closely related reaction chains.

Fast fusion neutrons have been proposed for isotope production including medical~\cite{engholm1986radioisotope,Bourque1988FAME,Leung2018_CompactNG,li2023feasibility,pereslavtsev2024potential,evitts2025theoretical,Parisi2025,Parisi2025IsotopeFusion,Parisi2025muCF,parisi2026neutronvalue} and nuclear battery radioisotopes~\cite{Parisi2026BetaBattery,Parisi2026FusionBattery}, and base metals~\cite{Bourque1988FAME,Rutkowski2025}. Most of the neutron routes we discuss here were recently introduced at fusion-power-plant scale for battery fuels in~\cite{Parisi2026FusionBattery}. This work applies them at the much smaller scales required to supply medical radioisotopes, and extends them to ${}^{225}$Ac production through the generator ${}^{229}$Th.

\begin{figure*}[t]
\centering
\includegraphics[width=\textwidth]{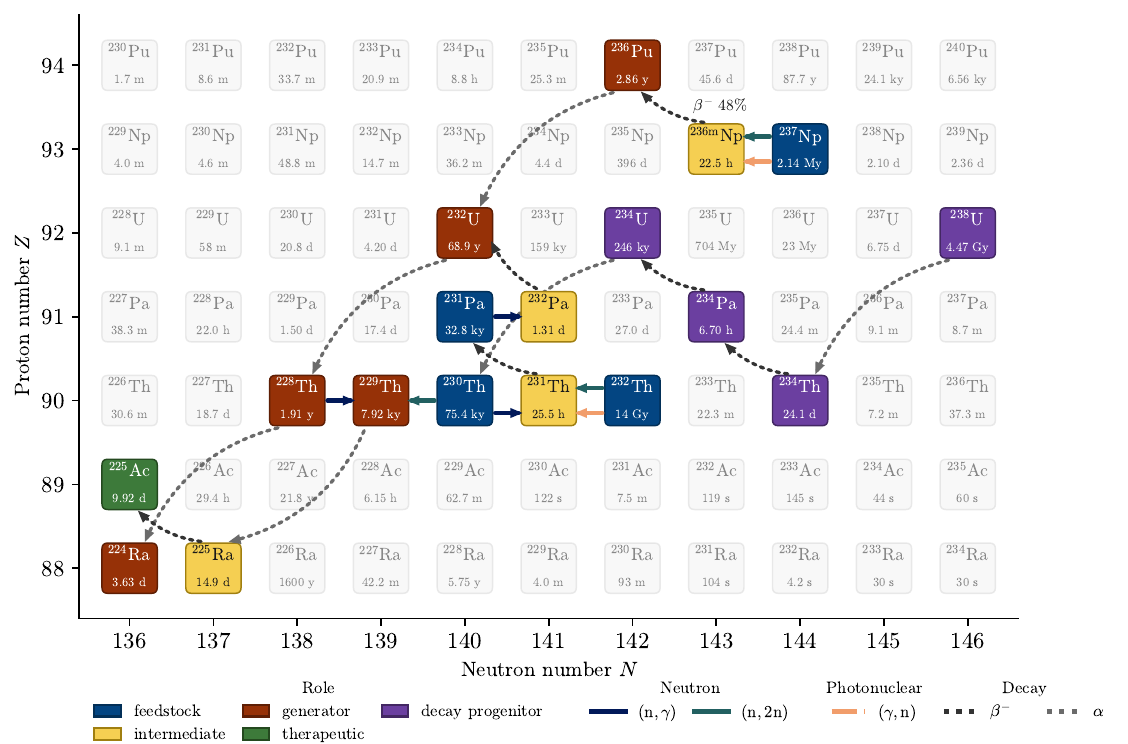}
\caption{Production pathways on the chart of nuclides.}
\label{fig:master_pathways}
\end{figure*}

In this work we propose several scalable fast-neutron-driven pathways to ${}^{229}$Th and ${}^{228}$Th. The first produces ${}^{229}$Th through ${}^{230}$Th(n,2n)${}^{229}$Th. Natural uranium decay continuously produces ${}^{230}$Th, so uranium-processing residues can contain thorium with a ${}^{230}$Th fraction of tens of percent. We show that several megawatt-years of D-T neutron irradiation of such material can produce enough ${}^{229}$Th for millions of annual ${}^{225}$Ac dose-equivalents.

We also propose using ${}^{231}$Pa made by fast-neutron irradiation of thorium to produce ${}^{232}$U via the thermal-neutron route ${}^{231}$Pa(n,$\gamma$)${}^{232}$Pa\,$\to$\,${}^{232}$U. At low neutron energies, capture dominates fission (\Cref{fig:pa231_xs}), so almost every absorption makes the desired product. The ${}^{231}$Pa can be bred from ${}^{232}$Th by ${}^{232}$Th(n,2n)${}^{231}$Th\,$\to$\,${}^{231}$Pa, in quantities of kilograms per megawatt year of D-T fusion power. In a high-flux fission reactor, the route breeds hundreds of grams of ${}^{232}$U per kilogram of ${}^{231}$Pa. The ${}^{228}$Th milked from the ${}^{232}$U feeds ${}^{212}$Pb generators. A second reactor irradiation converts it by ${}^{228}$Th(n,$\gamma$) into ${}^{229}$Th, the generator parent of ${}^{225}$Ac.

Based on the new production pathways in \cite{Parisi2026FusionBattery}, we also propose ${}^{236}$Pu as a ${}^{212}$Pb generator parent, made by ${}^{237}$Np(n,2n) or ${}^{237}$Np($\gamma$,n) with peak cross sections of a few hundred millibarns. The ${}^{237}$Np pathway has the advantage of directly producing ${}^{228}$Th with a single-step neutron/photonuclear reaction followed by decays, but suffers from limitations on ${}^{237}$Np material. The photon sources, rhodotrons and LINACs, are already used in industry and medicine~\cite{VanLancker1999,Starovoitova2014,hawkins2025cu}, and the D-T fusion neutron sources are being actively developed. Fusion-driven actinide blankets have already been proposed as waste burners~\cite{Freidberg2009,Stacey2008,Kuteev2015,Wu2006,NucNet2025xinghuo}, and the targets studied here are similar kinds of blankets. We show the decay chains of the two generator parents in \Cref{fig:decay}, and all routes on the chart of nuclides in \Cref{fig:master_pathways}.

The layout of this paper is as follows: we first discuss ${}^{228}$Th and ${}^{229}$Th generator yields of ${}^{212}$Pb and ${}^{225}$Ac in \Cref{sec:generator}. In \Cref{sec:th230} we show ${}^{229}$Th yields by the direct reaction ${}^{230}$Th(n,2n)${}^{229}$Th. In \Cref{sec:th230_thermal} we examine thermal-neutron production of ${}^{232}$U, and hence ${}^{228}$Th, through successive captures on ${}^{230}$Th and ${}^{231}$Pa. In \Cref{sec:pa231} we describe alternate pathways to ${}^{228}$Th and ${}^{229}$Th by following ${}^{232}$Th through ${}^{231}$Pa breeding, ${}^{231}$Pa irradiation to ${}^{232}$U and ${}^{228}$Th, and ${}^{228}$Th irradiation to ${}^{229}$Th. In \Cref{sec:np237_path} we describe the ${}^{237}$Np neutron and photonuclear routes to ${}^{236}$Pu, ${}^{232}$U, and ${}^{228}$Th. We conclude in \Cref{sec:discussion}.

\begin{figure*}[tb]
\centering
\includegraphics[width=0.8\textwidth]{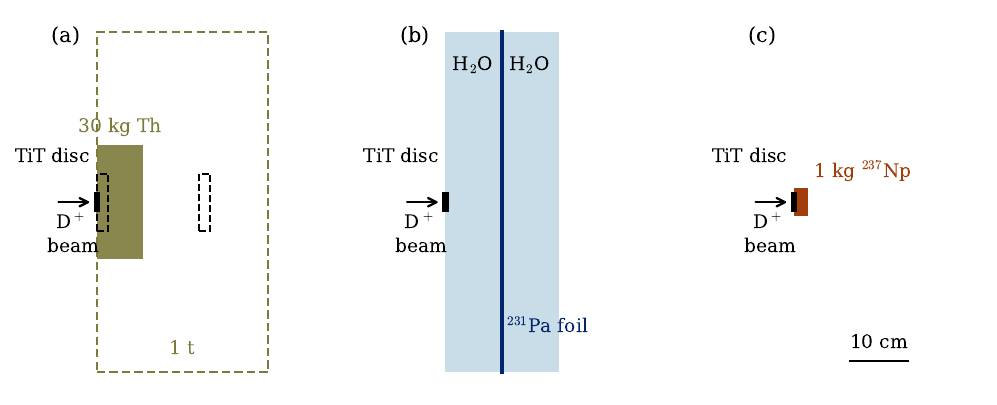}
\caption{Side view, to scale, of the three targets on the same 10 cm$^{2}$ TiT disc, which a deuteron beam strikes from the left: (a) the 30 kg thorium target, with the 1 t blanket of \Cref{sec:th230} outlined (\Cref{sec:pa231} uses 662 kg in the same arrangement), (b) a ${}^{231}$Pa foil between 10 cm layers of H$_2$O, (c) 1 kg ${}^{237}$Np. Dashed boxes in (a): the front and back regions used for tallying fluxes in OpenMC.}
\label{fig:layout}
\end{figure*}

\begin{figure}[tb]
\centering
\begin{subfigure}[t]{\columnwidth}
\centering
\includegraphics[width=0.95\linewidth]{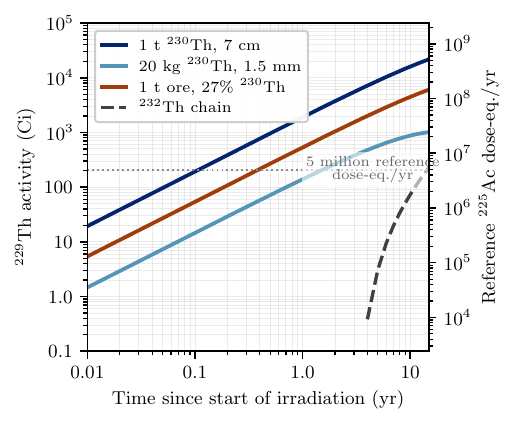}
\caption{${}^{229}$Th activity at 10 MW of D-T fusion power.}
\label{fig:th230_a}
\end{subfigure}

\begin{subfigure}[t]{\columnwidth}
\centering
\includegraphics[width=0.95\linewidth]{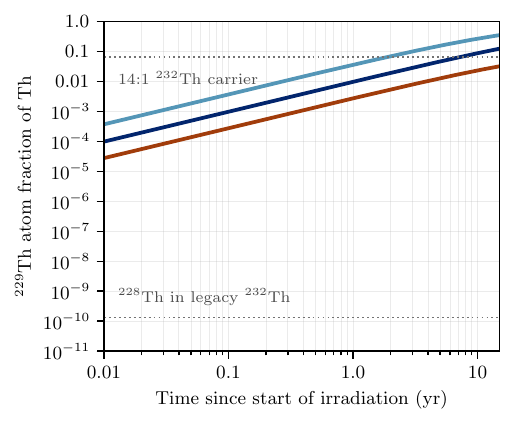}
\caption{Atom fraction of the thorium that is ${}^{229}$Th.}
\label{fig:th230_b}
\end{subfigure}
\caption{${}^{230}$Th(n,2n)${}^{229}$Th at 10 MW of D-T fusion power, compared with the ${}^{232}$Th chain of \Cref{sec:pa231}. Each target is a slab on the 1.2 m$^{2}$ source plane described in the text. ${}^{230}$Th depletion and ${}^{229}$Th burn-up are included.}
\label{fig:th230}
\end{figure}

\section{\texorpdfstring{${}^{228}$Th and ${}^{229}$Th generator yields}{Th-228 and Th-229 generator yields}}\label{sec:generator}

In this section, we first set the scale of the problem by calculating how many annual dose-equivalents $^{212}$Pb and $^{225}$Ac are produced by ${}^{228}$Th and ${}^{229}$Th decay. Useful quantities are summarized in \Cref{tab:doses_per_ci}, and in Appendix~\ref{app:pb212_doses} we derive the ${}^{212}$Pb conversion and show its dependence on operating schedule. We assume that one $^{212}$Pb administration is 2.7 mCi. Pure $^{228}$Th has a specific activity of \qty{820}{Ci\per\gram}. At its current activity, one Ci of $^{228}$Th produces $2.1 \cdot 10^5$ ideal annual dose-equivalents. Applying the 67\% reference recovery gives $1.4 \cdot 10^5$. Five million annual dose-equivalents therefore correspond to a working inventory of $\sim$ \qty{44}{mg} ($\sim$ \qty{36}{Ci}) of $^{228}$Th. The inventory required for a particular delivery system will depend on its collection interval and delivery delay.

$^{225}$Ac, the radionuclide chelated to a ligand, can be produced by $^{229}$Th decay. Note that $^{229}$Th is much longer lived than $^{228}$Th, so we require many more moles of $^{229}$Th to produce $^{225}$Ac at the same rate as $^{228}$Th to produce $^{212}$Pb. $^{225}$Ac administrations use much less activity than $^{212}$Pb administrations, and here we assume \qty{0.2}{mCi} of $^{225}$Ac per dose. Without losses, one curie of $^{229}$Th produces $1.3\cdot 10^5$ ideal annual dose-equivalents, while the ORNL operating benchmark is $2.4\cdot 10^4$ annual dose-equivalents per curie. Because of the long $^{229}$Th half-life, \qty{5}{g} corresponds to only \qty{1}{Ci}. Five million annual dose-equivalents therefore require \qty{208}{Ci} of $^{229}$Th, corresponding to \qty{1.04}{kg} of $^{229}$Th, far in excess of any existing inventory.

Compared with the $\sim$ \qty{44}{mg} (\qty{36}{Ci}) of $^{228}$Th needed to provide five million annual dose-equivalents, \qty{1.04}{kg} (\qty{208}{Ci}) of $^{229}$Th corresponds to $\sim25{,}000$ times as many moles. This requires much more transmutation capacity. Once produced, however, the ${}^{229}$Th can be retained on anion-exchange resin while Ra and Ac pass, and a guard column can return any parent breakthrough. Purified ${}^{225}$Ac has contained only $4\cdot10^{-5}$\% of its activity as ${}^{229}$Th~\cite{robertson2018ac225,Perron2020,morgenstern2018}. Radioactive decay alone leaves a 250 Ci stock above the 208 Ci reference level for millennia.

\section{\texorpdfstring{Making the ${}^{229}$Th generator by fast-neutron irradiation of ${}^{230}$Th}{Making the Th-229 generator by fast-neutron irradiation of Th-230}}\label{sec:th230}

In this section we show the fast pathway to stockpiling large quantities of ${}^{229}$Th. This is in contrast with ${}^{232}$Th routes we will discuss in \Cref{sec:pa231}, which reach ${}^{229}$Th much more slowly through ${}^{228}$Th milked from stored ${}^{232}$U (half-life 69 years). Fast neutrons on ${}^{230}$Th skip the ${}^{232}$U decay chain, making ${}^{229}$Th directly,
\begin{equation}
{}^{230}\mrm{Th}\,\ntn\,{}^{229}\mrm{Th},
\label{eq:th230_n2n}
\end{equation}
The evaluated nuclear cross section library ENDF/B-VIII.0 gives a cross section of 1.8 barn (b) at 14.1 MeV and a 2.1 b peak at 12 MeV above a 6.8 MeV threshold. A Department of Energy program irradiated 25 to 30 mg enriched ${}^{230}$Th targets with fast neutrons at Lawrence Berkeley National Laboratory~\cite{Molnar2019,Heilbronn2024}. These neutrons came from deuteron breakup at the 88-Inch Cyclotron, 8 to 40 MeV with fluxes up to $10^{12}$ n cm$^{-2}$ s$^{-1}$~\cite{Morrell2023}.

${}^{230}$Th is a relatively abundant feedstock in uranium ore, produced by ${}^{238}$U decay in uranium ore at 16.19 g per tonne of uranium~\cite{Figgins1966ionium,klm1971}, fifty times the abundance of ${}^{231}$Pa and ${}^{226}$Ra. The ore associated with one year of world uranium production contains about 950 kg of ${}^{230}$Th (\Cref{tab:th230_ore}). Identified uranium resources of 7.9 Mt contain 128 t of ${}^{230}$Th~\cite{RedBook2023}, up from the 81 t Kim and Born gave for the smaller reserves of 1968~\cite{klm1971}. However, only conventional uranium mills leave a residue to work from, because in-situ leach mining which forms roughly half of world production, leaves the thorium underground.

The isotopic mix of ${}^{230}$Th and ${}^{232}$Th varies hugely across uranium ores. Kim and Born measured eighteen uranium minerals (see \Cref{tab:ionium}), ranging from about 100\% ${}^{230}$Th in Joachimsthal pitchblende, whose thorium to uranium ratio is vanishingly small, down to 0.035\% $^{230}$Th in Swedish shale~\cite{klm1971}. Ionium is the historical name for ${}^{230}$Th. Ionium recovery campaigns have already been run in the United States, Canada, and Great Britain, all starting from uranium mill wastes~\cite{Figgins1966ionium}. The St.\ Louis Airport Residues, left by Mallinckrodt from Belgian Congo ore, held some 250 kg of ${}^{230}$Th at 11.6\% of the thorium, the richest of six sources above 1\% reported in~\cite{Figgins1966ionium}. Mallinckrodt, Argonne, and Mound recovered a thorium compound of 6 to 12\% ionium from that stockpile. In 1960, Rohrmann estimated that the mills he surveyed could recover over 40 kg of ionium per year at 70\% recovery. Extrapolating across mills with the preferred processes gave about 70 kg per year~\cite{Rohrman1960}. Canadian pitchblende residue reached 27\%, although only at milligram scale~\cite{klm1971}. The two Athabasca mines alone process 225 kg of ${}^{230}$Th a year, in thorium that is 18 to 33\% ${}^{230}$Th (\Cref{tab:th230_ore}).

We simulate fast-neutron irradiation of ${}^{230}$Th targets with OpenMC (see \Cref{tab:th230}) and report ${}^{229}$Th yields. A 1 t blanket makes 0.30 ${}^{229}$Th atoms per emitted neutron, whereas 1 t of the 27\% ${}^{230}$Th residue makes only 0.086 ${}^{229}$Th atoms per emitted neutron, and a smaller 30 kg target makes 0.19 ${}^{229}$Th atoms per emitted neutron. Each kilowatt-year of irradiation therefore accumulates 0.25, 0.072, and 0.16 Ci of ${}^{229}$Th, corresponding to 6100, 1700, and 3900 additional reference annual ${}^{225}$Ac dose-equivalents (\Cref{tab:doses_per_ci}).

For a megawatt-scale D-T fusion neutron source, the irradiated volume can be much larger, and hence much more $^{229}$Th can be produced. We represent a 10 MW device by an isotropic planar source distributed over 1.2 m$^{2}$, with an emitted-neutron areal rate of $3\cdot10^{14}$ n cm$^{-2}$ s$^{-1}$. Half of the emitted neutrons enter the one-sided target. The target is then a thin slab over that source-plane area, and its thickness will give the $^{229}$Th yield. One tonne of pure ${}^{230}$Th with 7 cm thickness and 1.2 m$^{2}$ surface area makes 0.23 ${}^{229}$Th atoms per source neutron, and 20 kg at 1.5 mm thickness only 0.018 ${}^{229}$Th atoms per source neutron. One tonne of the 27\% residue contains 275 kg of ${}^{230}$Th and makes 0.065 ${}^{229}$Th atoms per source neutron (see \Cref{tab:th230}). Five million reference annual ${}^{225}$Ac dose-equivalents require 208 Ci of ${}^{229}$Th generator.

With 10 MW of fusion-neutron power, each of the three Th targets described above produces 208 Ci of ${}^{229}$Th after a relatively short irradiation: six weeks for the pure $^{230}$Th tonne, five months for the tonne containing 27\% ${}^{230}$Th, and 18 months for the 20 kg foil (\Cref{fig:th230}a). These irradiation times use the source condition above: the target-facing current density is $1.5\cdot10^{14}$ n cm$^{-2}$ s$^{-1}$ and has 3.4 MW/m$^{2}$ of 14.1 MeV neutron energy. Because this pathway is more likely to be limited by ${}^{230}$Th availability than by neutron supply, the relevant objective is to maximize ${}^{229}$Th production per kilogram of ${}^{230}$Th by operating at the highest practical neutron flux. After 10 yr, the foil yields 44 Ci of ${}^{229}$Th per kilogram, whereas the pure-tonne target yields 16 Ci/kg. For the same source strength and emitted-neutron areal rate, the one-tonne inventory forms a 7 cm-thick target that attenuates the neutrons. By attenuation, we mean that the neutron spectrum slows as it moves through material. It consequently produces more ${}^{229}$Th per source neutron but less per gram of ${}^{230}$Th than the foil. Twenty kilograms of ${}^{230}$Th, less than one tenth of the 250 kg once held in the St.\ Louis Airport Residues~\cite{Figgins1966ionium}, could therefore support foreseeable $^{225}$Ac demand for millennia on the radioactive-decay timescale. The ${}^{232}$Th chain of \Cref{sec:pa231} reaches only 233 Ci in 15 yr at the same power.

There have also been proposals of driving $^{230}$Th(n,2n)$^{229}$Th using the fast neutron spectrum in fast reactors~\cite{Iwahashi2022}. While fast reactors drive the same (n,2n) reaction, they do so far less efficiently (see \Cref{tab:th229_per_mwyr} for a summary). Per megawatt-year, Joyo makes $2.3\cdot10^{-5}$ Ci of ${}^{229}$Th from a 50 g capsule~\cite{Iwahashi2022}, whereas the same 50 g on our disc makes 19 Ci, a factor $8\cdot10^{5}$. Part of that factor is the neutron birth spectrum. For the unmoderated ${}^{235}$U thermal-fission Watt spectrum used here, 1.5\% of birth neutrons exceed the 6.8 MeV threshold for $^{230}$Th(n,2n)$^{229}$Th, so even with each source at the center of an infinite ${}^{230}$Th medium, where geometry favors neither, D-T makes 522 times more ${}^{229}$Th per megawatt-year of thermal power. However, the factor of 522 is a lower bound because an unmoderated birth spectrum in thorium is the hardest spectrum a fission source can offer. The remaining factor of $\sim$1500 enhancement is due to reactor geometry, because a fission reactor irradiation capsule intercepts a tiny fraction of the core's neutrons whereas a fusion blanket surrounds the source, absorbing a large fraction, if not all the fusion-produced neutrons.

Thorium isotopes cannot typically be separated chemically (with some exceptions such ${}^{232}$Th decay to ${}^{228}$Th via ${}^{228}$Ra), so the product is ${}^{229}$Th diluted in ${}^{230}$Th, and only the daughter is milked. This is likely not an obstacle in practice. ${}^{228}$Ra in legacy ${}^{232}$Th is milked at an atom fraction of $1.3\cdot10^{-10}$, whereas the pure tonne reaches $10^{-2}$ in a year and the 20 kg foil $3.6\cdot10^{-2}$ (\Cref{fig:th230}b). Under the specified 10 MW D-T neutron source condition, about 1\% of the ${}^{230}$Th converts into ${}^{229}$Th each year. Because ${}^{229}$Th has a 7916 yr half-life, it accumulates during continued irradiation.

Thorium fission can produce power densities comparable to those in fission reactor fuel rods. Under 10 MW fusion-neutron irradiation, the tonne is at 132 MW/m$^{3}$ of fission heat and the 20 kg foil is at 367 MW/m$^{3}$. These power densities correspond to approximately 4.8 MW/m$^{2}$ per face for the 7 cm target and 0.27 MW/m$^{2}$ per face for the 1.5 mm foil, so the target designs require substantial cooling; however, since heating is volumetric not all heat must be removed through the surface. Irradiating ${}^{230}$Th also gives co-products. Fission is the largest, taking 26\% of the ${}^{230}$Th destroyed, or 0.43 kg for every kilogram of ${}^{229}$Th made, and the fission products stay in the target. In the pure ${}^{230}$Th tonne, ${}^{230}$Th(n,$\gamma$) breeds 0.24 kg of ${}^{231}$Pa per megawatt-year~\cite{klm1971}, and ${}^{230}$Th decay produces 8.9 Ci of ${}^{226}$Ra per year~\cite{Nagatsu2022}.

The ${}^{230}$Th(n,2n) cross section at 14.1 MeV is evaluated rather than measured experimentally. An experimental measurement is required to validate the calculated yields.

\begin{figure}[tb]
\centering
\includegraphics[width=\columnwidth]{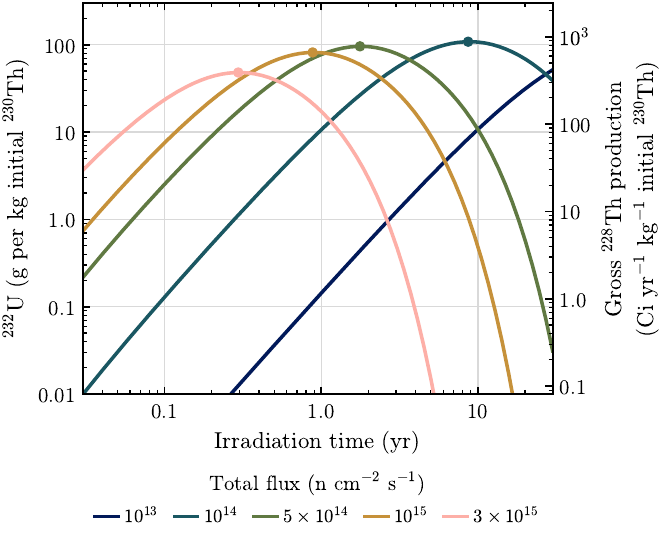}
\caption{${}^{232}$U produced per kilogram of initial ${}^{230}$Th during thermal-neutron irradiation of 10 at.\% ${}^{230}$Th pellets in the heavy-water position. The curves include the OpenMC neutron spectra and a 30-day post-irradiation cooling period. Circles show maxima. The right axis gives the corresponding gross ${}^{228}$Th production rate before decay during collection and chemical-recovery losses.}
\label{fig:th230_thermal_u232}
\end{figure}

\begin{table}[tb]
\centering
\caption{${}^{230}$Th targets from OpenMC. For the 1 kW case, the source is the 10 cm$^{2}$ disc shown in \Cref{fig:layout}. For the 10 MW case, an isotropic planar neutron source covers 1.2 m$^{2}$ at an emitted-neutron areal rate of $3\cdot10^{14}$ n cm$^{-2}$ s$^{-1}$, with half of the neutrons entering the target. The annual ${}^{229}$Th column gives the initial production during one full-power year, expressed as the activity of the produced atoms (it is calculated from the per-neutron yield and does not include target depletion or ${}^{229}$Th burn-up).}
\label{tab:th230}
\footnotesize
\setlength{\tabcolsep}{3pt}
\begin{tabular}{lccccc}
\toprule
Target & Depth & ${}^{230}$Th & ${}^{229}$Th & ${}^{229}$Th/yr & Heat \\
 & (cm) & (kg) & per n & (Ci) & (MW/m$^{3}$) \\[-1pt]
\midrule
\multicolumn{6}{l}{\emph{1 kW, 10 cm$^{2}$ disc}} \\
Blanket, pure & 30 & 999 & 0.303 & 0.255 & 0.02 \\
Blanket, 27\% & 30 & 275 & 0.086 & 0.072 & 0.006 \\
Compact & 8 & 30 & 0.193 & 0.162 & 0.35 \\
\midrule
\multicolumn{6}{l}{\emph{10 MW, 1.2 m$^{2}$ wall}} \\
Slab, pure & 0.15 & 20 & 0.018 & 147 & 367 \\
Slab, pure & 7 & 1000 & 0.231 & $1.94\cdot10^{3}$ & 132 \\
Slab, pure & 22 & 3000 & 0.304 & $2.55\cdot10^{3}$ & 68 \\
Slab, 27\% & 7 & 275 & 0.065 & 542 & 36 \\
\bottomrule
\end{tabular}
\end{table}

\begin{table}[tb]
\centering
\caption{${}^{229}$Th made per unit of source power. Joyo makes 14 MBq of ${}^{229}$Th per 60 day cycle from 50 g of ${}^{230}$Th at 100 MW of fission power~\cite{Iwahashi2022}. The fusion 50 g row is the same mass shaped to the disc, and gives 2.2 Ci per MW\,yr. The D-T power column is where each target sees the same neutron flux.}
\label{tab:th229_per_mwyr}
\footnotesize
\setlength{\tabcolsep}{3pt}
\begin{tabular}{llcc}
\toprule
Source & ${}^{230}$Th target & D-T power & ${}^{229}$Th (Ci/MW\,yr) \\
\midrule
D-T wall & 1 t slab, 7 cm & 10 MW & 194 \\
D-T wall & 20 kg foil, 1.5 mm & 10 MW & 15 \\
D-T disc & 50 g capsule & 8.4 kW & 19 \\
Joyo core & 50 g capsule & & $2.3\cdot10^{-5}$ \\
\bottomrule
\end{tabular}
\end{table}

\begin{table}[tb]
\centering
\caption{Estimated ${}^{230}$Th processed each year by the largest uranium operations, assuming 16.19 g of ${}^{230}$Th per tonne of uranium and 2024 uranium production figures. The Th values are estimated from host-rock thorium and are uncertain by a factor of two. In-situ leach mining leaves the thorium underground, so it yields no thorium.}
\label{tab:th230_ore}
\footnotesize
\setlength{\tabcolsep}{3.5pt}
\begin{tabular}{llcccc}
\toprule
Operation & Method & U & Th & ${}^{230}$Th & ${}^{230}$Th \\
 & & (\% ) & (ppm) & of Th & (kg/yr) \\
\midrule
Cigar Lake & underground & 15 & 5 & 33\% & 102 \\
McArthur River & underground & 6.7 & 5 & 18\% & 123 \\
Husab & open pit & 0.05 & 50 & 0.016\% & 70 \\
Olympic Dam & by-product & 0.03 & 50 & 0.010\% & 54 \\
\midrule
Kazakhstan, all & ISL & 0.05 & 10 & 0.08\% & (363) \\
\midrule
World total & & & & & 951 \\
\bottomrule
\end{tabular}
\end{table}

\begin{table*}[tb]
\centering
\caption{Measured ionium (${}^{230}$Th) content of uranium minerals, from Kim and Born~\cite{klm1971}. High-grade pitchblende contains thorium that is tens of percent ${}^{230}$Th, low-grade ore almost pure ${}^{232}$Th.}
\label{tab:ionium}
\footnotesize
\setlength{\tabcolsep}{4pt}
\begin{tabular}{llccc}
\toprule
Mineral & Source & U & Th/U & ${}^{230}$Th of Th \\
 & & (\%) & (ppm) & (\%) \\
\midrule
Pitchblende & Joachimsthal, Czechoslovakia & 45.7 & $\sim$0 & $\sim$100 \\
Pitchblende & Cinch Lake, Canada & & 14 & 57 \\
Pitchblende & Eldorado Mine, Great Bear Lake, Canada & 52 & 25 to 29 & 38 to 43 \\
Pitchblende residue & Eldorado Mine, Great Bear Lake, Canada & & & 28 \\
Pitchblende & Great Bear Lake region, Canada & & 51 & 25 to 26 \\
Pitchblende & Katanga, Congo & 75 & 139 to 178 & 9.2 to 12 \\
Pitchblende & Kirk Mine, Gilpin County, Colorado & 39 & 47 & 3.5 to 3.7 \\
Uraninite & Wilberforce, Canada & 60 & 969 & 1.9 \\
Carnotite & Colorado and Utah region & & 512 & 3.5 \\
Uraninite & Germany & 0.72 & 844 & 2.2 \\
Carnotite & U.S.A. & 0.36 & 956 & 1.8 \\
Uraninite & Australia & 0.32 & 1210 & 1.3 \\
Torbernite & Australia & 0.41 & 2250 & 0.73 \\
Torbernite & Spain & 0.26 & 3890 & 0.42 \\
Ore & Ellweiler, Germany & 0.25 to 3.0 & & 0.40 \\
Torbernite & Spain & 0.049 & 18900 & 0.08 \\
Samarskite & Mozambique & 6.2 & 23800 & 0.077 \\
Ore & Ranstad, Sweden & 0.030 & 42300 & 0.035 \\
\bottomrule
\end{tabular}
\end{table*}

\section{\texorpdfstring{Making the ${}^{228}$Th generator through ${}^{232}$U by neutron capture on ${}^{230}$Th}{Making the Th-228 generator through U-232 by neutron capture on Th-230}}\label{sec:th230_thermal}

Thermal-neutron irradiation provides an alternative route from the same ionium feedstock to the ${}^{212}$Pb generator. Two neutron captures and two beta decays produce ${}^{232}$U through ${}^{230}$Th(n,$\gamma$)${}^{231}$Th\,$\to$\,${}^{231}$Pa(n,$\gamma$)${}^{232}$Pa\,$\to$\,${}^{232}$U. The ${}^{232}$U then alpha decays to ${}^{228}$Th. Hanford proposed this route for ionium recovered from mill residues~\cite{hw63239,hw66600}, and Mound later operated a pilot plant that recovered both ${}^{231}$Pa and ${}^{230}$Th from Cotter concentrate~\cite{mlm2985}. Both capture reactions have large cross sections at low neutron energies (\Cref{fig:pa231_xs}). For 10 at.\% ${}^{230}$Th ionium pellets in a heavy-water irradiation position, one kilogram of initial ${}^{230}$Th produces 80 to 110 g of ${}^{232}$U (\Cref{fig:th230_thermal_u232}).

We calculate that the $^{232}$U inventory peaks after 8.7 yr at a total neutron flux of $10^{14}$ n cm$^{-2}$ s$^{-1}$ and after 0.88 yr at $10^{15}$ n cm$^{-2}$ s$^{-1}$. Once separated and stored, this ${}^{232}$U initially produces roughly 650 to 900 Ci of new ${}^{228}$Th inventory per year per kilogram of irradiated ${}^{230}$Th, before collection and chemical-recovery losses. Irradiation must end near the ${}^{232}$U peak because neutron reactions destroy ${}^{232}$U up to 371 times faster than radioactive decay does. Thus, fast-neutron irradiation of ionium produces the ${}^{225}$Ac parent in one reaction, whereas thermal-neutron irradiation produces the ${}^{212}$Pb parent through a four-step chain. At the calculated peak yield, irradiating only 40 to 60 g of ${}^{230}$Th in a fission reactor produces roughly 3 to 7 g of ${}^{232}$U, whose initial gross ${}^{228}$Th production is about 26 to 53 Ci/yr. A 36 Ci ${}^{228}$Th working inventory supports five million annual ${}^{212}$Pb dose-equivalents at the reference rate in \Cref{tab:doses_per_ci} and requires about 13 Ci/yr of replacement to offset radioactive decay. The calculated ${}^{232}$U stock can remain above that replacement rate for roughly 50 to 100 yr after allowing for recovery losses, although the collection schedule must be demonstrated experimentally. As a scale comparison, Daily and McDuffee modeled an optimized HFIR production scenario that required approximately 21 kg of ${}^{237}$Np target throughput per year to produce 1.05 to 1.32 kg of ${}^{238}$Pu~\cite{Daily2020Pu238}. The proposed 40 to 60 g ${}^{230}$Th irradiation is therefore very modest by high-flux reactor isotope-production standards.

\section{\texorpdfstring{Making ${}^{228}$Th and ${}^{229}$Th generators from ${}^{232}$Th}{Making Th-228 and Th-229 generators from Th-232}}\label{sec:pa231}

In this section we follow one feedstock, ${}^{232}$Th, through the production chains of \Cref{fig:master_pathways}, all the way to generators ${}^{228}$Th and ${}^{229}$Th.

Production starts with fast-neutron irradiation of ${}^{232}$Th to produce ${}^{231}$Pa via (n,2n) reactions. Thermal neutrons then convert the ${}^{231}$Pa to ${}^{232}$U via (n,$\gamma$) reactions. The ${}^{232}$U subsequently decays to the ${}^{212}$Pb generator ${}^{228}$Th. As a final option, thermal neutrons convert part of the ${}^{228}$Th to the ${}^{225}$Ac generator ${}^{229}$Th.

\subsection{\texorpdfstring{Making ${}^{231}$Pa by fast-neutron irradiation of ${}^{232}$Th}{Making Pa-231 by fast-neutron irradiation of Th-232}}
\label{sec:pa231_two_stage}

The first step to ${}^{228}$Th and ${}^{229}$Th requires producing ${}^{231}$Pa by fast-neutron reactions on ${}^{232}$Th.

${}^{231}$Pa is scarce: the world's declared separated stock is the \qty{125}{g} that the UK Atomic Energy Authority recovered in 1961~\cite{CEN1961protactinium,Collins1962protactinium,kirby2006protactinium}. Fast neutrons breed it from abundant ${}^{232}$Th, via the reaction
\begin{equation}
{}^{232}\mathrm{Th}\,\ntn\,{}^{231}\mathrm{Th}\xrightarrow{\betam,\,25.5\,\mathrm{h}}{}^{231}\mathrm{Pa},
\label{eq:stage1}
\end{equation}
with a \qty{6.4}{MeV} threshold and \qty{1.5}{b} cross section at \qty{14.1}{MeV}.
The neutron-multiplying (n,2n) reaction can also help tritium self-sufficiency in a fusion power plant.

\begin{figure*}[t]
\centering
\includegraphics[width=\textwidth]{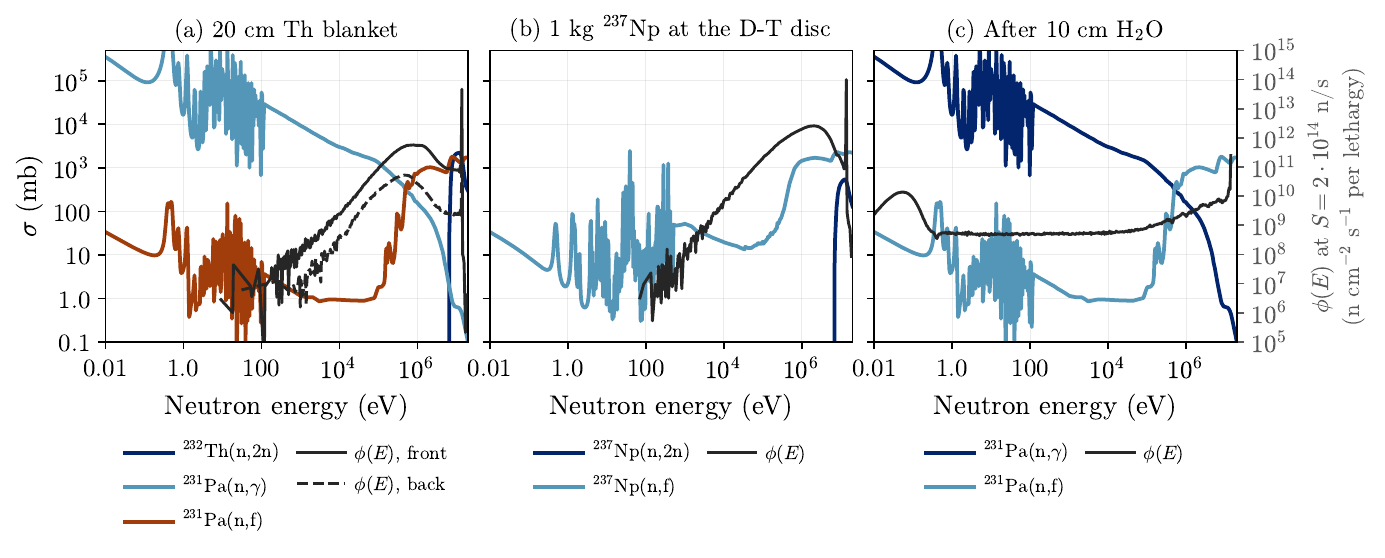}
\caption{Cross sections (mb) and OpenMC flux spectra in the geometry of \Cref{fig:layout}: (a) inside the 20 cm Th blanket behind the D-T disc, (b) inside the 1 kg ${}^{237}$Np sample at the disc, and (c) behind 10 cm of H$_2$O. We integrate each cross section over its overlaid OpenMC spectrum to calculate the reaction rates used for the ${}^{228}$Th yields in \Cref{fig:flux_mass_scan}. Grey curves (right axis) are the flux per unit lethargy at $S=2\cdot10^{14}$ n/s.}
\label{fig:flux_mass_spectra}
\end{figure*}

\begin{figure}[tb]
\centering
\begin{subfigure}[t]{\columnwidth}
\centering
\includegraphics[width=0.95\linewidth]{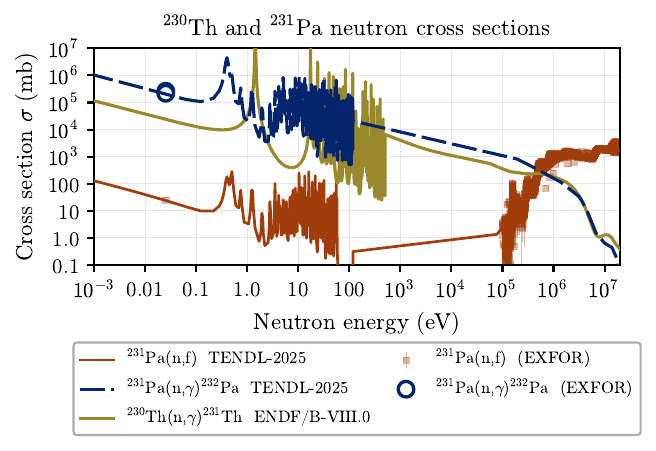}
\caption{${}^{230}$Th capture from ENDF/B-VIII.0 and ${}^{231}$Pa capture and fission from TENDL-2025, with EXFOR measurements for ${}^{231}$Pa capture (circles) and fission (squares).}
\label{fig:pa231_xs}
\end{subfigure}
\begin{subfigure}[t]{\columnwidth}
\centering
\includegraphics[width=0.95\linewidth]{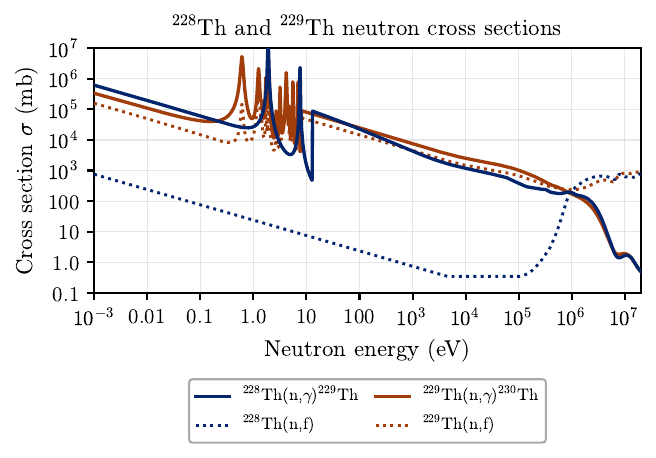}
\caption{${}^{228}$Th and ${}^{229}$Th, ENDF/B-VIII.0: capture (solid) and fission (dotted).}
\label{fig:th228_th229_xs}
\end{subfigure}
\caption{Neutron cross sections of the thermal capture steps.}
\label{fig:xs_thermal}
\end{figure}

We calculate the breeding of ${}^{231}$Pa by simulating neutron irradiation of a thorium blanket using OpenMC~\cite{openmc}: \qty{20}{cm} of thorium metal (\qty{662}{kg}) behind a \qty{10}{\cm^{2}} D-T disc emitting \qty{2e14}{n/s} isotropically, half of it into the blanket (\Cref{fig:layout}a). This corresponds to a fast neutron flux of $\phi \simeq \qty{e13}{n.cm^{-2}.s^{-1}}$, a value already achieved by the Rotating Target Neutron Source-II~\cite{davis1986rtns}. The blanket makes 0.26 ${}^{231}$Pa atoms per emitted neutron: \qty{0.62}{g} per year, or \qty{1.1}{g} per year per kilowatt of D-T fusion power. The flux falls from \qty{3.5e12}{n.cm^{-2}.s^{-1}} (59\% above the \qty{6.4}{MeV} threshold) in the front tally slab to \qty{1.2e11}{n.cm^{-2}.s^{-1}} (8\%) in the back slab (\Cref{fig:flux_mass_spectra}a), and half of all (n,2n) reactions occur within \qty{10}{cm} of the axis and of the disc.
We assume yearly Pa/Th/U separation~\cite{Radchenko2016} with 95\% recovery, since ${}^{231}$Pa left in the blanket is lost to fast fission (\Cref{fig:pa231_xs}). Protactinium chemistry is difficult: it hydrolyses and adsorbs on surfaces, and the UKAEA campaign recovered 63\% of the ${}^{231}$Pa from uranium~\cite{Collins1962protactinium}. A full fusion-plant blanket breeds \qty{2}{kg} of ${}^{231}$Pa per megawatt-year~\cite{Parisi2026FusionBattery}.

\begin{figure}[tb]
\centering
\includegraphics[width=0.95\columnwidth]{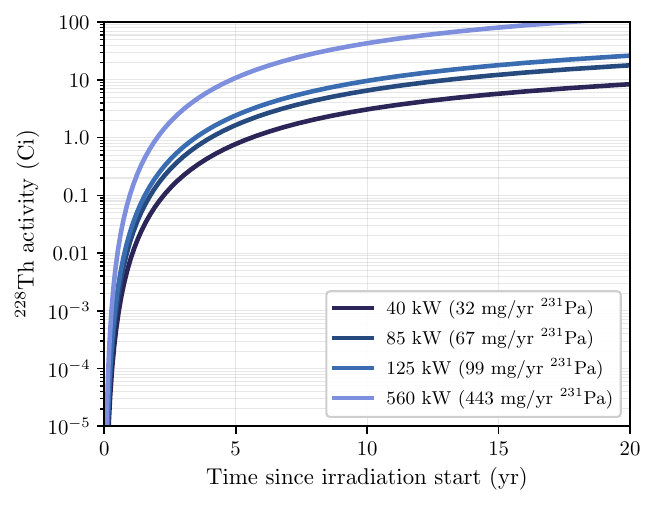}
\caption{${}^{228}$Th activity from ${}^{232}$Th($\gamma$,n) on a Rhodotron: 22.5 g of thorium, ${}^{231}$Pa extracted yearly to a reactor at $\phi_{\rm th}=\qty{e14}{n.cm^{-2}.s^{-1}}$, ${}^{232}$U extracted yearly. Rescaled from the ${}^{67}$Cu benchmark~\cite{hawkins2025cu} as in \ref{app:rhodotron_prod_rates}.}
\label{fig:rhodotron_th}
\end{figure}

Photons breed ${}^{231}$Pa as well, by ${}^{232}$Th($\gamma$,n)${}^{231}$Th~\cite{Caldwell1980}. Using the Rhodotron benchmark that we derive in Appendix~\ref{app:rhodotron_prod_rates}, \qty{22.5}{g} of thorium gives \qty{0.79}{mg} of ${}^{231}$Pa per year per kW, three times the ${}^{236}$Pu from the same mass of ${}^{237}$Np. With the reactor stage of \Cref{sec:pa231_irrad} this gives \qty{6.6}{Ci} of ${}^{228}$Th after \qty{10}{yr} at \qty{85}{kW} (\Cref{fig:rhodotron_th}), uncertain by a factor of two through the ${}^{68}$Zn($\gamma$,p) cross section uncertainty.

\subsection{\texorpdfstring{Making the ${}^{228}$Th generator through ${}^{232}$U by neutron capture on ${}^{231}$Pa}{Making the Th-228 generator through U-232 by neutron capture on Pa-231}}
\label{sec:pa231_irrad}

The second step is producing ${}^{232}$U from ${}^{231}$Pa. Kim and Born describe this (n,$\gamma$) route to ${}^{231}$Pa and ${}^{232}$U, and describe the reactor conditions that optimize it~\cite{klm1971}. Thermal neutrons convert ${}^{231}$Pa to ${}^{232}$U via neutron capture,
\begin{equation}
{}^{231}\mrm{Pa}\ngamma\,{}^{232}\mrm{Pa}\xrightarrow{\betam,\,1.31\,\mrm{d}}{}^{232}\mrm{U}.
\label{eq:path_pa231}
\end{equation}
At low neutron energies, ${}^{231}$Pa capture dominates fission~\cite{Mughabghab2018,Aleksandrov1972,Yurova1984,klm1971} (\Cref{fig:pa231_xs}). ${}^{232}$U decays with a 68.9 yr half-life to ${}^{228}$Th (\Cref{fig:decay_pu236}).
Each gram of stored ${}^{232}$U gives \qty{9.9}{mg} of ${}^{228}$Th per year, and each gram of ${}^{228}$Th (\qty{820}{Ci}) supports $1.1\cdot10^{8}$ reference ${}^{212}$Pb dose-equivalents per year (\Cref{tab:doses_per_ci}).

\begin{figure*}[t]
\centering
\includegraphics[width=\textwidth]{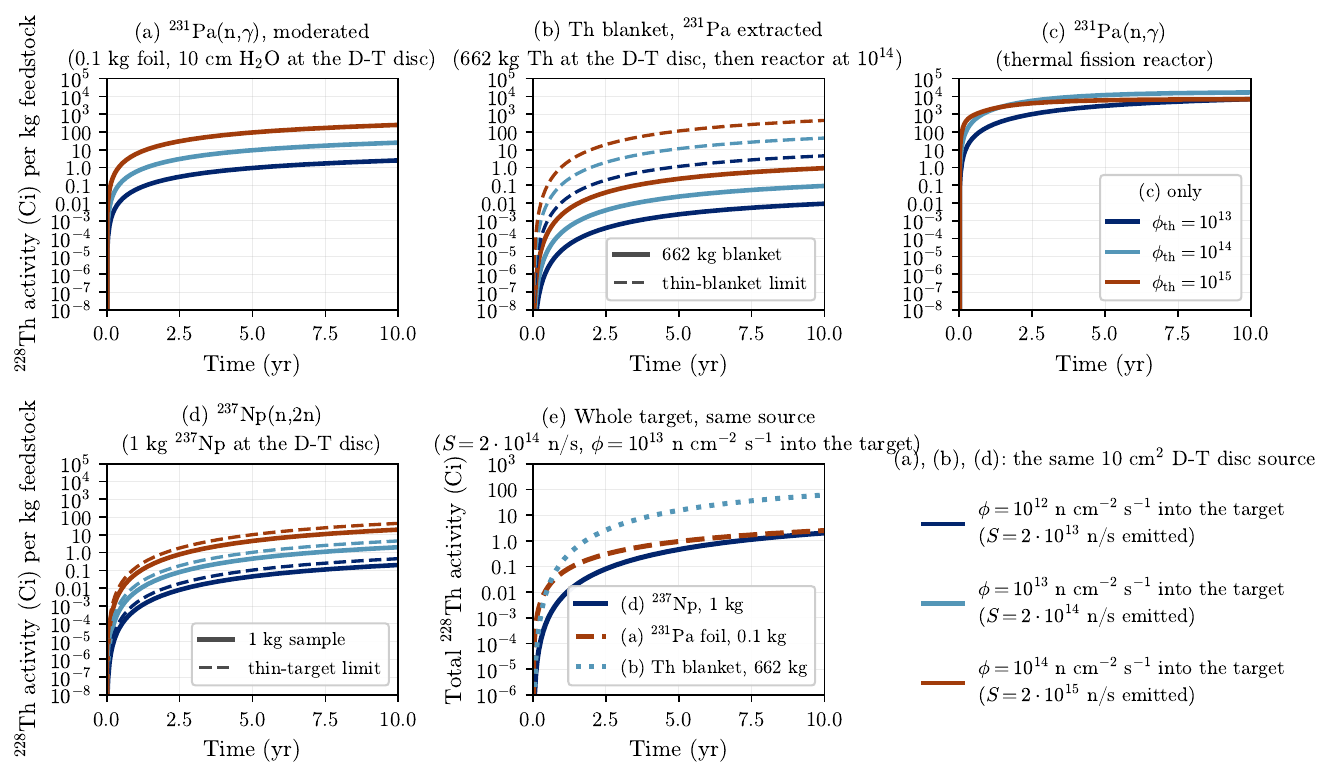}
\caption{${}^{228}$Th activity per kg of feedstock (target plus extracted cow) for the same D-T disc in (a), (b), and (d): $S=2\cdot10^{13}$ to $2\cdot10^{15}$ n/s emitted, half into the target, i.e.\ $\phi=10^{12}$ to $10^{14}$ n cm$^{-2}$ s$^{-1}$ (\Cref{fig:layout}). (a) \qty{100}{g} of pure ${}^{231}$Pa metal behind \qty{10}{cm} of \ce{H2O}. (b) the \qty{662}{kg} thorium blanket with yearly ${}^{231}$Pa extraction to a reactor at $\phi_{\rm th}=\qty{e14}{n.cm^{-2}.s^{-1}}$, dashed: thin-blanket limit. (c) ${}^{231}$Pa in a thermal reactor. (d) \qty{1}{kg} ${}^{237}$Np at the disc, dashed: thin-target limit. (e) the whole targets of (a), (b), and (d) at $S=\qty{2e14}{n/s}$. Reaction rates are from OpenMC. For the reactor stages, we use the model that we describe in Appendix~\ref{app:openmc_reactor}.}
\label{fig:flux_mass_scan}
\end{figure*}

We consider two thermal neutron sources.
The first is a D-T source with a \qty{10}{cm} \ce{H2O} moderator (\Cref{fig:flux_mass_spectra}c), where capture dominates fission.
\Cref{fig:flux_mass_scan}(a) shows the ${}^{228}$Th activity per kilogram for a \qty{100}{g} foil of pure ${}^{231}$Pa metal behind the water at the same disc: \qty{0.56}{Ci} after \qty{1}{yr} and \qty{25}{Ci} after \qty{10}{yr} at $S=\qty{2e14}{n.s^{-1}}$.

The second is a research reactor at $\phi_{\rm th}=\qtyrange{e13}{e15}{n.cm^{-2}.s^{-1}}$, where capture and fission both remove ${}^{232}$U.
OpenMC depletion of a 10 wt\% ${}^{231}$Pa cermet pin (see Appendix~\ref{app:openmc_reactor}, \Cref{fig:reactor}a for more details) gives a ${}^{232}$U peak of \qty{0.41}{g} per gram of ${}^{231}$Pa after 2.0 yr at $\phi_{\rm th}=\qty{e14}{n.cm^{-2}.s^{-1}}$, and \qty{0.33}{g} at $\phi_{\rm th}=\qty{e15}{n.cm^{-2}.s^{-1}}$, where ${}^{232}$Pa (\qty{1.31}{d}) absorbs neutrons before it decays. The uranium is 11\% ${}^{233}$U, but likely self-protected from a proliferation perspective by the ${}^{232}$U~\cite{KangVonHippel2001} due to the hard gamma from $^{208}$Tl decay.

Fission of the in-grown uranium heats the pin to \qty{120}{MW\per\cubic\meter} at the peak, the power density of light water reactor fuel, so significant cooling is required. The stored ${}^{232}$U therefore yields \qty{4.1}{g} (3.4 kCi) of ${}^{228}$Th per year per kilogram of ${}^{231}$Pa, equivalent to $4.7\cdot10^{5}$ reference ${}^{212}$Pb dose-equivalents per year per gram of ${}^{231}$Pa. This is nearly five times the $10^{5}$ doses per year planned for 2030~\cite{Zimmermann2024}. With the blanket of \Cref{sec:pa231_two_stage} as the ${}^{231}$Pa source (\Cref{fig:flux_mass_scan}b) the yield is \qty{9.2e-2}{Ci} per kilogram of thorium after \qty{10}{yr}, or \qty{61}{Ci} for the whole blanket, 30 times the \qty{1}{kg} ${}^{237}$Np sample (2.0 Ci) at the same source (\Cref{fig:flux_mass_scan}e). Per kilogram the thick blanket is the worst target in terms of $^{228}$Th per kilogram of feedstock (here, thorium), but thorium is cheap. The thin-blanket limit (dashed) reaches much higher values: \qty{46}{Ci} per kilogram. In the thin-target limit thorium gives ten times the ${}^{228}$Th per kilogram of ${}^{237}$Np (\qty{46}{Ci} versus \qty{4.6}{Ci}, dashed in \Cref{fig:flux_mass_scan}d), because at \qty{14.1}{MeV} ${}^{232}$Th(n,2n) is \qty{1.5}{b} versus \qty{0.4}{b} to ${}^{236\mathrm{m}}$Np, of which 48\% reaches ${}^{236}$Pu. Continuous reactor irradiation of separated ${}^{231}$Pa (\Cref{fig:flux_mass_scan}c) gives $1.7\cdot10^{4}$ Ci per kilogram of ${}^{231}$Pa after \qty{10}{yr} at $\phi_{\rm th}=10^{14}$ n cm$^{-2}$ s$^{-1}$.

\subsection{\texorpdfstring{Making the ${}^{229}$Th generator by neutron capture on ${}^{228}$Th}{Making the Th-229 generator by neutron capture on Th-228}}
\label{sec:reactor}

We now describe how to make $^{229}$Th from $^{232}$U decay products.

\begin{figure*}[tb]
\centering
\begin{subfigure}[t]{0.48\textwidth}
\centering
\includegraphics[width=\linewidth]{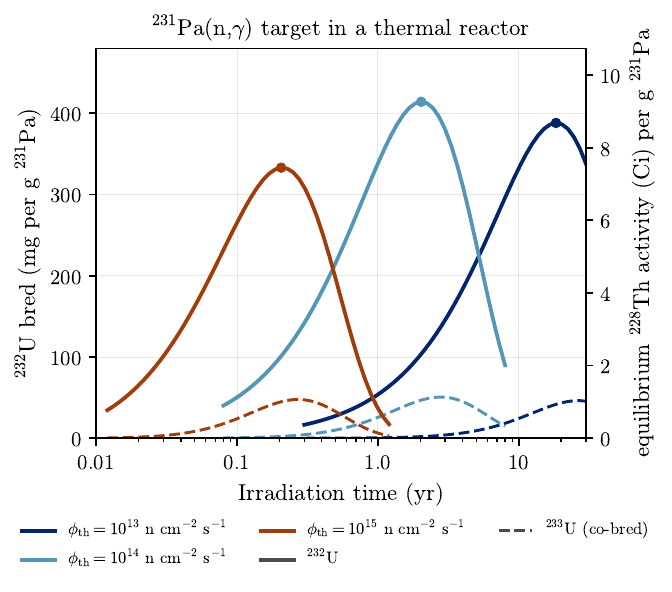}
\caption{${}^{232}$U per gram of ${}^{231}$Pa. Dashed: co-bred ${}^{233}$U.}
\label{fig:reactor_a}
\end{subfigure}\hfill
\begin{subfigure}[t]{0.48\textwidth}
\centering
\includegraphics[width=\linewidth]{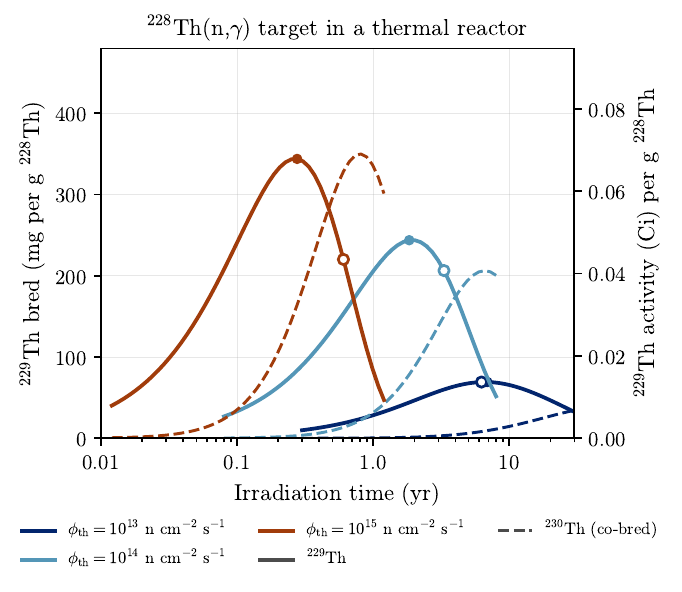}
\caption{${}^{229}$Th per gram of ${}^{228}$Th. Filled dots: maxima. Open dots: 10\% of the ${}^{228}$Th left. Dashed: co-bred ${}^{230}$Th.}
\label{fig:reactor_b}
\end{subfigure}
\caption{OpenMC depletion of 10 wt\% cermet pins in a water-moderated irradiation position (\ref{app:openmc_reactor}). $\phi_{\rm th}$ is the thermal flux next to the pin.}
\label{fig:reactor}
\end{figure*}

\begin{table*}[t]
\centering
\caption{${}^{228}$Th and ${}^{229}$Th yields from 1 kW and 10 MW of D-T fusion power. The ${}^{237}$Np chain continues through ${}^{236\mathrm{m}}$Np$\to{}^{236}$Pu$\to{}^{232}$U$\to{}^{228}$Th. ${}^{231}$Pa capture continues through ${}^{232}$Pa$\to{}^{232}$U$\to{}^{228}$Th. Converted rows finish with ${}^{228}$Th(n,$\gamma$)${}^{229}$Th. The direct ${}^{230}$Th row uses the 1 t target containing 27\% ${}^{230}$Th from \Cref{tab:th230}. Each ${}^{231}$Pa batch spends 2 yr in a reactor at $\phi_{\rm th}=\qty{e14}{n.cm^{-2}.s^{-1}}$ making 0.41 g ${}^{232}$U per gram. Kept-in-reactor rows hold each yearly ${}^{231}$Pa batch in core for 6 yr. Dose-equivalents from the reference values in \Cref{tab:doses_per_ci}. Activities are the inventory present at that time, not a yearly addition. The 1\,kW ${}^{231}$Pa row is the foil in \Cref{fig:layout}(b) at 1 kW, i.e.\ \qty{1.8e14}{n/s} into the target. Neutron fluxes have units of \unit{\cm^{-2} s^{-1}}.}
\label{tab:scenarios}
\footnotesize
\begin{tabular}{llcrrrrr}
\toprule
Source & Route & Product & 1 yr & 3 yr & 5 yr & 10 yr & 15 yr \\
\midrule
1 kW & ${}^{237}$Np(n,2n)${}^{236\mathrm{m}}$Np, 1 kg at $\phi=10^{13}$ & ${}^{228}$Th (Ci) & $1.5\cdot10^{-2}$ & 0.30 & 1.0 & 4.6 & 9.3 \\
 & & ${}^{212}$Pb dose-eq./yr & $2.0\cdot10^{3}$ & $4.1\cdot10^{4}$ & $1.5\cdot10^{5}$ & $6.4\cdot10^{5}$ & $1.3\cdot10^{6}$ \\
\addlinespace[1pt]
1 kW & ${}^{231}$Pa(n,$\gamma$)${}^{232}$Pa, 1 kg behind 10 cm H$_2$O & ${}^{228}$Th (Ci) & 1.0 & 7.3 & 17 & 45 & 73 \\
 & & ${}^{212}$Pb dose-eq./yr & $1.4\cdot10^{5}$ & $1.0\cdot10^{6}$ & $2.3\cdot10^{6}$ & $6.2\cdot10^{6}$ & $1.0\cdot10^{7}$ \\
\addlinespace[1pt]
1 kW & ${}^{232}$Th(n,2n)${}^{231}$Th$\to{}^{231}$Pa, reactor, ${}^{228}$Th kept & ${}^{228}$Th (Ci) & 0 & 3 & 21 & 96 & 178 \\
 & & ${}^{212}$Pb dose-eq./yr & 0 & $4.2\cdot10^{5}$ & $2.9\cdot10^{6}$ & $1.3\cdot10^{7}$ & $2.5\cdot10^{7}$ \\
\addlinespace[1pt]
\midrule
10 MW & ${}^{230}$Th(n,2n)${}^{229}$Th, 1 t with 27\% ${}^{230}$Th & ${}^{229}$Th (Ci) & 532 & $1.53\cdot10^{3}$ & $2.46\cdot10^{3}$ & $4.46\cdot10^{3}$ & $6.06\cdot10^{3}$ \\
 & & ${}^{225}$Ac dose-eq./yr & $1.3\cdot10^{7}$ & $3.7\cdot10^{7}$ & $5.9\cdot10^{7}$ & $1.1\cdot10^{8}$ & $1.5\cdot10^{8}$ \\
\addlinespace[1pt]
10 MW & ${}^{232}$Th(n,2n)${}^{231}$Th$\to{}^{231}$Pa, reactor, ${}^{228}$Th kept & ${}^{228}$Th (Ci) & 0 & $3.0\cdot10^{4}$ & $2.1\cdot10^{5}$ & $9.6\cdot10^{5}$ & $1.8\cdot10^{6}$ \\
 & & ${}^{212}$Pb dose-eq./yr & 0 & $4.2\cdot10^{9}$ & $2.9\cdot10^{10}$ & $1.3\cdot10^{11}$ & $2.5\cdot10^{11}$ \\
\addlinespace[1pt]
10 MW & ${}^{232}$Th(n,2n)${}^{231}$Th$\to{}^{231}$Pa, ${}^{228}$Th(n,$\gamma$)${}^{229}$Th & ${}^{229}$Th (Ci) & 0 & 0 & 3.07 & 73 & 233 \\
 & & ${}^{225}$Ac dose-eq./yr & 0 & 0 & $7.4\cdot10^{4}$ & $1.7\cdot10^{6}$ & $5.6\cdot10^{6}$ \\
\addlinespace[1pt]
10 MW & ${}^{232}$Th(n,2n)${}^{231}$Th$\to{}^{231}$Pa(n,$\gamma$)${}^{232}$Pa, kept in core & ${}^{229}$Th (Ci) & 0 & 4.9 & 25 & 95 & 166 \\
 & & ${}^{225}$Ac dose-eq./yr & 0 & $1.2\cdot10^{5}$ & $5.9\cdot10^{5}$ & $2.3\cdot10^{6}$ & $4.0\cdot10^{6}$ \\
\bottomrule
\end{tabular}
\end{table*}

\begin{table*}[t]
\centering
\caption{Accelerated yield schedules for 1 kW and 10 MW of D-T fusion power, with goal of ${}^{212}$Pb and ${}^{225}$Ac production in the first year. ${}^{231}$Pa is extracted from the blanket monthly and spends 0.2 yr in a reactor at $\phi_{\rm th}=\qty{e15}{n.cm^{-2}.s^{-1}}$ (0.33 g ${}^{232}$U per gram of ${}^{231}$Pa). Kept rows store the ${}^{232}$U and milk its ${}^{228}$Th. Converted rows: stored ${}^{232}$U milked monthly, and the ${}^{228}$Th converted at $\phi_{\rm th}=10^{15}$ for 0.1 yr (0.25 g ${}^{229}$Th per gram, 0.15 ${}^{230}$Th per ${}^{229}$Th). Neutron fluxes have units of \unit{\cm^{-2} s^{-1}}.}
\label{tab:scenarios_acc}
\footnotesize
\setlength{\tabcolsep}{1pt}
\begin{tabular}{llcrrrrrr}
\toprule
Source & Route & Product & 0.5 yr & 1 yr & 3 yr & 5 yr & 10 yr & 15 yr \\
\midrule
1 kW & ${}^{232}$Th(n,2n)${}^{231}$Th$\to{}^{231}$Pa, reactor, ${}^{228}$Th kept & ${}^{228}$Th (Ci) & 0.25 & 1.42 & 15 & 36 & 100 & 166 \\
 & & ${}^{212}$Pb dose-eq./yr & $3.6\cdot10^{4}$ & $2.0\cdot10^{5}$ & $2.0\cdot10^{6}$ & $5.0\cdot10^{6}$ & $1.4\cdot10^{7}$ & $2.3\cdot10^{7}$ \\
\midrule
10 MW & ${}^{232}$Th(n,2n)${}^{231}$Th$\to{}^{231}$Pa, reactor, ${}^{228}$Th kept & ${}^{228}$Th (Ci) & $2.5\cdot10^{3}$ & $1.4\cdot10^{4}$ & $1.5\cdot10^{5}$ & $3.6\cdot10^{5}$ & $1.0\cdot10^{6}$ & $1.7\cdot10^{6}$ \\
 & & ${}^{212}$Pb dose-eq./yr & $3.6\cdot10^{8}$ & $2.0\cdot10^{9}$ & $2.0\cdot10^{10}$ & $5.0\cdot10^{10}$ & $1.4\cdot10^{11}$ & $2.3\cdot10^{11}$ \\
\addlinespace[1pt]
10 MW & ${}^{232}$Th(n,2n)${}^{231}$Th$\to{}^{231}$Pa, ${}^{228}$Th(n,$\gamma$)${}^{229}$Th & ${}^{229}$Th (Ci) & 0.071 & 0.61 & 10 & 32 & 140 & 319 \\
 & & ${}^{225}$Ac dose-eq./yr & $1.7\cdot10^{3}$ & $1.5\cdot10^{4}$ & $2.5\cdot10^{5}$ & $7.7\cdot10^{5}$ & $3.3\cdot10^{6}$ & $7.6\cdot10^{6}$ \\
\addlinespace[1pt]
10 MW & ${}^{232}$Th(n,2n)${}^{231}$Th$\to{}^{231}$Pa(n,$\gamma$)${}^{232}$Pa, kept in core & ${}^{229}$Th (Ci) & 0.26 & 1.14 & 4.76 & 8.38 & 17 & 26 \\
 & & ${}^{225}$Ac dose-eq./yr & $6.3\cdot10^{3}$ & $2.7\cdot10^{4}$ & $1.1\cdot10^{5}$ & $2.0\cdot10^{5}$ & $4.2\cdot10^{5}$ & $6.4\cdot10^{5}$ \\
\addlinespace[1pt]
10 MW & ${}^{232}$Th(n,2n)${}^{231}$Th$\to{}^{231}$Pa(n,$\gamma$)${}^{232}$Pa, kept in core & ${}^{229}$Th (Ci) & 0.013 & 0.20 & 1.9 & 3.6 & 7.85 & 12 \\
 & & ${}^{225}$Ac dose-eq./yr & 302 & $4.8\cdot10^{3}$ & $4.6\cdot10^{4}$ & $8.6\cdot10^{4}$ & $1.9\cdot10^{5}$ & $2.9\cdot10^{5}$ \\
\bottomrule
\end{tabular}
\end{table*}

${}^{229}$Th ($\thalf=7916$ yr) is the parent of ${}^{225}$Ac through ${}^{225}$Ra (\Cref{fig:decay_th229}). The world supply is a few grams, separated from ${}^{233}$U~\cite{robertson2018ac225,morgenstern2018,isotek2025}. Thermal neutrons convert the milked ${}^{228}$Th into ${}^{229}$Th, via the reaction
\begin{equation}
{}^{228}\mrm{Th}\ngamma\,{}^{229}\mrm{Th}.
\label{eq:path_th228}
\end{equation}
At low neutron energies, ${}^{228}$Th capture dominates fission (\Cref{fig:th228_th229_xs}). Two losses limit the yield: ${}^{228}$Th decays (1.91 yr), and ${}^{229}$Th absorbs neutrons. Measurements constrain only part of the relevant ${}^{229}$Th reaction data~\cite{Molnar2019}, so the projected yield requires validation under the intended irradiation conditions. At $\phi=\qty{e13}{n.cm^{-2}.s^{-1}}$ only 1\% of ${}^{228}$Th converts before the ${}^{228}$Th decays, so this step needs a reactor with higher neutron flux.
The OpenMC result (\Cref{fig:reactor}b) peaks at \qtylist{69;244;344}{mg} of ${}^{229}$Th per gram of ${}^{228}$Th at neutron fluxes of $\phi_{\rm th}=\qtylist{e13;e14;e15}{cm^{-2}s^{-1}}$, respectively, after \qtylist{6.9;1.8;0.28}{yr}, respectively.

Stopping at the ${}^{229}$Th maximum leaves 27\% to 34\% of the loaded ${}^{228}$Th unconverted and undecayed. Thorium isotopes cannot be separated chemically, so this ${}^{228}$Th stays with the ${}^{229}$Th product.
At \qty{820}{Ci/g} versus \qty{0.2}{Ci/g}, it dominates the activity of the mixture until it decays away over about \qty{10}{yr}. However, because neither the ${}^{228}$Th nor the ${}^{230}$Th decay chain contains actinium, eluting Ac from the thorium mixture should give isotopically pure ${}^{225}$Ac.
Running on until 10\% of the loaded ${}^{228}$Th remains takes 3.3 yr at $\phi_{\rm th}=10^{14}$ n cm$^{-2}$ s$^{-1}$ and keeps 84\% of the maximum ${}^{229}$Th (open markers). The co-bred ${}^{230}$Th (dashed) dilutes the thorium.
Its ratio to ${}^{229}$Th depends on the fluence alone, so a shorter irradiation increases  ${}^{229}$Th purity: at $\phi_{\rm th}=10^{14}$ n cm$^{-2}$ s$^{-1}$, \qty{1}{yr} gives \qty{0.21}{g} of ${}^{229}$Th per gram with 0.16 ${}^{230}$Th per ${}^{229}$Th, versus 0.73 at \qty{3.3}{yr}.
After the ${}^{228}$Th decays the thorium is 86\% ${}^{229}$Th instead of 58\%.

The mixed cow yields ${}^{224}$Ra and ${}^{225}$Ra together. Lead elution gives ${}^{212}$Pb, while actinium separation gives ${}^{225}$Ac free of ${}^{227}$Ac. Converting the \qty{4.1}{g.yr^{-1}} of ${}^{228}$Th from \qty{1}{kg} of ${}^{231}$Pa gives \qty{1.0}{g.yr^{-1}} of ${}^{229}$Th. After a decade the \qty{10}{g} stock (\qty{2.0}{Ci}) supports $4.8\cdot10^{4}$ reference annual ${}^{225}$Ac dose-equivalents (\Cref{tab:doses_per_ci}), compared with a world supply of about $8\cdot10^{3}$~\cite{morgenstern2018,robertson2018ac225}.
The whole chain from ${}^{232}$Th to ${}^{225}$Ac uses only neutrons.

A faster option leaves the ${}^{231}$Pa in the reactor, where ${}^{228}$Th born from ${}^{232}$U decay converts into ${}^{229}$Th. At $\phi_{\rm th}=\qty{e14}{n.cm^{-2}.s^{-1}}$ this gives \qty{1.0}{g} of ${}^{229}$Th per kilogram of ${}^{231}$Pa after \qty{2}{yr} and a maximum of \qty{3.6}{g} after \qty{6}{yr}, with 0.7 ${}^{230}$Th per ${}^{229}$Th.
The cost is the ${}^{232}$U, which fissions in core: \qty{0.18}{g} per gram remains after \qty{6}{yr} instead of \qty{0.41}{g}.
We compare the two schedules in \Cref{tab:scenarios}. The accelerated schedules in \Cref{tab:scenarios_acc} produce ${}^{225}$Ac within the first year by extracting ${}^{231}$Pa monthly, irradiating it for 0.2 yr at $\phi_{\rm th}=\qty{e15}{n.cm^{-2}.s^{-1}}$, milking ${}^{228}$Th monthly, and converting it for \qty{0.1}{yr}.
At \qty{10}{MW} of fusion power this gives $1.5\cdot10^{4}$ reference annual ${}^{225}$Ac dose-equivalents after \qty{1}{yr} and $7.7\cdot10^{6}$ after \qty{15}{yr}, because the ${}^{232}$U forms in \qty{0.2}{yr} instead of 2 and each ${}^{228}$Th atom converts as soon as it is born. Keeping the ${}^{231}$Pa in a $10^{15}$ core is faster in the first year ($2.8\cdot10^{4}$ reference annual dose-equivalents) but saturates at $6.4\cdot10^{5}$, since the ${}^{232}$U fissions before it decays. Both need an irradiation position with a flux of $\phi_\mathrm{th} \sim 10^{15}$ , i.e., a high-flux reactor, and monthly extractions. \Cref{tab:time_to_doses} gives the years from the start of ${}^{232}$Th irradiation to each ${}^{225}$Ac reference annual dose-equivalent rate. \Cref{tab:time_to_doses_pb} gives the same for ${}^{212}$Pb.

\begin{table*}[t]
\centering
\caption{Years from the start of ${}^{232}$Th irradiation at 10 MW of D-T fusion power to a given ${}^{225}$Ac reference annual dose-equivalent rate, using $2.4\cdot10^{4}$ annual dose-equivalents per Ci of ${}^{229}$Th (\Cref{tab:doses_per_ci}). A rate of \num{5e6} annual dose-equivalents needs \qty{208}{Ci}, about \qty{1}{kg}. Schedules as in \Cref{tab:scenarios,tab:scenarios_acc}. Neutron fluxes have units of \unit{\cm^{-2} s^{-1}}.}
\label{tab:time_to_doses}
\small
\begin{tabular}{lrrrrr}
\toprule
Route (10 MW, from ${}^{232}$Th) & \multicolumn{5}{c}{Target reference annual dose-eq./yr} \\
\cmidrule(lr){2-6}
& $10^{4}$ & $10^{5}$ & $5\cdot10^{5}$ & $10^{6}$ & $5\cdot10^{6}$ \\
\midrule
Batch: 2 yr reactor, ${}^{228}$Th converted & 4.0 & 5.2 & 7.1 & 8.4 & 14.4 \\
${}^{231}$Pa kept in core at $10^{14}$, 6 yr & 1.5 & 2.8 & 4.7 & 6.2 & 17.9 \\
Accelerated: monthly ${}^{231}$Pa, $10^{15}$, converted & 0.9 & 2.0 & 4.1 & 5.6 & 12.2 \\
${}^{231}$Pa kept in core at $10^{15}$, 0.67 yr & 0.6 & 2.7 & 11.9 & 23.4 & $>40$ \\
${}^{231}$Pa kept in core at $10^{14}$, 1 yr & 1.3 & 5.7 & 25.3 & $>40$ & $>40$ \\
\bottomrule
\end{tabular}
\end{table*}

\begin{table*}[t]
\centering
\caption{Years from the start of irradiation to a given ${}^{212}$Pb reference annual dose-equivalent rate, using $1.4\cdot10^{5}$ annual dose-equivalents per Ci of ${}^{228}$Th (\Cref{tab:doses_per_ci}). A rate of \num{5e6} annual dose-equivalents needs \qty{36}{Ci} (\qty{44}{mg}). Routes follow \Cref{tab:scenarios,tab:scenarios_acc}. The ${}^{237}$Np and ${}^{231}$Pa rows start from those feedstocks, while the blanket rows start from ${}^{232}$Th. At \qty{10}{MW} the first reactor batch alone exceeds all five rates, so the time is the wait for that batch: \qty{2}{yr} in the reactor, or \qty{0.2}{yr} on the accelerated schedule. Neutron fluxes have units of \unit{\cm^{-2} s^{-1}}.}
\label{tab:time_to_doses_pb}
\small
\begin{tabular}{llrrrrr}
\toprule
Source & Route & \multicolumn{5}{c}{Target reference annual dose-eq./yr} \\
\cmidrule(lr){3-7}
& & $10^{4}$ & $10^{5}$ & $5\cdot10^{5}$ & $10^{6}$ & $5\cdot10^{6}$ \\
\midrule
1 kW & ${}^{237}$Np(n,2n), 1 kg thin target, $\phi=10^{13}$ & 1.8 & 4.3 & 8.8 & 12.8 & $>40$ \\
1 kW & ${}^{231}$Pa(n,$\gamma$), 1 kg foil behind 10 cm H$_2$O & 0.3 & 0.8 & 2.0 & 3.0 & 8.5 \\
1 kW & Th blanket, batch: 2 yr reactor, ${}^{228}$Th kept & 2.1 & 2.5 & 3.1 & 3.6 & 6.1 \\
1 kW & Th blanket, accelerated: monthly ${}^{231}$Pa, $10^{15}$ & 0.3 & 0.8 & 1.5 & 2.1 & 5.0 \\
\midrule
10 MW & Th blanket, batch: 2 yr reactor, ${}^{228}$Th kept & 2.0 & 2.0 & 2.0 & 2.0 & 2.0 \\
10 MW & Th blanket, accelerated: monthly ${}^{231}$Pa, $10^{15}$ & 0.1 & 0.1 & 0.1 & 0.1 & 0.1 \\
\bottomrule
\end{tabular}
\end{table*}

${}^{231}$Pa also feeds ${}^{230}$U and ${}^{226}$Th by (n,2n)~\cite{Parisi2026FusionBattery} and, by its own decay, ${}^{227}$Ac, the parent of ${}^{223}$Ra and ${}^{211}$Pb. ${}^{231}$Pa(n,2n)${}^{230}$Pa is also a pathway for generating $^{230}$U/$^{226}$Th, other potential Targeted Alpha Therapy candidates. We leave detailed discussion of these chains to future work.

In natural uranium, ${}^{231}$Pa is in equilibrium with ${}^{235}$U at 0.33 g per tonne of uranium. World mines produce about 59 kt of uranium per year, and identified resources below USD 260 per kg are 8 Mt~\cite{RedBook2023}. The ore milled each year therefore contains \qty{19}{kg} of ${}^{231}$Pa, and the resources \qty{2.6}{t}. ${}^{226}$Ra, the feedstock of the radium routes to ${}^{225}$Ac, is equally abundant at \qty{0.34}{g} per tonne, in equilibrium with ${}^{238}$U, and leaves the mill in the same residues. Today's ${}^{226}$Ra comes from recycled legacy sources, about \qty{2.5}{kg} ever extracted in total~\cite{Terrill1954,Nagatsu2022,ANS2026radium}, while mills discard both isotopes to tailings. Protactinium comes from mill residues rather than the uranium product, and the UKAEA effort in the 1960s recovered 63\% of it from such a residue~\cite{Collins1962protactinium}. That effort took 127 g of 99.9\% pure ${}^{231}$Pa from 60 t of siliceous sludge at 4 ppm~\cite{Morss2010}. The sludge had built up on the walls of the uranium ether-extraction plant, ten times richer in protactinium than the ore. At that recovery rate, mining gives \qty{12}{kg} of ${}^{231}$Pa per year, about 100 times the declared stock.
Each kilogram, irradiated as in \Cref{sec:pa231_irrad}, yields \qty{4.1}{g} of ${}^{228}$Th per year, or $4.7\cdot10^{8}$ reference annual ${}^{212}$Pb dose-equivalents at the rate of \Cref{tab:doses_per_ci}. One year of mined ${}^{231}$Pa thus exceeds five million annual ${}^{212}$Pb dose-equivalents by roughly three orders of magnitude.
For ${}^{225}$Ac the limit is the conversion step, not the ore.
Each \qty{12}{kg} of ${}^{231}$Pa gives \qty{12}{g} of ${}^{229}$Th per year (\qty{2.4}{Ci}, or $5.8\cdot10^{4}$ reference annual dose-equivalents), so with \qty{12}{kg} added each year the \qty{1}{kg} of ${}^{229}$Th for five million annual dose-equivalents takes about \qty{13}{yr} to accumulate. This could be achieved faster with higher neutron flux irradiation of $^{228}$Th in a reactor.

\section{\texorpdfstring{Making the ${}^{228}$Th generator from ${}^{237}$Np}{Making the Th-228 generator from Np-237}}\label{sec:np237_path}

\begin{figure}[b]
\centering
\includegraphics[width=\columnwidth]{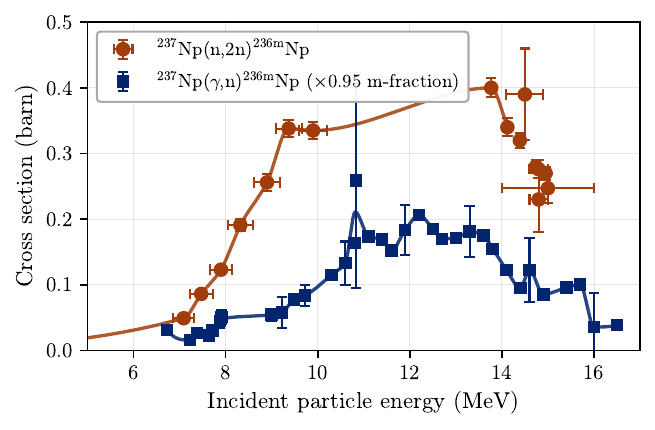}
\caption{${}^{237}$Np(n,2n) and ($\gamma$,n) cross sections to ${}^{236\mathrm{m}}$Np: EXFOR data~\cite{Perkin1961,Landrum1973,Lindeke1975,Veyssiere1973,Berman1986}, the ($\gamma$,n) points scaled by the 95\% isomer branch~\cite{Gardner1989}, and the splines used for the yields.}
\label{fig:XS}
\end{figure}

${}^{237}$Np offers a second path to ${}^{232}$U and ${}^{228}$Th, through the ${}^{236}$Pu decay chain (\Cref{fig:decay_pu236}). We consider two reactions that make ${}^{236}$Pu from ${}^{237}$Np. Fast neutrons drive
\begin{equation}
{}^{237} \mathrm{Np} (\mathrm{n,2n}) {}^{236 \mathrm{m} } \mathrm{Np} \to {}^{236} \mathrm{Pu}.
\label{eq:path1}
\end{equation}
The \qty{22.5}{h} isomer ${}^{236 \mathrm{m} }$Np decays to ${}^{236}$Pu with a 48\% branch~\cite{Chechev2006Np236}, the rest by electron capture to ${}^{236}$U (\Cref{fig:decay_pu236}). The second route,
\begin{equation}
{}^{237} \mathrm{Np} (\gamma,\mathrm{n}) {}^{236 \mathrm{m} } \mathrm{Np} \to {}^{236} \mathrm{Pu},
\label{eq:path2}
\end{equation}
uses photons, again relying on the \qty{22.5}{h} isomer ${}^{236 \mathrm{m} }$Np. Both cross sections are measured (\Cref{fig:XS}). The (n,2n) cross section to the isomer is \qtyrange{0.35}{0.40}{b} at \qty{14}{MeV}~\cite{Perkin1961,Landrum1973,Lindeke1975}, 75\% to 85\% of the \qty{0.47}{b} total in ENDF/B-VIII.0. The ($\gamma$,n) reaction feeds the isomer 95\% of the time~\cite{Veyssiere1973,Berman1986,Gardner1989}. ${}^{236}$Pu decays (2.86 yr) to ${}^{232}$U, and the chain continues as in \Cref{sec:pa231_irrad}, including the ${}^{229}$Th step of \Cref{sec:reactor}.

\begin{figure}[tb]
    \centering
    \begin{subfigure}[t]{\columnwidth}
    \centering
    \includegraphics[width=1.0\textwidth]{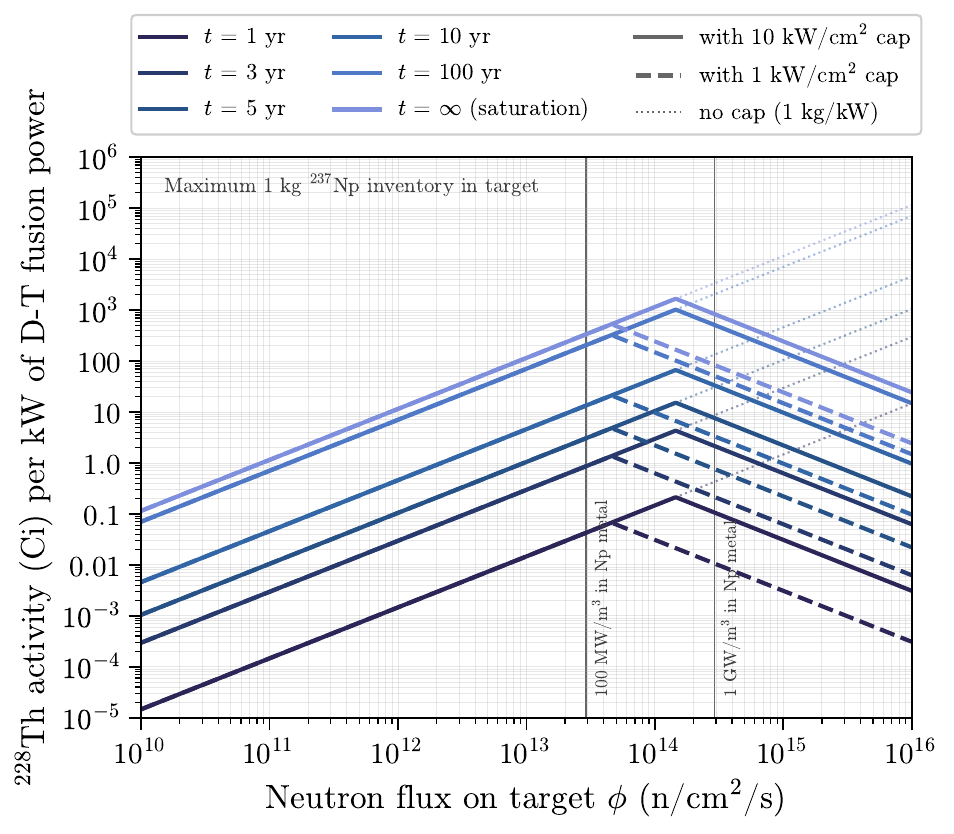}
    \caption{D-T: ${}^{228}$Th activity per kW of D-T fusion power versus on-target flux for \qty{1}{kg} of ${}^{237}$Np at several irradiation times. Solid, dashed, and dotted curves apply a \qty{10}{kW\per\centi\meter\squared}, \qty{1}{kW/cm^2}, and no cooling cap. Vertical lines mark 100 MW/m$^{3}$ and 1 GW/m$^{3}$ of fission heat in solid Np metal.}
    \end{subfigure}
    \centering
    \begin{subfigure}[t]{\columnwidth}
    \centering
    \includegraphics[width=1.0\textwidth]{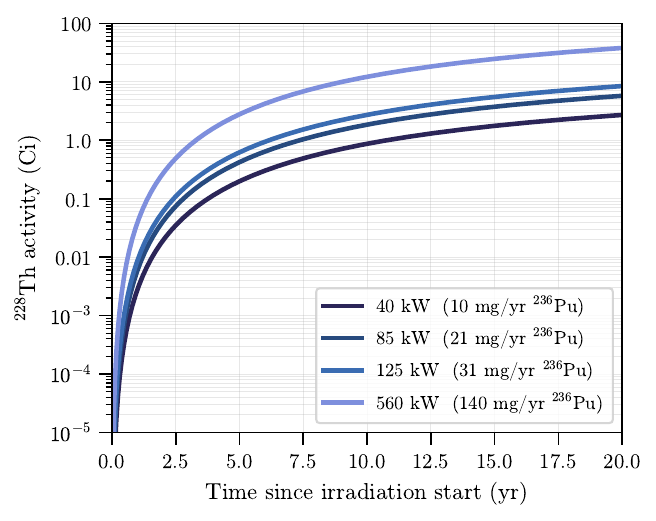}
    \caption{Rhodotron: ${}^{228}$Th activity versus time for 22.5 g of ${}^{237}$Np, rescaled from the ${}^{67}$Cu benchmark~\cite{hawkins2025cu}.}
    \end{subfigure}
    \caption{${}^{228}$Th activity from the ${}^{237}$Np routes.}
    \label{fig:production_rates}
\end{figure}

We use OpenMC to simulate \qty{1}{kg} of ${}^{237}$Np inventory. Fission heat limits the areal density, so above a cooling-limited flux the yield per kilowatt falls with $\phi$ (the kink in \Cref{fig:production_rates}a). However, volumetric heating likely limits the neutron flux instead: solid ${}^{237}$Np metal reaches \qty{100}{MW\per\cubic\meter} at $\phi=\qty{2.9e13}{n.cm^{-2}s^{-1}}$ and \qty{1}{GW\per\cubic\meter} at \qty{2.9e14}{n.cm^{-2}s^{-1}} (vertical lines), so higher fluxes need a diluted target, or a very advanced volumetric cooling system. For photons (\Cref{fig:production_rates}b) we calculate production yields by benchmarking to NorthStar's Rhodotron ${}^{67}$Cu output~\cite{hawkins2025cu} (in Appendix \ref{app:rhodotron_prod_rates}): \qtyrange{10}{31}{mg/yr} of ${}^{236}$Pu from \qty{22.5}{g} of ${}^{237}$Np at \qtyrange{40}{125}{kW}, uncertain by a factor of two.

While the \qty{68.9}{yr} ${}^{232}$U half-life sets the chain transient timescale (see Appendix \ref{app:time_ev_equations}), useful ${}^{212}$Pb output starts within the first year. \Cref{tab:production_rates} lists the ${}^{228}$Th activity from ${}^{237}$Np(n,2n) on a D-T source and from ${}^{237}$Np($\gamma$,n) on a Rhodotron, with the ${}^{232}$Th($\gamma$,n) route of \Cref{sec:pa231_two_stage} for comparison.

\begin{table}[b]
\centering
\caption{${}^{228}$Th activity (Ci) at three irradiation times, for \qty{1}{kg} of ${}^{237}$Np at \qty{1}{kW} of D-T fusion power ($3.6\cdot10^{14}$ n/s) at the stated flux $\phi$ (\unit{n.cm^{-2}.s^{-1}}), and for \qty{22.5}{g} of ${}^{237}$Np or ${}^{232}$Th on an \qty{85}{kW} Rhodotron. The thorium column is ${}^{232}$Th($\gamma$,n) followed by the reactor stage of \Cref{sec:pa231_irrad} (\Cref{fig:rhodotron_th}).}
\label{tab:production_rates}
\small
\begin{tabular}{lcccc}
\toprule
$t / \unit{yr}$ & \multicolumn{2}{c}{D-T, 1 kW} & \multicolumn{2}{c}{Rhodotron, 85 kW} \\
    & $\phi = 10^{13}$ & $\phi = 10^{14}$ & ${}^{237}$Np & ${}^{232}$Th \\
\midrule
1   & $1.5\cdot 10^{-2}$ & $0.15$ & $5.9\cdot 10^{-3}$ & $1.7\cdot10^{-2}$ \\
3   & $0.30$              & $3.0$  & $0.12$ & $0.47$ \\
10  & $4.6$               & $46$   & $1.9$ & $6.6$ \\
\bottomrule
\end{tabular}
\end{table}

\section{Discussion}\label{sec:discussion}

We have modeled transmutation pathways that make the $^{228}$Th and $^{229}$Th generators of $^{212}$Pb and $^{225}$Ac for targeted alpha therapy. The routes use fast neutrons, thermal neutrons, and/or energetic photons. Our modeled yields indicate that modest feedstock inventories can support millions of annual dose-equivalents.

Each production route has relative advantages and disadvantages. Because $^{228}$Th is much shorter-lived than $^{229}$Th, modest neutron or photon irradiations through the $^{237}$Np or $^{232}$Th pathways can support substantial annual ${}^{212}$Pb dose-equivalents. Building a kilogram-scale ${}^{229}$Th stockpile through the indirect ${}^{232}$Th chain requires of order 100 megawatt-years of D-T fusion exposure, followed by lengthy thermal-neutron irradiations. The direct ${}^{230}$Th(n,2n) route requires only about 5 megawatt-years ($\sim$2 gigawatt-days) of fusion exposure because it requires neither a reactor stage nor ${}^{232}$U decay. Using 10 MW of D-T fusion power, a tonne of thorium containing 27\% ${}^{230}$Th reaches the 208 Ci reference inventory in five months, while 20 kg of pure ${}^{230}$Th reaches it in 18 months (\Cref{sec:th230}). Thermal-neutron capture can instead convert the same ionium feedstock into ${}^{228}$Th (\Cref{sec:th230_thermal}), allowing one feedstock to supply either generator. The long ${}^{229}$Th half-life then preserves the stock on a millennial timescale. Finally, if recovering $^{231}$Pa from uranium ore were economically viable, this may also be an attractive route to $^{228}$Th and $^{229}$Th supply.

Abundant generator parents would also increase the supply of their shorter-lived daughters. ${}^{229}$Th can supply ${}^{213}$Bi through ${}^{225}$Ac, and ${}^{228}$Th supplies ${}^{212}$Bi through ${}^{212}$Pb. ${}^{231}$Pa decays to ${}^{227}$Ac, the generator of ${}^{227}$Th and clinically used ${}^{223}$Ra. Further decay produces $^{211}$Pb, another potential short-lived alpha-generator for TAT.

Experimental measurements of the key reaction rates under the proposed irradiation conditions are required to validate the predicted yields.

\begin{acknowledgments}
We are grateful for conversations with L. A. Bernstein, J. W. Engle, P. F. Peterson, J. A. Schwartz, and J. S. Wexler.
\end{acknowledgments}

\section*{AI-Usage Statement}
We had extensive assistance from Claude Fable and OpenAI Astra in producing and validating the results in this work, particularly for running large numbers of OpenMC simulations and for figure preparation. All outputs remain the responsibility of the publishing authors. Upon publication, all data used to produce this work will be uploaded to an open-source repository.

\appendix

\section{\texorpdfstring{Calculating ${}^{212}$Pb doses per curie of ${}^{228}$Th}{Calculating Pb-212 doses per curie of Th-228}}\label{app:pb212_doses}

In this Appendix we estimate the annual ${}^{212}$Pb dose-equivalents that one current curie of ${}^{228}$Th can produce under different elution schedules and delays.

An important starting heuristic is that in equilibrium, the ${}^{212}$Pb activity equals the ${}^{228}$Th activity, so a single elution returns at most 1 Ci of ${}^{212}$Pb per Ci of ${}^{228}$Th. In practice, this will be much lower. Each ${}^{228}$Th decay produces exactly one ${}^{212}$Pb atom (via ${}^{224}$Ra, ${}^{220}$Rn, and ${}^{216}$Po, none of which branch), so ${}^{212}$Pb atoms are created at a rate equal to the ${}^{228}$Th activity $A_\mrm{Th}$, and are lost by their own decay. After stripping the ${}^{228}$Th of ${}^{212}$Pb, the number of ${}^{212}$Pb atoms $N_\mrm{Pb}$ obeys
\begin{equation}
\frac{\mrm{d}N_\mrm{Pb}}{\mrm{d}t} = A_\mrm{Th} - \lambda_\mrm{Pb} N_\mrm{Pb},
\end{equation}
where $\lambda_\mrm{Pb} = \ln 2/\thalf^{(\mrm{Pb})}$ is the ${}^{212}$Pb decay constant, with $\thalf^{(\mrm{Pb})} = \qty{10.64}{h}$. For a generator with current activity $A_\mrm{Th}$, the ideal annualized output is $A_\mrm{Th}T_\mrm{yr}$ daughter atoms. Expressed as ${}^{212}$Pb activity at administration, one current curie of ${}^{228}$Th therefore gives $\lambda_\mrm{Pb}T_\mrm{yr}=571$ Ci of ideal annual ${}^{212}$Pb activity-equivalent, where $T_\mrm{yr}=\qty{1}{yr}$. Over the complete decay of an initial 1 Ci ${}^{228}$Th inventory, the absolute theoretical maximum is $\lambda_\mrm{Pb}/\lambda_\mrm{Th}\simeq1.57\cdot10^3$ Ci of cumulative ${}^{212}$Pb activity-equivalent. We use 571 Ci for the annual comparison because it gives the annualized output per current curie rather than the lifetime output from an initial curie. Dividing by the \qty{2.7}{mCi} activity per administration gives $2.1\cdot10^5$ ideal annual dose-equivalents per current curie, as listed in \Cref{tab:doses_per_ci}.

\begin{figure}[tb]
\centering
\includegraphics[width=\columnwidth]{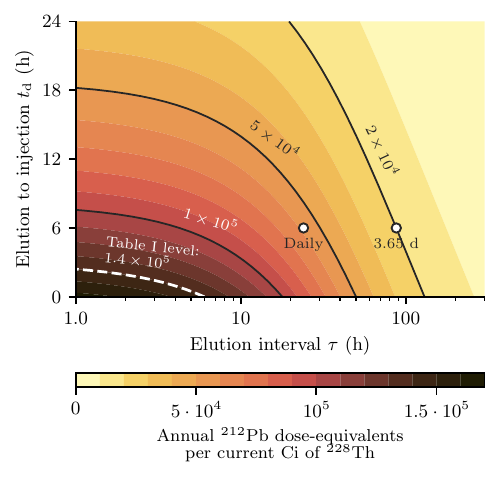}
\caption{Annual ${}^{212}$Pb dose-equivalents delivered per current curie of ${}^{228}$Th, from \Cref{eq:pb212_yield} at $\epsilon=0.8$. The dashed line is the reference conversion in \Cref{tab:doses_per_ci}, and the circles are the two schedules quoted in the text.}
\label{fig:pb212_doses_map}
\end{figure}

Only part of this activity reaches patients. If the generator is eluted at intervals of $\tau$ hours, a ${}^{212}$Pb atom born at time $t'$ after the preceding elution survives until the next elution with probability $\exp[-\lambda_\mrm{Pb}(\tau-t')]$. The ${}^{212}$Pb production rate changes negligibly within one cycle, giving a surviving fraction
\begin{equation}
\eta(\tau) = \frac{1}{\tau} \int_0^\tau \exp[-\lambda_\mrm{Pb}(\tau - t')] \, \mrm{d}t' = \frac{1 - \exp(-\lambda_\mrm{Pb} \tau)}{\lambda_\mrm{Pb} \tau}.
\label{eq:pb212_eta}
\end{equation}
The resulting values are $\eta=$0.69 for $\tau = \qty{12}{h}$, 0.51 for \qty{24}{h}, and $\eta=$0.18 for \qty{3.65}{d}. Of the activity collected, a fraction $\exp(-\lambda_\mrm{Pb} t_\mrm{d})$ survives the delay $t_\mrm{d}$ between elution and injection. If $\epsilon$ is the chemical recovery fraction, the annual activity-equivalent delivered per current curie of ${}^{228}$Th is
\begin{equation}
Y_\mrm{yr} = \lambda_\mrm{Pb}T_\mrm{yr}\, \eta(\tau)\, \epsilon\, \exp(-\lambda_\mrm{Pb} t_\mrm{d}).
\label{eq:pb212_yield}
\end{equation}
The corresponding annual dose-equivalents per current $^{228}$Th curie are
\begin{equation}
y_\mrm{yr}=Y_\mrm{yr}/a_\mrm{Pb},
\end{equation}
where $a_\mrm{Pb}=\qty{2.7}{mCi}$ is the ${}^{212}$Pb activity administered per dose (\Cref{tab:doses_per_ci}). We plot $y_\mrm{yr}$ in \Cref{fig:pb212_doses_map} for $\epsilon=0.8$. With a \qty{6}{h} delivery delay, daily elution gives $Y_\mrm{yr}=156$ Ci/yr and $y_\mrm{yr}=5.8\cdot10^{4}$ annual dose-equivalents per current curie. Elution every \qty{3.65}{d} gives 54 Ci/yr and $2.0\cdot10^{4}$ dose-equivalents. The $1.4\cdot10^{5}$ reference value in \Cref{tab:doses_per_ci} instead treats 67\% of all daughter atoms produced during the year as recovered and administered promptly. It provides a common conversion for comparing production routes, while the figure shows how an operating schedule can lower the delivered rate. We neglect the very small reported ${}^{228}$Th breakthrough during harvesting~\cite{Li2023generator}.

\section{Calculating photonuclear production from the Rhodotron benchmark}\label{app:rhodotron_prod_rates}

In this Appendix we calculate photonuclear production rates by benchmarking to NorthStar's reported yields on an IBA Rhodotron ~\cite{hawkins2025cu}. NorthStar reported target masses of 7.5 to 22.5 g, beam powers of 40 to 85 kW at 40 MeV, and batch yields of 74 to 300 GBq over six days. We use the upper endpoints of these production figures as our scale-up benchmark, i.e.\ $R_\mrm{Cu}=3.8\cdot10^{11}$ atoms/s. We estimate the ${}^{237}$Np($\gamma$,n)${}^{236\mrm{m}}$Np rate for $N_\mrm{Np}$ atoms at beam power $P$ by rescaling,
\begin{equation}
R_\mrm{Np} = R_\mrm{Cu}\,\frac{P}{85 \mrm{kW}}\,\frac{N_\mrm{Np}}{N_\mrm{Zn}}\,F, \qquad
F = \frac{\int w(k)\,\sigma_\mrm{Np}(k)\,dk}{\int w(k)\,\sigma_\mrm{Zn}(k)\,dk},
\label{eq:rhodotron_scaling}
\end{equation}
with $N_\mrm{Zn}$ the ${}^{68}$Zn atoms in the NorthStar target, $w(k)\propto(1/k)(1-k/E_0)$ the bremsstrahlung spectrum ($E_0=40$ MeV), $\sigma_\mrm{Np}$ the EXFOR spline scaled by the 95\% isomer branching ratio~\cite{Gardner1989}, and we model $\sigma_\mrm{Zn}$ as a Lorentzian with a \qty{10}{mb} peak at \qty{22}{MeV}.
These give $F=33$ in \Cref{eq:rhodotron_scaling}. With $b_\mrm{Pu}=0.48$ this is 0.25\,mg/yr/kW of ${}^{236}$Pu for \qty{22.5}{g} of ${}^{237}$Np (0.17 at a \qty{15}{mb} peak and 0.50 at a \qty{5}{mb} peak).

\section{Decay-chain time-evolution equations}\label{app:time_ev_equations}

\begin{figure*}[t]
\centering
\includegraphics[width=0.85\textwidth]{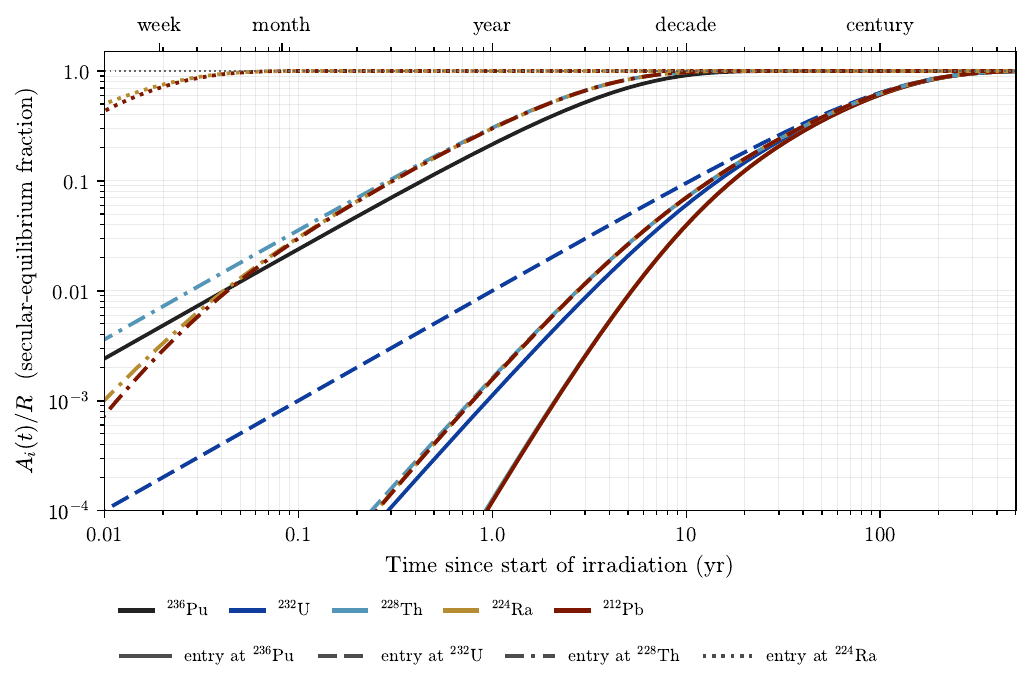}
\caption{Activity ratios $A_i/R$ along the ${}^{236}$Pu decay chain to ${}^{212}$Pb for a constant production rate $R$ of the entry nuclide.}
\label{fig:pu236_chain_population}
\end{figure*}

In this Appendix, we briefly describe the decay equations for the  ${}^{236}$Pu decay chain. Given a ${}^{236}$Pu production rate $R(t)$, the decay chain to ${}^{212}$Bi is
\begin{equation}
\begin{split}
{}^{236}\mrm{Pu} \xrightarrow{\alpha} {}^{232}\mrm{U} \xrightarrow{\alpha} {}^{228}\mrm{Th} \xrightarrow{\alpha} {}^{224}\mrm{Ra} \xrightarrow{\alpha} {}^{220}\mrm{Rn} \\
\xrightarrow{\alpha} {}^{216}\mrm{Po} \xrightarrow{\alpha} {}^{212}\mrm{Pb} \xrightarrow{\betam} {}^{212}\mrm{Bi}.
\end{split}
\label{eq:chain}
\end{equation}
${}^{220}$Rn and ${}^{216}$Po reach equilibrium with ${}^{224}$Ra within minutes, so we treat ${}^{224}\mrm{Ra}\to{}^{212}\mrm{Pb}$ as one decay. With $N_i$ the inventory and $\lambda_i=\ln 2/\thalf^{(i)}$ the decay constant of species $i\in\{\mrm{Pu},\mrm{U},\mrm{Th},\mrm{Ra},\mrm{Pb},\mrm{Bi}\}$, the Bateman system~\cite{Bateman1910} is
\begin{align}
\dot N_\mrm{Pu} &= R(t) - \lambda_\mrm{Pu}\, N_\mrm{Pu}, \\
\dot N_\mrm{U}  &= \lambda_\mrm{Pu}\, N_\mrm{Pu} - \lambda_\mrm{U}\, N_\mrm{U}, \\
\dot N_\mrm{Th} &= \lambda_\mrm{U}\, N_\mrm{U}  - \lambda_\mrm{Th}\, N_\mrm{Th}, \\
\dot N_\mrm{Ra} &= \lambda_\mrm{Th}\, N_\mrm{Th} - \lambda_\mrm{Ra}\, N_\mrm{Ra}, \\
\dot N_\mrm{Pb} &= \lambda_\mrm{Ra}\, N_\mrm{Ra} - \lambda_\mrm{Pb}\, N_\mrm{Pb}, \\
\dot N_\mrm{Bi} &= \lambda_\mrm{Pb}\, N_\mrm{Pb} - \lambda_\mrm{Bi}\, N_\mrm{Bi},
\label{eq:bateman}
\end{align}
with $N_i(0) = 0$ for all $i$.

The ${}^{228}$Th activity is $A_\mrm{Th}=\lambda_\mrm{Th}N_\mrm{Th}$. ${}^{224}$Ra and ${}^{212}$Pb follow it within weeks. For constant $R$, equilibrium ($A_\mrm{Th}\to R$) takes $\sim5\thalf^{(\mrm{U})}\approx\qty{350}{yr}$. \Cref{fig:pu236_chain_population} shows the activity ratios $A_i/R$ for four entry points. For \qty{1}{kg} of ${}^{237}$Np at \qty{1}{kW} of D-T fusion power and $\phi=\qty{e13}{n.cm^{-2}.s^{-1}}$, $R=\qty{115}{Ci}$, with 4\% of it reached after \qty{10}{yr}. We do not plot any activities after $^{212}$Pb because all downstream radionuclides are relatively short-lived.

\section{OpenMC reactor-irradiation models}\label{app:openmc_reactor}

\begin{figure}[tb]
\centering
\begin{subfigure}[t]{\columnwidth}
\centering
\includegraphics[width=0.95\linewidth]{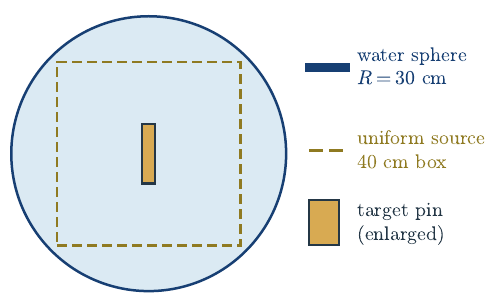}
\caption{Water moderator and uniform Watt-spectrum source volume. The pin is enlarged for visibility.}
\label{fig:openmc_reactor_pin_geometry_a}
\end{subfigure}

\begin{subfigure}[t]{\columnwidth}
\centering
\includegraphics[width=0.95\linewidth]{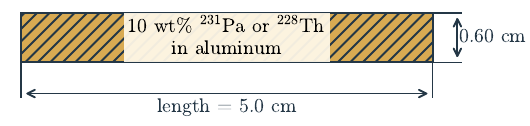}
\caption{Target-pin dimensions and composition.}
\label{fig:openmc_reactor_pin_geometry_b}
\end{subfigure}

\begin{subfigure}[t]{\columnwidth}
\centering
\includegraphics[width=0.95\linewidth]{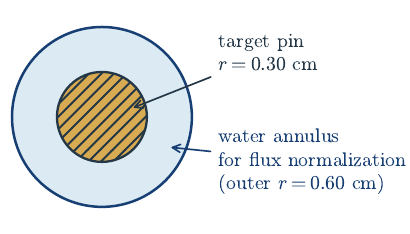}
\caption{Pin and adjacent water annulus used to normalize the thermal flux.}
\label{fig:openmc_reactor_pin_geometry_c}
\end{subfigure}
\caption{OpenMC irradiation geometry used in the reactor-depletion calculations.}
\label{fig:openmc_reactor_pin_geometry}
\end{figure}

We briefly describe the reactor irradiation geometry simulated with OpenMC~\cite{openmc}. We show the modeled geometry in \Cref{fig:openmc_reactor_pin_geometry}. The irradiated material is a \qty{0.3}{cm} radius, \qty{5}{cm} long pin of ${}^{231}$Pa or ${}^{228}$Th at 10 wt\% in aluminum (\qty{0.42}{g}) in a \qty{30}{cm} water sphere fed by Watt-spectrum neutrons. $\phi_{\rm th}$ is the flux below the \qty{0.625}{eV} cadmium cut-off next to the pin. We average each energy-dependent cross section over the neutron spectrum in the pin and use the resulting effective cross section in the depletion calculation.

\bibliography{references}

%apsrev4-2.bst 2019-01-14 (MD) hand-edited version of apsrev4-1.bst
%Control: key (0)
%Control: author (8) initials jnrlst
%Control: editor formatted (1) identically to author
%Control: production of article title (0) allowed
%Control: page (0) single
%Control: year (1) truncated
%Control: production of eprint (0) enabled
\begin{thebibliography}{79}%
\makeatletter
\providecommand \@ifxundefined [1]{%
 \@ifx{#1\undefined}
}%
\providecommand \@ifnum [1]{%
 \ifnum #1\expandafter \@firstoftwo
 \else \expandafter \@secondoftwo
 \fi
}%
\providecommand \@ifx [1]{%
 \ifx #1\expandafter \@firstoftwo
 \else \expandafter \@secondoftwo
 \fi
}%
\providecommand \natexlab [1]{#1}%
\providecommand \enquote  [1]{``#1''}%
\providecommand \bibnamefont  [1]{#1}%
\providecommand \bibfnamefont [1]{#1}%
\providecommand \citenamefont [1]{#1}%
\providecommand \href@noop [0]{\@secondoftwo}%
\providecommand \href [0]{\begingroup \@sanitize@url \@href}%
\providecommand \@href[1]{\@@startlink{#1}\@@href}%
\providecommand \@@href[1]{\endgroup#1\@@endlink}%
\providecommand \@sanitize@url [0]{\catcode `\\12\catcode `\$12\catcode
  `\&12\catcode `\#12\catcode `\^12\catcode `\_12\catcode `\%12\relax}%
\providecommand \@@startlink[1]{}%
\providecommand \@@endlink[0]{}%
\providecommand \url  [0]{\begingroup\@sanitize@url \@url }%
\providecommand \@url [1]{\endgroup\@href {#1}{\urlprefix }}%
\providecommand \urlprefix  [0]{URL }%
\providecommand \Eprint [0]{\href }%
\providecommand \doibase [0]{https://doi.org/}%
\providecommand \selectlanguage [0]{\@gobble}%
\providecommand \bibinfo  [0]{\@secondoftwo}%
\providecommand \bibfield  [0]{\@secondoftwo}%
\providecommand \translation [1]{[#1]}%
\providecommand \BibitemOpen [0]{}%
\providecommand \bibitemStop [0]{}%
\providecommand \bibitemNoStop [0]{.\EOS\space}%
\providecommand \EOS [0]{\spacefactor3000\relax}%
\providecommand \BibitemShut  [1]{\csname bibitem#1\endcsname}%
\let\auto@bib@innerbib\@empty
%</preamble>
\bibitem [{\citenamefont {Yong}\ and\ \citenamefont
  {Brechbiel}(2015)}]{Yong2015}%
  \BibitemOpen
  \bibfield  {author} {\bibinfo {author} {\bibfnamefont {K.}~\bibnamefont
  {Yong}}\ and\ \bibinfo {author} {\bibfnamefont {M.~W.}\ \bibnamefont
  {Brechbiel}},\ }\bibfield  {title} {\bibinfo {title} {Application of
  ${}^{212}${Pb} for targeted $\alpha$-particle therapy ({TAT}): pre-clinical
  and mechanistic understanding through to clinical translation},\ }\href
  {https://doi.org/10.3934/medsci.2015.3.228} {\bibfield  {journal} {\bibinfo
  {journal} {AIMS Med. Sci.}\ }\textbf {\bibinfo {volume} {2}},\ \bibinfo
  {pages} {228} (\bibinfo {year} {2015})}\BibitemShut {NoStop}%
\bibitem [{\citenamefont {Pedersen}\ \emph {et~al.}(2026)\citenamefont
  {Pedersen}, \citenamefont {Straathof}, \citenamefont {Elvas}, \citenamefont
  {Herth},\ and\ \citenamefont {Battisti}}]{Pedersen2026TAT}%
  \BibitemOpen
  \bibfield  {author} {\bibinfo {author} {\bibfnamefont {N.~B.}\ \bibnamefont
  {Pedersen}}, \bibinfo {author} {\bibfnamefont {N.~J.~W.}\ \bibnamefont
  {Straathof}}, \bibinfo {author} {\bibfnamefont {F.}~\bibnamefont {Elvas}},
  \bibinfo {author} {\bibfnamefont {M.~M.}\ \bibnamefont {Herth}},\ and\
  \bibinfo {author} {\bibfnamefont {U.~M.}\ \bibnamefont {Battisti}},\
  }\bibfield  {title} {\bibinfo {title} {Targeted alpha therapy (r)evolution:
  emerging nuclides for clinical applications},\ }\href
  {https://doi.org/10.1016/j.tips.2026.01.001} {\bibfield  {journal} {\bibinfo
  {journal} {Trends in Pharmacological Sciences}\ }\textbf {\bibinfo {volume}
  {47}},\ \bibinfo {pages} {263} (\bibinfo {year} {2026})}\BibitemShut
  {NoStop}%
\bibitem [{\citenamefont {{RayzeBio, Inc.}}(2022)}]{CTGovAcNET3}%
  \BibitemOpen
  \bibfield  {author} {\bibinfo {author} {\bibnamefont {{RayzeBio, Inc.}}},\
  }\href {https://clinicaltrials.gov/study/NCT05477576} {\bibinfo {title}
  {Study of {RYZ101} compared with standard of care in patients with inoperable
  {SSTR}-positive well-differentiated gastroenteropancreatic neuroendocrine
  tumors}},\ \bibinfo {howpublished} {ClinicalTrials.gov, NCT05477576}
  (\bibinfo {year} {2022}),\ \bibinfo {note} {registry record verified 5
  September 2026}\BibitemShut {NoStop}%
\bibitem [{\citenamefont {{Novartis
  Pharmaceuticals}}(2025)}]{CTGovAcProstate3}%
  \BibitemOpen
  \bibfield  {author} {\bibinfo {author} {\bibnamefont {{Novartis
  Pharmaceuticals}}},\ }\href {https://clinicaltrials.gov/study/NCT06855277}
  {\bibinfo {title} {Study comparing {AAA817} plus androgen-receptor pathway
  inhibitor versus standard of care in adults with {PSMA}-positive metastatic
  castration-resistant prostate cancer}},\ \bibinfo {howpublished}
  {ClinicalTrials.gov, NCT06855277} (\bibinfo {year} {2025}),\ \bibinfo {note}
  {registry record verified 5 September 2026}\BibitemShut {NoStop}%
\bibitem [{\citenamefont {{Orano Med LLC}}(2021)}]{CTGovPbNET2}%
  \BibitemOpen
  \bibfield  {author} {\bibinfo {author} {\bibnamefont {{Orano Med LLC}}},\
  }\href {https://clinicaltrials.gov/study/NCT05153772} {\bibinfo {title}
  {Targeted alpha-emitter therapy of {PRRT}-naive and previously treated
  neuroendocrine tumor patients}},\ \bibinfo {howpublished}
  {ClinicalTrials.gov, NCT05153772} (\bibinfo {year} {2021}),\ \bibinfo {note}
  {registry record verified 5 September 2026}\BibitemShut {NoStop}%
\bibitem [{\citenamefont {{Fred Hutchinson Cancer
  Center}}(2017)}]{CTGovAtLeukemia}%
  \BibitemOpen
  \bibfield  {author} {\bibinfo {author} {\bibnamefont {{Fred Hutchinson Cancer
  Center}}},\ }\href {https://clinicaltrials.gov/study/NCT03128034} {\bibinfo
  {title} {${}^{211}$at-{BC8-B10} before donor stem-cell transplant in patients
  with high-risk acute leukemia or myelodysplastic syndrome}},\ \bibinfo
  {howpublished} {ClinicalTrials.gov, NCT03128034} (\bibinfo {year} {2017}),\
  \bibinfo {note} {registry record verified 5 September 2026}\BibitemShut
  {NoStop}%
\bibitem [{\citenamefont {{Osaka University}}(2024)}]{CTGovAtPSMA}%
  \BibitemOpen
  \bibfield  {author} {\bibinfo {author} {\bibnamefont {{Osaka University}}},\
  }\href {https://clinicaltrials.gov/study/NCT06441994} {\bibinfo {title}
  {Targeted alpha therapy using astatine-211 {PSMA-5} for prostate cancer}},\
  \bibinfo {howpublished} {ClinicalTrials.gov, NCT06441994} (\bibinfo {year}
  {2024}),\ \bibinfo {note} {registry record verified 5 September
  2026}\BibitemShut {NoStop}%
\bibitem [{\citenamefont {Radchenko}\ \emph {et~al.}(2021)\citenamefont
  {Radchenko}, \citenamefont {Morgenstern}, \citenamefont {Jalilian},
  \citenamefont {Ramogida}, \citenamefont {Cutler}, \citenamefont {Duchemin},
  \citenamefont {Hoehr}, \citenamefont {Haddad}, \citenamefont
  {Bruchertseifer}, \citenamefont {Gausemel}, \citenamefont {Yang},
  \citenamefont {Osso}, \citenamefont {Washiyama}, \citenamefont {Czerwinski},
  \citenamefont {Leufgen}, \citenamefont {Pruszy{\'n}ski}, \citenamefont
  {Valzdorf}, \citenamefont {Causey}, \citenamefont {Schaffer}, \citenamefont
  {Perron}, \citenamefont {Maxim}, \citenamefont {Wilbur}, \citenamefont
  {Stora},\ and\ \citenamefont {Li}}]{Radchenko2021Supply}%
  \BibitemOpen
  \bibfield  {author} {\bibinfo {author} {\bibfnamefont {V.}~\bibnamefont
  {Radchenko}}, \bibinfo {author} {\bibfnamefont {A.}~\bibnamefont
  {Morgenstern}}, \bibinfo {author} {\bibfnamefont {A.~R.}\ \bibnamefont
  {Jalilian}}, \bibinfo {author} {\bibfnamefont {C.~F.}\ \bibnamefont
  {Ramogida}}, \bibinfo {author} {\bibfnamefont {C.}~\bibnamefont {Cutler}},
  \bibinfo {author} {\bibfnamefont {C.}~\bibnamefont {Duchemin}}, \bibinfo
  {author} {\bibfnamefont {C.}~\bibnamefont {Hoehr}}, \bibinfo {author}
  {\bibfnamefont {F.}~\bibnamefont {Haddad}}, \bibinfo {author} {\bibfnamefont
  {F.}~\bibnamefont {Bruchertseifer}}, \bibinfo {author} {\bibfnamefont
  {H.}~\bibnamefont {Gausemel}}, \bibinfo {author} {\bibfnamefont
  {H.}~\bibnamefont {Yang}}, \bibinfo {author} {\bibfnamefont {J.~A.}\
  \bibnamefont {Osso}}, \bibinfo {author} {\bibfnamefont {K.}~\bibnamefont
  {Washiyama}}, \bibinfo {author} {\bibfnamefont {K.}~\bibnamefont
  {Czerwinski}}, \bibinfo {author} {\bibfnamefont {K.}~\bibnamefont {Leufgen}},
  \bibinfo {author} {\bibfnamefont {M.}~\bibnamefont {Pruszy{\'n}ski}},
  \bibinfo {author} {\bibfnamefont {O.}~\bibnamefont {Valzdorf}}, \bibinfo
  {author} {\bibfnamefont {P.}~\bibnamefont {Causey}}, \bibinfo {author}
  {\bibfnamefont {P.}~\bibnamefont {Schaffer}}, \bibinfo {author}
  {\bibfnamefont {R.}~\bibnamefont {Perron}}, \bibinfo {author} {\bibfnamefont
  {S.}~\bibnamefont {Maxim}}, \bibinfo {author} {\bibfnamefont {D.~S.}\
  \bibnamefont {Wilbur}}, \bibinfo {author} {\bibfnamefont {T.}~\bibnamefont
  {Stora}},\ and\ \bibinfo {author} {\bibfnamefont {Y.}~\bibnamefont {Li}},\
  }\bibfield  {title} {\bibinfo {title} {Production and supply of
  alpha-particle-emitting radionuclides for targeted alpha-therapy},\ }\href
  {https://doi.org/10.2967/jnumed.120.261016} {\bibfield  {journal} {\bibinfo
  {journal} {Journal of Nuclear Medicine}\ }\textbf {\bibinfo {volume} {62}},\
  \bibinfo {pages} {1495} (\bibinfo {year} {2021})}\BibitemShut {NoStop}%
\bibitem [{\citenamefont {Kokov}\ \emph {et~al.}(2022)\citenamefont {Kokov},
  \citenamefont {Egorova}, \citenamefont {German}, \citenamefont {Klabukov},
  \citenamefont {Krasheninnikov}, \citenamefont {Larkin-Kondrov}, \citenamefont
  {Makoveeva}, \citenamefont {Ovchinnikov}, \citenamefont {Sidorova},\ and\
  \citenamefont {Chuvilin}}]{Kokov2022}%
  \BibitemOpen
  \bibfield  {author} {\bibinfo {author} {\bibfnamefont {K.~V.}\ \bibnamefont
  {Kokov}}, \bibinfo {author} {\bibfnamefont {B.~V.}\ \bibnamefont {Egorova}},
  \bibinfo {author} {\bibfnamefont {M.~N.}\ \bibnamefont {German}}, \bibinfo
  {author} {\bibfnamefont {I.~D.}\ \bibnamefont {Klabukov}}, \bibinfo {author}
  {\bibfnamefont {M.~E.}\ \bibnamefont {Krasheninnikov}}, \bibinfo {author}
  {\bibfnamefont {A.~A.}\ \bibnamefont {Larkin-Kondrov}}, \bibinfo {author}
  {\bibfnamefont {K.~A.}\ \bibnamefont {Makoveeva}}, \bibinfo {author}
  {\bibfnamefont {M.~V.}\ \bibnamefont {Ovchinnikov}}, \bibinfo {author}
  {\bibfnamefont {M.~V.}\ \bibnamefont {Sidorova}},\ and\ \bibinfo {author}
  {\bibfnamefont {D.~Y.}\ \bibnamefont {Chuvilin}},\ }\bibfield  {title}
  {\bibinfo {title} {${}^{212}${Pb}: production approaches and targeted therapy
  applications},\ }\href {https://doi.org/10.3390/pharmaceutics14010189}
  {\bibfield  {journal} {\bibinfo  {journal} {Pharmaceutics}\ }\textbf
  {\bibinfo {volume} {14}},\ \bibinfo {pages} {189} (\bibinfo {year}
  {2022})}\BibitemShut {NoStop}%
\bibitem [{\citenamefont {Robertson}\ \emph {et~al.}(2018)\citenamefont
  {Robertson}, \citenamefont {Ramogida}, \citenamefont {Schaffer},\ and\
  \citenamefont {Radchenko}}]{robertson2018ac225}%
  \BibitemOpen
  \bibfield  {author} {\bibinfo {author} {\bibfnamefont {A.~K.~H.}\
  \bibnamefont {Robertson}}, \bibinfo {author} {\bibfnamefont {C.~F.}\
  \bibnamefont {Ramogida}}, \bibinfo {author} {\bibfnamefont {P.}~\bibnamefont
  {Schaffer}},\ and\ \bibinfo {author} {\bibfnamefont {V.}~\bibnamefont
  {Radchenko}},\ }\bibfield  {title} {\bibinfo {title} {Development of
  ${}^{225}${Ac} radiopharmaceuticals: {TRIUMF} perspectives and experiences},\
  }\href {https://doi.org/10.2174/1874471011666180416161908} {\bibfield
  {journal} {\bibinfo  {journal} {Current Radiopharmaceuticals}\ }\textbf
  {\bibinfo {volume} {11}},\ \bibinfo {pages} {156} (\bibinfo {year}
  {2018})}\BibitemShut {NoStop}%
\bibitem [{\citenamefont {Morgenstern}\ \emph {et~al.}(2018)\citenamefont
  {Morgenstern}, \citenamefont {Apostolidis}, \citenamefont {Kratochwil},
  \citenamefont {Sathekge}, \citenamefont {Krolicki},\ and\ \citenamefont
  {Bruchertseifer}}]{morgenstern2018}%
  \BibitemOpen
  \bibfield  {author} {\bibinfo {author} {\bibfnamefont {A.}~\bibnamefont
  {Morgenstern}}, \bibinfo {author} {\bibfnamefont {C.}~\bibnamefont
  {Apostolidis}}, \bibinfo {author} {\bibfnamefont {C.}~\bibnamefont
  {Kratochwil}}, \bibinfo {author} {\bibfnamefont {M.}~\bibnamefont
  {Sathekge}}, \bibinfo {author} {\bibfnamefont {L.}~\bibnamefont {Krolicki}},\
  and\ \bibinfo {author} {\bibfnamefont {F.}~\bibnamefont {Bruchertseifer}},\
  }\bibfield  {title} {\bibinfo {title} {An overview of targeted alpha therapy
  with $^{225}$ac and $^{213}$bi},\ }\href
  {https://doi.org/10.2174/1874471011666180502104524} {\bibfield  {journal}
  {\bibinfo  {journal} {Curr. Radiopharm.}\ }\textbf {\bibinfo {volume} {11}},\
  \bibinfo {pages} {200} (\bibinfo {year} {2018})}\BibitemShut {NoStop}%
\bibitem [{\citenamefont {Nagatsu}\ \emph {et~al.}(2022)\citenamefont
  {Nagatsu}, \citenamefont {Suzuki}, \citenamefont {Fukada}, \citenamefont
  {Ito}, \citenamefont {Ichinose}, \citenamefont {Honda}, \citenamefont
  {Minegishi}, \citenamefont {Higashi},\ and\ \citenamefont
  {Zhang}}]{Nagatsu2022}%
  \BibitemOpen
  \bibfield  {author} {\bibinfo {author} {\bibfnamefont {K.}~\bibnamefont
  {Nagatsu}}, \bibinfo {author} {\bibfnamefont {H.}~\bibnamefont {Suzuki}},
  \bibinfo {author} {\bibfnamefont {M.}~\bibnamefont {Fukada}}, \bibinfo
  {author} {\bibfnamefont {T.}~\bibnamefont {Ito}}, \bibinfo {author}
  {\bibfnamefont {J.}~\bibnamefont {Ichinose}}, \bibinfo {author}
  {\bibfnamefont {Y.}~\bibnamefont {Honda}}, \bibinfo {author} {\bibfnamefont
  {K.}~\bibnamefont {Minegishi}}, \bibinfo {author} {\bibfnamefont
  {T.}~\bibnamefont {Higashi}},\ and\ \bibinfo {author} {\bibfnamefont {M.-R.}\
  \bibnamefont {Zhang}},\ }\bibfield  {title} {\bibinfo {title} {Cyclotron
  production of $^{225}$ac from an electroplated $^{226}$ra target},\ }\href
  {https://doi.org/10.1007/s00259-021-05460-7} {\bibfield  {journal} {\bibinfo
  {journal} {European Journal of Nuclear Medicine and Molecular Imaging}\
  }\textbf {\bibinfo {volume} {49}},\ \bibinfo {pages} {279} (\bibinfo {year}
  {2022})}\BibitemShut {NoStop}%
\bibitem [{\citenamefont {Diamond}\ and\ \citenamefont
  {Ross}(2021)}]{diamond2021actinium}%
  \BibitemOpen
  \bibfield  {author} {\bibinfo {author} {\bibfnamefont {W.}~\bibnamefont
  {Diamond}}\ and\ \bibinfo {author} {\bibfnamefont {C.}~\bibnamefont {Ross}},\
  }\bibfield  {title} {\bibinfo {title} {Actinium-225 production with an
  electron accelerator},\ }\bibfield  {journal} {\bibinfo  {journal} {Journal
  of Applied Physics}\ }\textbf {\bibinfo {volume} {129}},\ \href
  {https://doi.org/10.1063/5.0043509} {10.1063/5.0043509} (\bibinfo {year}
  {2021})\BibitemShut {NoStop}%
\bibitem [{\citenamefont {{American Nuclear Society}}(2026)}]{ANS2026radium}%
  \BibitemOpen
  \bibfield  {author} {\bibinfo {author} {\bibnamefont {{American Nuclear
  Society}}},\ }\href
  {https://www.ans.org/news/2026-01-14/article-7675/iaeas-global-radium226-management-initiative-continues-to-progress/}
  {\bibinfo {title} {Radium sources yield cancer-fighting {Ac-225} in {IAEA}
  program}},\ \bibinfo {howpublished} {Nuclear Newswire, 14 January 2026}
  (\bibinfo {year} {2026})\BibitemShut {NoStop}%
\bibitem [{\citenamefont {Morrell}(2021)}]{morrell2021next}%
  \BibitemOpen
  \bibfield  {author} {\bibinfo {author} {\bibfnamefont {J.}~\bibnamefont
  {Morrell}},\ }\href {https://escholarship.org/uc/item/1cj6716s} {\emph
  {\bibinfo {title} {Next-generation isotope production via deuteron
  breakup}}}\ (\bibinfo  {publisher} {University of California, Berkeley},\
  \bibinfo {year} {2021})\BibitemShut {NoStop}%
\bibitem [{\citenamefont {McAlister}\ and\ \citenamefont
  {Horwitz}(2018)}]{McAlister2018}%
  \BibitemOpen
  \bibfield  {author} {\bibinfo {author} {\bibfnamefont {D.~R.}\ \bibnamefont
  {McAlister}}\ and\ \bibinfo {author} {\bibfnamefont {E.~P.}\ \bibnamefont
  {Horwitz}},\ }\bibfield  {title} {\bibinfo {title} {Selective separation of
  radium and actinium from bulk thorium target material on strong acid cation
  exchange resin from sulfate media},\ }\href
  {https://doi.org/10.1016/j.apradiso.2018.06.008} {\bibfield  {journal}
  {\bibinfo  {journal} {Appl. Radiat. Isot.}\ }\textbf {\bibinfo {volume}
  {140}},\ \bibinfo {pages} {18} (\bibinfo {year} {2018})}\BibitemShut
  {NoStop}%
\bibitem [{\citenamefont {Pruszy{\'n}ski}\ \emph {et~al.}(2021)\citenamefont
  {Pruszy{\'n}ski}, \citenamefont {Walczak}, \citenamefont {Rodak},
  \citenamefont {Bruchertseifer}, \citenamefont {Morgenstern},\ and\
  \citenamefont {Bilewicz}}]{pruszynski2021radiochemical}%
  \BibitemOpen
  \bibfield  {author} {\bibinfo {author} {\bibfnamefont {M.}~\bibnamefont
  {Pruszy{\'n}ski}}, \bibinfo {author} {\bibfnamefont {R.}~\bibnamefont
  {Walczak}}, \bibinfo {author} {\bibfnamefont {M.}~\bibnamefont {Rodak}},
  \bibinfo {author} {\bibfnamefont {F.}~\bibnamefont {Bruchertseifer}},
  \bibinfo {author} {\bibfnamefont {A.}~\bibnamefont {Morgenstern}},\ and\
  \bibinfo {author} {\bibfnamefont {A.}~\bibnamefont {Bilewicz}},\ }\bibfield
  {title} {\bibinfo {title} {Radiochemical separation of 224ra from 232u and
  228th sources for 224ra/212pb/212bi generator},\ }\href
  {https://doi.org/10.1016/j.apradiso.2021.109655} {\bibfield  {journal}
  {\bibinfo  {journal} {Applied Radiation and Isotopes}\ }\textbf {\bibinfo
  {volume} {172}},\ \bibinfo {pages} {109655} (\bibinfo {year}
  {2021})}\BibitemShut {NoStop}%
\bibitem [{\citenamefont {Kuznetsov}\ \emph {et~al.}(2012)\citenamefont
  {Kuznetsov}, \citenamefont {Butkalyuk} \emph {et~al.}}]{Kuznetsov2012}%
  \BibitemOpen
  \bibfield  {author} {\bibinfo {author} {\bibfnamefont {R.~A.}\ \bibnamefont
  {Kuznetsov}}, \bibinfo {author} {\bibfnamefont {P.~S.}\ \bibnamefont
  {Butkalyuk}}, \emph {et~al.},\ }\bibfield  {title} {\bibinfo {title} {Yields
  of activation products in ${}^{226}${Ra} irradiation in the high-flux {SM}
  reactor},\ }\href {https://doi.org/10.1134/S1066362212040121} {\bibfield
  {journal} {\bibinfo  {journal} {Radiochemistry}\ }\textbf {\bibinfo {volume}
  {54}},\ \bibinfo {pages} {383} (\bibinfo {year} {2012})}\BibitemShut
  {NoStop}%
\bibitem [{\citenamefont {Melville}\ and\ \citenamefont
  {Melville}(2013)}]{melville2013theoretical}%
  \BibitemOpen
  \bibfield  {author} {\bibinfo {author} {\bibfnamefont {G.}~\bibnamefont
  {Melville}}\ and\ \bibinfo {author} {\bibfnamefont {P.}~\bibnamefont
  {Melville}},\ }\bibfield  {title} {\bibinfo {title} {A theoretical model for
  the production of ac-225 for cancer therapy by neutron capture transmutation
  of ra-226},\ }\href {https://doi.org/10.1016/j.apradiso.2012.09.019}
  {\bibfield  {journal} {\bibinfo  {journal} {Applied Radiation and Isotopes}\
  }\textbf {\bibinfo {volume} {72}},\ \bibinfo {pages} {152} (\bibinfo {year}
  {2013})}\BibitemShut {NoStop}%
\bibitem [{\citenamefont {Zimmermann}(2024)}]{Zimmermann2024}%
  \BibitemOpen
  \bibfield  {author} {\bibinfo {author} {\bibfnamefont {R.}~\bibnamefont
  {Zimmermann}},\ }\bibfield  {title} {\bibinfo {title} {Is $^{212}$pb really
  happening? the post-$^{177}$lu/$^{225}$ac blockbuster?},\ }\href
  {https://doi.org/10.2967/jnumed.123.266774} {\bibfield  {journal} {\bibinfo
  {journal} {J. Nucl. Med.}\ }\textbf {\bibinfo {volume} {65}},\ \bibinfo
  {pages} {176} (\bibinfo {year} {2024})}\BibitemShut {NoStop}%
\bibitem [{\citenamefont {Morgenstern}\ \emph {et~al.}(2003)\citenamefont
  {Morgenstern}, \citenamefont {Lebeda}, \citenamefont {Stursa}, \citenamefont
  {Capote}, \citenamefont {Sintes}, \citenamefont {Bruchertseifer},
  \citenamefont {Schwarzbach},\ and\ \citenamefont
  {Apostolidis}}]{Morgenstern2003}%
  \BibitemOpen
  \bibfield  {author} {\bibinfo {author} {\bibfnamefont {A.}~\bibnamefont
  {Morgenstern}}, \bibinfo {author} {\bibfnamefont {O.}~\bibnamefont {Lebeda}},
  \bibinfo {author} {\bibfnamefont {J.}~\bibnamefont {Stursa}}, \bibinfo
  {author} {\bibfnamefont {R.}~\bibnamefont {Capote}}, \bibinfo {author}
  {\bibfnamefont {B.}~\bibnamefont {Sintes}}, \bibinfo {author} {\bibfnamefont
  {F.}~\bibnamefont {Bruchertseifer}}, \bibinfo {author} {\bibfnamefont
  {R.}~\bibnamefont {Schwarzbach}},\ and\ \bibinfo {author} {\bibfnamefont
  {C.}~\bibnamefont {Apostolidis}},\ }\bibfield  {title} {\bibinfo {title}
  {Production of ${}^{235}${Np}, ${}^{236}${Pu} and ${}^{237}${Pu} via nuclear
  reactions on ${}^{235,236,238}${U} and ${}^{237}${Np} targets},\ }\href
  {https://doi.org/10.1524/ract.91.10.557.22475} {\bibfield  {journal}
  {\bibinfo  {journal} {Radiochim. Acta}\ }\textbf {\bibinfo {volume} {91}},\
  \bibinfo {pages} {557} (\bibinfo {year} {2003})}\BibitemShut {NoStop}%
\bibitem [{\citenamefont {Aaltonen}\ \emph {et~al.}(1993)\citenamefont
  {Aaltonen}, \citenamefont {Brenner}, \citenamefont {Dmitriev} \emph
  {et~al.}}]{Aaltonen1993}%
  \BibitemOpen
  \bibfield  {author} {\bibinfo {author} {\bibfnamefont {J.}~\bibnamefont
  {Aaltonen}}, \bibinfo {author} {\bibfnamefont {M.}~\bibnamefont {Brenner}},
  \bibinfo {author} {\bibfnamefont {V.~D.}\ \bibnamefont {Dmitriev}}, \emph
  {et~al.},\ }\bibfield  {title} {\bibinfo {title} {Formation of ${}^{236}${Pu}
  and ${}^{237}${Pu} by proton-induced reactions on ${}^{237}${Np}},\ }\href
  {https://doi.org/10.1016/0969-8043(93)90024-5} {\bibfield  {journal}
  {\bibinfo  {journal} {Applied Radiation and Isotopes}\ }\textbf {\bibinfo
  {volume} {44}},\ \bibinfo {pages} {831} (\bibinfo {year} {1993})}\BibitemShut
  {NoStop}%
\bibitem [{\citenamefont {Li}\ \emph {et~al.}(2023)\citenamefont {Li},
  \citenamefont {Stenberg},\ and\ \citenamefont {Larsen}}]{Li2023generator}%
  \BibitemOpen
  \bibfield  {author} {\bibinfo {author} {\bibfnamefont {R.~G.}\ \bibnamefont
  {Li}}, \bibinfo {author} {\bibfnamefont {V.~Y.}\ \bibnamefont {Stenberg}},\
  and\ \bibinfo {author} {\bibfnamefont {R.~H.}\ \bibnamefont {Larsen}},\
  }\bibfield  {title} {\bibinfo {title} {An experimental generator for
  production of high-purity $^{212}${Pb} for use in radiopharmaceuticals},\
  }\href {https://doi.org/10.2967/jnumed.122.264009} {\bibfield  {journal}
  {\bibinfo  {journal} {Journal of Nuclear Medicine}\ }\textbf {\bibinfo
  {volume} {64}},\ \bibinfo {pages} {173} (\bibinfo {year} {2023})}\BibitemShut
  {NoStop}%
\bibitem [{\citenamefont {Engholm}\ \emph {et~al.}(1986)\citenamefont
  {Engholm}, \citenamefont {Cheng},\ and\ \citenamefont
  {Schultz}}]{engholm1986radioisotope}%
  \BibitemOpen
  \bibfield  {author} {\bibinfo {author} {\bibfnamefont {B.~A.}\ \bibnamefont
  {Engholm}}, \bibinfo {author} {\bibfnamefont {E.~T.}\ \bibnamefont {Cheng}},\
  and\ \bibinfo {author} {\bibfnamefont {K.~R.}\ \bibnamefont {Schultz}},\
  }\bibfield  {title} {\bibinfo {title} {Radioisotope production in fusion
  reactors},\ }\href {https://doi.org/10.13182/FST86-A24908} {\bibfield
  {journal} {\bibinfo  {journal} {Fusion technology}\ }\textbf {\bibinfo
  {volume} {10}},\ \bibinfo {pages} {1290} (\bibinfo {year}
  {1986})}\BibitemShut {NoStop}%
\bibitem [{\citenamefont {Bourque}\ \emph {et~al.}(1988)\citenamefont
  {Bourque}, \citenamefont {Schultz},\ and\ \citenamefont
  {Staff}}]{Bourque1988FAME}%
  \BibitemOpen
  \bibfield  {author} {\bibinfo {author} {\bibfnamefont {R.}~\bibnamefont
  {Bourque}}, \bibinfo {author} {\bibfnamefont {K.}~\bibnamefont {Schultz}},\
  and\ \bibinfo {author} {\bibfnamefont {P.}~\bibnamefont {Staff}},\ }\href
  {https://archive.org/download/DTIC_ADA243768/DTIC_ADA243768.pdf} {\emph
  {\bibinfo {title} {Fusion Applications and Market Evaluation (FAME)
  Study}}},\ \bibinfo {type} {Technical Report}\ \bibinfo {number}
  {GA\mbox{-}A18658 / UCRL\mbox{-}21073 / UC\mbox{-}420 / UC\mbox{-}424 /
  UC\mbox{-}712}\ (\bibinfo  {institution} {GA Technologies, Inc.\ (General
  Atomics)},\ \bibinfo {address} {San Diego, CA},\ \bibinfo {year} {1988})\
  \bibinfo {note} {prepared under Subcontract 8236305 for Lawrence Livermore
  National Laboratory; DTIC accession AD\mbox{-}A243 768}\BibitemShut {NoStop}%
\bibitem [{\citenamefont {Leung}\ \emph {et~al.}(2018)\citenamefont {Leung},
  \citenamefont {Leung},\ and\ \citenamefont {Melville}}]{Leung2018_CompactNG}%
  \BibitemOpen
  \bibfield  {author} {\bibinfo {author} {\bibfnamefont {K.~N.}\ \bibnamefont
  {Leung}}, \bibinfo {author} {\bibfnamefont {J.~K.}\ \bibnamefont {Leung}},\
  and\ \bibinfo {author} {\bibfnamefont {G.}~\bibnamefont {Melville}},\
  }\bibfield  {title} {\bibinfo {title} {Feasibility study on medical isotope
  production using a compact neutron generator},\ }\href
  {https://doi.org/10.1016/j.apradiso.2018.02.026} {\bibfield  {journal}
  {\bibinfo  {journal} {Applied Radiation and Isotopes}\ }\textbf {\bibinfo
  {volume} {137}},\ \bibinfo {pages} {23} (\bibinfo {year} {2018})}\BibitemShut
  {NoStop}%
\bibitem [{\citenamefont {Li}\ and\ \citenamefont
  {Zheng}(2023)}]{li2023feasibility}%
  \BibitemOpen
  \bibfield  {author} {\bibinfo {author} {\bibfnamefont {J.}~\bibnamefont
  {Li}}\ and\ \bibinfo {author} {\bibfnamefont {S.}~\bibnamefont {Zheng}},\
  }\bibfield  {title} {\bibinfo {title} {Feasibility study to byproduce medical
  radioisotopes in a fusion reactor},\ }\href
  {https://doi.org/10.3390/molecules28052040} {\bibfield  {journal} {\bibinfo
  {journal} {Molecules}\ }\textbf {\bibinfo {volume} {28}},\ \bibinfo {pages}
  {2040} (\bibinfo {year} {2023})}\BibitemShut {NoStop}%
\bibitem [{\citenamefont {Pereslavtsev}\ \emph {et~al.}(2024)\citenamefont
  {Pereslavtsev}, \citenamefont {Bachmann}, \citenamefont {Elbez-Uzan},\ and\
  \citenamefont {Park}}]{pereslavtsev2024potential}%
  \BibitemOpen
  \bibfield  {author} {\bibinfo {author} {\bibfnamefont {P.}~\bibnamefont
  {Pereslavtsev}}, \bibinfo {author} {\bibfnamefont {C.}~\bibnamefont
  {Bachmann}}, \bibinfo {author} {\bibfnamefont {J.}~\bibnamefont
  {Elbez-Uzan}},\ and\ \bibinfo {author} {\bibfnamefont {J.~H.}\ \bibnamefont
  {Park}},\ }\bibfield  {title} {\bibinfo {title} {Potential of radioactive
  isotopes production in demo for commercial use},\ }\href
  {https://doi.org/10.3390/app14010442} {\bibfield  {journal} {\bibinfo
  {journal} {Applied Sciences}\ }\textbf {\bibinfo {volume} {14}},\ \bibinfo
  {pages} {442} (\bibinfo {year} {2024})}\BibitemShut {NoStop}%
\bibitem [{\citenamefont {Evitts}\ \emph {et~al.}(2025)\citenamefont {Evitts},
  \citenamefont {Miller}, \citenamefont {{Da Pieve}}, \citenamefont {Turner},\
  and\ \citenamefont {Borini}}]{evitts2025theoretical}%
  \BibitemOpen
  \bibfield  {author} {\bibinfo {author} {\bibfnamefont {L.~J.}\ \bibnamefont
  {Evitts}}, \bibinfo {author} {\bibfnamefont {P.~W.}\ \bibnamefont {Miller}},
  \bibinfo {author} {\bibfnamefont {C.}~\bibnamefont {{Da Pieve}}}, \bibinfo
  {author} {\bibfnamefont {A.}~\bibnamefont {Turner}},\ and\ \bibinfo {author}
  {\bibfnamefont {S.}~\bibnamefont {Borini}},\ }\bibfield  {title} {\bibinfo
  {title} {Theoretical novel medical isotope production with deuterium-tritium
  fusion technology},\ }\href {https://doi.org/10.1016/j.apradiso.2025.112163}
  {\bibfield  {journal} {\bibinfo  {journal} {Applied Radiation and Isotopes}\
  }\textbf {\bibinfo {volume} {226}},\ \bibinfo {pages} {112163} (\bibinfo
  {year} {2025})}\BibitemShut {NoStop}%
\bibitem [{\citenamefont {Parisi}\ \emph
  {et~al.}(2025{\natexlab{a}})\citenamefont {Parisi}, \citenamefont
  {Rutkowski}, \citenamefont {Harter}, \citenamefont {Schwartz},\ and\
  \citenamefont {Chen}}]{Parisi2025}%
  \BibitemOpen
  \bibfield  {author} {\bibinfo {author} {\bibfnamefont {J.~F.}\ \bibnamefont
  {Parisi}}, \bibinfo {author} {\bibfnamefont {A.}~\bibnamefont {Rutkowski}},
  \bibinfo {author} {\bibfnamefont {J.}~\bibnamefont {Harter}}, \bibinfo
  {author} {\bibfnamefont {J.~A.}\ \bibnamefont {Schwartz}},\ and\ \bibinfo
  {author} {\bibfnamefont {S.}~\bibnamefont {Chen}},\ }\bibfield  {title}
  {\bibinfo {title} {Production of high-specific-activity radioisotopes using
  high-energy fusion neutrons},\ }\href {https://arxiv.org/abs/2511.02814}
  {\bibfield  {journal} {\bibinfo  {journal} {arXiv preprint}\ } (\bibinfo
  {year} {2025}{\natexlab{a}})},\ \Eprint {https://arxiv.org/abs/2511.02814}
  {2511.02814} \BibitemShut {NoStop}%
\bibitem [{\citenamefont {Parisi}\ \emph
  {et~al.}(2025{\natexlab{b}})\citenamefont {Parisi}, \citenamefont {Schwartz},
  \citenamefont {Wurzel}, \citenamefont {Rutkowski},\ and\ \citenamefont
  {Harter}}]{Parisi2025IsotopeFusion}%
  \BibitemOpen
  \bibfield  {author} {\bibinfo {author} {\bibfnamefont {J.~F.}\ \bibnamefont
  {Parisi}}, \bibinfo {author} {\bibfnamefont {J.~A.}\ \bibnamefont
  {Schwartz}}, \bibinfo {author} {\bibfnamefont {S.~E.}\ \bibnamefont
  {Wurzel}}, \bibinfo {author} {\bibfnamefont {A.}~\bibnamefont {Rutkowski}},\
  and\ \bibinfo {author} {\bibfnamefont {J.}~\bibnamefont {Harter}},\
  }\bibfield  {title} {\bibinfo {title} {Isotope production in fusion
  systems},\ }\href {https://arxiv.org/abs/2512.09242} {\bibfield  {journal}
  {\bibinfo  {journal} {arXiv preprint arXiv:2512.09242}\ } (\bibinfo {year}
  {2025}{\natexlab{b}})}\BibitemShut {NoStop}%
\bibitem [{\citenamefont {Parisi}\ and\ \citenamefont
  {Rutkowski}(2025)}]{Parisi2025muCF}%
  \BibitemOpen
  \bibfield  {author} {\bibinfo {author} {\bibfnamefont {J.~F.}\ \bibnamefont
  {Parisi}}\ and\ \bibinfo {author} {\bibfnamefont {A.}~\bibnamefont
  {Rutkowski}},\ }\bibfield  {title} {\bibinfo {title} {Isotope production in
  muon-catalyzed-fusion systems},\ }\href {https://arxiv.org/abs/2511.20951}
  {\bibfield  {journal} {\bibinfo  {journal} {arXiv preprint arXiv:2511.20951}\
  } (\bibinfo {year} {2025})}\BibitemShut {NoStop}%
\bibitem [{\citenamefont {Parisi}\ and\ \citenamefont
  {Schiller}(2026)}]{parisi2026neutronvalue}%
  \BibitemOpen
  \bibfield  {author} {\bibinfo {author} {\bibfnamefont {J.~F.}\ \bibnamefont
  {Parisi}}\ and\ \bibinfo {author} {\bibfnamefont {K.}~\bibnamefont
  {Schiller}},\ }\bibfield  {title} {\bibinfo {title} {The value and cost of
  fusion neutrons},\ }\href {https://arxiv.org/abs/2603.00835} {\bibfield
  {journal} {\bibinfo  {journal} {arXiv preprint arXiv:2603.00835}\ } (\bibinfo
  {year} {2026})}\BibitemShut {NoStop}%
\bibitem [{\citenamefont {Parisi}(2026{\natexlab{a}})}]{Parisi2026BetaBattery}%
  \BibitemOpen
  \bibfield  {author} {\bibinfo {author} {\bibfnamefont {J.~F.}\ \bibnamefont
  {Parisi}},\ }\bibfield  {title} {\bibinfo {title} {Production of nuclear
  battery $\beta^{-}$ emitters driven by fusion neutrons},\ }\href
  {https://arxiv.org/abs/2605.20260} {\bibfield  {journal} {\bibinfo  {journal}
  {arXiv preprint arXiv:2605.20260}\ } (\bibinfo {year}
  {2026}{\natexlab{a}})}\BibitemShut {NoStop}%
\bibitem [{\citenamefont
  {Parisi}(2026{\natexlab{b}})}]{Parisi2026FusionBattery}%
  \BibitemOpen
  \bibfield  {author} {\bibinfo {author} {\bibfnamefont {J.~F.}\ \bibnamefont
  {Parisi}},\ }\bibfield  {title} {\bibinfo {title} {Scalable production of
  nuclear battery alpha emitters using fusion neutrons},\ }\href
  {https://arxiv.org/abs/2608.12963} {\bibfield  {journal} {\bibinfo  {journal}
  {arXiv preprint arXiv:2608.12963}\ } (\bibinfo {year}
  {2026}{\natexlab{b}})}\BibitemShut {NoStop}%
\bibitem [{\citenamefont {Rutkowski}\ \emph {et~al.}(2025)\citenamefont
  {Rutkowski}, \citenamefont {Harter},\ and\ \citenamefont
  {Parisi}}]{Rutkowski2025}%
  \BibitemOpen
  \bibfield  {author} {\bibinfo {author} {\bibfnamefont {A.}~\bibnamefont
  {Rutkowski}}, \bibinfo {author} {\bibfnamefont {J.}~\bibnamefont {Harter}},\
  and\ \bibinfo {author} {\bibfnamefont {J.~F.}\ \bibnamefont {Parisi}},\
  }\bibfield  {title} {\bibinfo {title} {Scalable chrysopoeia via $(n,2n)$
  reactions driven by deuterium-tritium fusion neutrons},\ }\bibfield
  {journal} {\bibinfo  {journal} {arXiv preprint arXiv:2507.13461}\ }\href
  {https://doi.org/10.48550/arXiv.2507.13461} {10.48550/arXiv.2507.13461}
  (\bibinfo {year} {2025}),\ \Eprint {https://arxiv.org/abs/2507.13461}
  {2507.13461 [physics.plasm-ph]} \BibitemShut {NoStop}%
\bibitem [{\citenamefont {Lancker}\ \emph {et~al.}(1999)\citenamefont
  {Lancker}, \citenamefont {Herer}, \citenamefont {Cleland}, \citenamefont
  {Jongen},\ and\ \citenamefont {Abs}}]{VanLancker1999}%
  \BibitemOpen
  \bibfield  {author} {\bibinfo {author} {\bibfnamefont {M.~V.}\ \bibnamefont
  {Lancker}}, \bibinfo {author} {\bibfnamefont {A.}~\bibnamefont {Herer}},
  \bibinfo {author} {\bibfnamefont {M.~R.}\ \bibnamefont {Cleland}}, \bibinfo
  {author} {\bibfnamefont {Y.}~\bibnamefont {Jongen}},\ and\ \bibinfo {author}
  {\bibfnamefont {M.}~\bibnamefont {Abs}},\ }\bibfield  {title} {\bibinfo
  {title} {The {IBA} rhodotron: an industrial high-voltage high-powered
  electron beam accelerator for polymers radiation processing},\ }\href
  {https://doi.org/10.1016/S0168-583X(99)00099-3} {\bibfield  {journal}
  {\bibinfo  {journal} {Nucl. Instrum. Methods Phys. Res. B}\ }\textbf
  {\bibinfo {volume} {151}},\ \bibinfo {pages} {242} (\bibinfo {year}
  {1999})}\BibitemShut {NoStop}%
\bibitem [{\citenamefont {Starovoitova}\ \emph {et~al.}(2014)\citenamefont
  {Starovoitova}, \citenamefont {Tchelidze},\ and\ \citenamefont
  {Wells}}]{Starovoitova2014}%
  \BibitemOpen
  \bibfield  {author} {\bibinfo {author} {\bibfnamefont {V.~N.}\ \bibnamefont
  {Starovoitova}}, \bibinfo {author} {\bibfnamefont {L.}~\bibnamefont
  {Tchelidze}},\ and\ \bibinfo {author} {\bibfnamefont {D.~P.}\ \bibnamefont
  {Wells}},\ }\bibfield  {title} {\bibinfo {title} {Production of medical
  radioisotopes with linear accelerators},\ }\href
  {https://doi.org/10.1016/j.apradiso.2013.11.122} {\bibfield  {journal}
  {\bibinfo  {journal} {Appl. Radiat. Isot.}\ }\textbf {\bibinfo {volume}
  {85}},\ \bibinfo {pages} {39} (\bibinfo {year} {2014})}\BibitemShut {NoStop}%
\bibitem [{\citenamefont {Hawkins}\ \emph {et~al.}(2025)\citenamefont
  {Hawkins}, \citenamefont {Drain}, \citenamefont {Baumeister}, \citenamefont
  {Gonzalez}, \citenamefont {Wheeless},\ and\ \citenamefont
  {Devries}}]{hawkins2025cu}%
  \BibitemOpen
  \bibfield  {author} {\bibinfo {author} {\bibfnamefont {C.}~\bibnamefont
  {Hawkins}}, \bibinfo {author} {\bibfnamefont {T.}~\bibnamefont {Drain}},
  \bibinfo {author} {\bibfnamefont {J.}~\bibnamefont {Baumeister}}, \bibinfo
  {author} {\bibfnamefont {M.}~\bibnamefont {Gonzalez}}, \bibinfo {author}
  {\bibfnamefont {L.}~\bibnamefont {Wheeless}},\ and\ \bibinfo {author}
  {\bibfnamefont {D.}~\bibnamefont {Devries}},\ }\href
  {https://jnm.snmjournals.org/content/66/supplement_1/251892} {\bibinfo
  {title} {Cu-67 production at northstar-progress toward scale-up}} (\bibinfo
  {year} {2025})\BibitemShut {NoStop}%
\bibitem [{\citenamefont {Freidberg}\ and\ \citenamefont
  {Kadak}(2009)}]{Freidberg2009}%
  \BibitemOpen
  \bibfield  {author} {\bibinfo {author} {\bibfnamefont {J.~P.}\ \bibnamefont
  {Freidberg}}\ and\ \bibinfo {author} {\bibfnamefont {A.~C.}\ \bibnamefont
  {Kadak}},\ }\bibfield  {title} {\bibinfo {title} {Fusion--fission hybrids
  revisited},\ }\href {https://doi.org/10.1038/nphys1288} {\bibfield  {journal}
  {\bibinfo  {journal} {Nature Physics}\ }\textbf {\bibinfo {volume} {5}},\
  \bibinfo {pages} {370} (\bibinfo {year} {2009})}\BibitemShut {NoStop}%
\bibitem [{\citenamefont {Stacey}\ \emph {et~al.}(2008)\citenamefont {Stacey}
  \emph {et~al.}}]{Stacey2008}%
  \BibitemOpen
  \bibfield  {author} {\bibinfo {author} {\bibfnamefont {W.~M.}\ \bibnamefont
  {Stacey}} \emph {et~al.},\ }\bibfield  {title} {\bibinfo {title} {A {TRU-Zr}
  metal-fuel sodium-cooled fast subcritical advanced burner reactor},\ }\href
  {https://doi.org/10.13182/NT08-A3933} {\bibfield  {journal} {\bibinfo
  {journal} {Nuclear Technology}\ }\textbf {\bibinfo {volume} {162}},\ \bibinfo
  {pages} {53} (\bibinfo {year} {2008})}\BibitemShut {NoStop}%
\bibitem [{\citenamefont {Kuteev}\ \emph {et~al.}(2015)\citenamefont {Kuteev},
  \citenamefont {Azizov}, \citenamefont {Alexeev}, \citenamefont {Ignatiev},
  \citenamefont {Subbotin},\ and\ \citenamefont {Tsibulskiy}}]{Kuteev2015}%
  \BibitemOpen
  \bibfield  {author} {\bibinfo {author} {\bibfnamefont {B.~V.}\ \bibnamefont
  {Kuteev}}, \bibinfo {author} {\bibfnamefont {E.~A.}\ \bibnamefont {Azizov}},
  \bibinfo {author} {\bibfnamefont {P.~N.}\ \bibnamefont {Alexeev}}, \bibinfo
  {author} {\bibfnamefont {V.~V.}\ \bibnamefont {Ignatiev}}, \bibinfo {author}
  {\bibfnamefont {S.~A.}\ \bibnamefont {Subbotin}},\ and\ \bibinfo {author}
  {\bibfnamefont {V.~F.}\ \bibnamefont {Tsibulskiy}},\ }\bibfield  {title}
  {\bibinfo {title} {Development of {DEMO-FNS} tokamak for fusion and hybrid
  technologies},\ }\href {https://doi.org/10.1088/0029-5515/55/7/073035}
  {\bibfield  {journal} {\bibinfo  {journal} {Nuclear Fusion}\ }\textbf
  {\bibinfo {volume} {55}},\ \bibinfo {pages} {073035} (\bibinfo {year}
  {2015})}\BibitemShut {NoStop}%
\bibitem [{\citenamefont {Wu}\ and\ \citenamefont {{FDS Team}}(2006)}]{Wu2006}%
  \BibitemOpen
  \bibfield  {author} {\bibinfo {author} {\bibfnamefont {Y.}~\bibnamefont
  {Wu}}\ and\ \bibinfo {author} {\bibnamefont {{FDS Team}}},\ }\bibfield
  {title} {\bibinfo {title} {Conceptual design of the {China} fusion-driven
  subcritical system {FDS-I}},\ }\href
  {https://doi.org/10.1016/j.fusengdes.2005.10.015} {\bibfield  {journal}
  {\bibinfo  {journal} {Fusion Engineering and Design}\ }\textbf {\bibinfo
  {volume} {81}},\ \bibinfo {pages} {1305} (\bibinfo {year}
  {2006})}\BibitemShut {NoStop}%
\bibitem [{\citenamefont {{NucNet}}(2025)}]{NucNet2025xinghuo}%
  \BibitemOpen
  \bibfield  {author} {\bibinfo {author} {\bibnamefont {{NucNet}}},\
  }\href@noop {} {\bibinfo {title} {China aims to operate world's first hybrid
  fusion-fission nuclear plant by 2030}},\ \bibinfo {howpublished} {NucNet
  news, \url{https://www.nucnet.org}} (\bibinfo {year} {2025}),\ \bibinfo
  {note} {5 March 2025}\BibitemShut {NoStop}%
\bibitem [{\citenamefont {Perron}\ \emph {et~al.}(2020)\citenamefont {Perron},
  \citenamefont {Gendron},\ and\ \citenamefont {Causey}}]{Perron2020}%
  \BibitemOpen
  \bibfield  {author} {\bibinfo {author} {\bibfnamefont {R.}~\bibnamefont
  {Perron}}, \bibinfo {author} {\bibfnamefont {D.}~\bibnamefont {Gendron}},\
  and\ \bibinfo {author} {\bibfnamefont {P.~W.}\ \bibnamefont {Causey}},\
  }\bibfield  {title} {\bibinfo {title} {Construction of a thorium/actinium
  generator at the {Canadian Nuclear Laboratories}},\ }\href
  {https://doi.org/10.1016/j.apradiso.2020.109262} {\bibfield  {journal}
  {\bibinfo  {journal} {Appl. Radiat. Isot.}\ }\textbf {\bibinfo {volume}
  {164}},\ \bibinfo {pages} {109262} (\bibinfo {year} {2020})}\BibitemShut
  {NoStop}%
\bibitem [{\citenamefont {Molnar}(2019)}]{Molnar2019}%
  \BibitemOpen
  \bibfield  {author} {\bibinfo {author} {\bibfnamefont {M.~M.}\ \bibnamefont
  {Molnar}},\ }\emph {\bibinfo {title} {Experimental Approaches for the
  Production of Thorium-229}},\ \href {https://hdl.handle.net/2142/106285}
  {Master's thesis},\ \bibinfo  {school} {University of Illinois at
  Urbana-Champaign} (\bibinfo {year} {2019})\BibitemShut {NoStop}%
\bibitem [{\citenamefont {Heilbronn}(2024)}]{Heilbronn2024}%
  \BibitemOpen
  \bibfield  {author} {\bibinfo {author} {\bibfnamefont {L.~H.}\ \bibnamefont
  {Heilbronn}},\ }\href {https://doi.org/10.2172/2282305} {\emph {\bibinfo
  {title} {Novel Methods for {Th-229} Production through Fast Neutron
  Irradiation of {Th-230} and Charged Particle Irradiation of {Th-230} and
  {Th-232}. Final Report {DE-SC0020140}}}},\ \bibinfo {type} {Tech. Rep.}\
  \bibinfo {number} {DOE-SC0020140}\ (\bibinfo  {institution} {University of
  Tennessee, Knoxville},\ \bibinfo {year} {2024})\BibitemShut {NoStop}%
\bibitem [{\citenamefont {Morrell}\ \emph {et~al.}(2023)\citenamefont
  {Morrell}, \citenamefont {Voyles}, \citenamefont {Batchelder}, \citenamefont
  {Brown},\ and\ \citenamefont {Bernstein}}]{Morrell2023}%
  \BibitemOpen
  \bibfield  {author} {\bibinfo {author} {\bibfnamefont {J.~T.}\ \bibnamefont
  {Morrell}}, \bibinfo {author} {\bibfnamefont {A.~S.}\ \bibnamefont {Voyles}},
  \bibinfo {author} {\bibfnamefont {J.~C.}\ \bibnamefont {Batchelder}},
  \bibinfo {author} {\bibfnamefont {J.~A.}\ \bibnamefont {Brown}},\ and\
  \bibinfo {author} {\bibfnamefont {L.~A.}\ \bibnamefont {Bernstein}},\
  }\bibfield  {title} {\bibinfo {title} {Secondary neutron production from
  thick target deuteron breakup},\ }\href
  {https://doi.org/10.1103/PhysRevC.108.024616} {\bibfield  {journal} {\bibinfo
   {journal} {Physical Review C}\ }\textbf {\bibinfo {volume} {108}},\ \bibinfo
  {pages} {024616} (\bibinfo {year} {2023})}\BibitemShut {NoStop}%
\bibitem [{\citenamefont {Figgins}\ and\ \citenamefont
  {Kirby}(1966)}]{Figgins1966ionium}%
  \BibitemOpen
  \bibfield  {author} {\bibinfo {author} {\bibfnamefont {P.~E.}\ \bibnamefont
  {Figgins}}\ and\ \bibinfo {author} {\bibfnamefont {H.~W.}\ \bibnamefont
  {Kirby}},\ }\href {https://doi.org/10.2172/4478230} {\emph {\bibinfo {title}
  {Survey of Sources of Ionium (Thorium-230)}}},\ \bibinfo {type} {Tech. Rep.}\
  \bibinfo {number} {MLM-1349}\ (\bibinfo  {institution} {Mound Laboratory,
  Monsanto Research Corporation},\ \bibinfo {year} {1966})\BibitemShut
  {NoStop}%
\bibitem [{\citenamefont {Kim}\ and\ \citenamefont {Born}(1971)}]{klm1971}%
  \BibitemOpen
  \bibfield  {author} {\bibinfo {author} {\bibfnamefont {J.~I.}\ \bibnamefont
  {Kim}}\ and\ \bibinfo {author} {\bibfnamefont {H.-J.}\ \bibnamefont {Born}},\
  }\bibfield  {title} {\bibinfo {title} {The production of ${}^{231}$pa and
  ${}^{232}$u by reactor neutron irradiation of ${}^{230}$th and ${}^{232}$th
  mixture},\ }\href {https://doi.org/10.1524/ract.1971.16.34.160} {\bibfield
  {journal} {\bibinfo  {journal} {Radiochimica Acta}\ }\textbf {\bibinfo
  {volume} {16}},\ \bibinfo {pages} {160} (\bibinfo {year} {1971})}\BibitemShut
  {NoStop}%
\bibitem [{\citenamefont {{OECD Nuclear Energy Agency and International Atomic
  Energy Agency}}(2023)}]{RedBook2023}%
  \BibitemOpen
  \bibfield  {author} {\bibinfo {author} {\bibnamefont {{OECD Nuclear Energy
  Agency and International Atomic Energy Agency}}},\ }\href
  {https://doi.org/10.1787/2c4e111b-en} {\emph {\bibinfo {title} {Uranium 2022:
  Resources, Production and Demand}}},\ \bibinfo {type} {Tech. Rep.}\ \bibinfo
  {number} {NEA No. 7634}\ (\bibinfo  {institution} {OECD Publishing},\
  \bibinfo {address} {Paris},\ \bibinfo {year} {2023})\BibitemShut {NoStop}%
\bibitem [{\citenamefont {Rohrmann}(1960{\natexlab{a}})}]{Rohrman1960}%
  \BibitemOpen
  \bibfield  {author} {\bibinfo {author} {\bibfnamefont {C.~A.}\ \bibnamefont
  {Rohrmann}},\ }\href@noop {} {\emph {\bibinfo {title} {A study of the
  feasibility for the large scale recovery of ionium (thorium-230) from the
  uranium ore milling industry in the {United States}}}},\ \bibinfo {type}
  {Tech. Rep.}\ \bibinfo {number} {HW-65518}\ (\bibinfo  {institution} {Hanford
  Laboratories Operation, General Electric Company},\ \bibinfo {year}
  {1960})\BibitemShut {NoStop}%
\bibitem [{\citenamefont {Iwahashi}\ \emph {et~al.}(2022)\citenamefont
  {Iwahashi}, \citenamefont {Kawamoto}, \citenamefont {Sasaki},\ and\
  \citenamefont {Takaki}}]{Iwahashi2022}%
  \BibitemOpen
  \bibfield  {author} {\bibinfo {author} {\bibfnamefont {D.}~\bibnamefont
  {Iwahashi}}, \bibinfo {author} {\bibfnamefont {K.}~\bibnamefont {Kawamoto}},
  \bibinfo {author} {\bibfnamefont {Y.}~\bibnamefont {Sasaki}},\ and\ \bibinfo
  {author} {\bibfnamefont {N.}~\bibnamefont {Takaki}},\ }\bibfield  {title}
  {\bibinfo {title} {Neutronic study on {Ac-225} production for cancer therapy
  by (n,2n) reaction of {Ra-226} or {Th-230} using fast reactor {Joyo}},\
  }\href {https://doi.org/10.3390/pr10071239} {\bibfield  {journal} {\bibinfo
  {journal} {Processes}\ }\textbf {\bibinfo {volume} {10}},\ \bibinfo {pages}
  {1239} (\bibinfo {year} {2022})}\BibitemShut {NoStop}%
\bibitem [{\citenamefont {Coppinger}\ and\ \citenamefont
  {Rohrmann}(1959)}]{hw63239}%
  \BibitemOpen
  \bibfield  {author} {\bibinfo {author} {\bibfnamefont {E.~A.}\ \bibnamefont
  {Coppinger}}\ and\ \bibinfo {author} {\bibfnamefont {C.~A.}\ \bibnamefont
  {Rohrmann}},\ }\href {https://doi.org/10.2172/4093076} {\emph {\bibinfo
  {title} {Ionium (Thorium-230) for radioisotope preparation (status
  report)}}},\ \bibinfo {type} {Tech. Rep.}\ \bibinfo {number} {HW-63239}\
  (\bibinfo  {institution} {General Electric, Hanford Atomic Products
  Operation},\ \bibinfo {year} {1959})\BibitemShut {NoStop}%
\bibitem [{\citenamefont {Rohrmann}(1960{\natexlab{b}})}]{hw66600}%
  \BibitemOpen
  \bibfield  {author} {\bibinfo {author} {\bibfnamefont {C.~A.}\ \bibnamefont
  {Rohrmann}},\ }\href {https://doi.org/10.2172/4802354} {\emph {\bibinfo
  {title} {Ionium, uranium-232 and thorium-228: properties, applications and
  availability}}},\ \bibinfo {type} {Tech. Rep.}\ \bibinfo {number} {HW-66600}\
  (\bibinfo  {institution} {General Electric, Hanford Atomic Products
  Operation},\ \bibinfo {year} {1960})\BibitemShut {NoStop}%
\bibitem [{\citenamefont {Hertz}\ \emph {et~al.}(1983)\citenamefont {Hertz},
  \citenamefont {Figgins},\ and\ \citenamefont {Deal}}]{mlm2985}%
  \BibitemOpen
  \bibfield  {author} {\bibinfo {author} {\bibfnamefont {M.~R.}\ \bibnamefont
  {Hertz}}, \bibinfo {author} {\bibfnamefont {P.~E.}\ \bibnamefont {Figgins}},\
  and\ \bibinfo {author} {\bibfnamefont {W.~R.}\ \bibnamefont {Deal}},\ }\href
  {https://doi.org/10.2172/6462739} {\emph {\bibinfo {title} {Recovery of
  protactinium-231 and thorium-230 from {Cotter} concentrate: pilot plant
  operations and process development}}},\ \bibinfo {type} {Tech. Rep.}\
  \bibinfo {number} {MLM-2985}\ (\bibinfo  {institution} {Mound Laboratory},\
  \bibinfo {year} {1983})\BibitemShut {NoStop}%
\bibitem [{\citenamefont {Daily}\ and\ \citenamefont
  {McDuffee}(2020)}]{Daily2020Pu238}%
  \BibitemOpen
  \bibfield  {author} {\bibinfo {author} {\bibfnamefont {C.~R.}\ \bibnamefont
  {Daily}}\ and\ \bibinfo {author} {\bibfnamefont {J.~L.}\ \bibnamefont
  {McDuffee}},\ }\bibfield  {title} {\bibinfo {title} {Design studies for the
  optimization of ${}^{238}$pu production in {NpO}$_2$ targets irradiated at
  the high flux isotope reactor},\ }\href
  {https://doi.org/10.1080/00295450.2019.1674594} {\bibfield  {journal}
  {\bibinfo  {journal} {Nuclear Technology}\ }\textbf {\bibinfo {volume}
  {206}},\ \bibinfo {pages} {1182} (\bibinfo {year} {2020})}\BibitemShut
  {NoStop}%
\bibitem [{\citenamefont {{Chemical and Engineering
  News}}(1961)}]{CEN1961protactinium}%
  \BibitemOpen
  \bibfield  {author} {\bibinfo {author} {\bibnamefont {{Chemical and
  Engineering News}}},\ }\bibfield  {title} {\bibinfo {title} {Britain
  declassifies protactinium extraction},\ }\href
  {https://doi.org/10.1021/cen-v039n032.p048} {\bibfield  {journal} {\bibinfo
  {journal} {Chem. Eng. News}\ }\textbf {\bibinfo {volume} {39}},\ \bibinfo
  {pages} {48} (\bibinfo {year} {1961})}\BibitemShut {NoStop}%
\bibitem [{\citenamefont {Collins}\ \emph {et~al.}(1962)\citenamefont
  {Collins}, \citenamefont {Hillary}, \citenamefont {Nairn},\ and\
  \citenamefont {Phillips}}]{Collins1962protactinium}%
  \BibitemOpen
  \bibfield  {author} {\bibinfo {author} {\bibfnamefont {D.~A.}\ \bibnamefont
  {Collins}}, \bibinfo {author} {\bibfnamefont {J.~J.}\ \bibnamefont
  {Hillary}}, \bibinfo {author} {\bibfnamefont {J.~S.}\ \bibnamefont {Nairn}},\
  and\ \bibinfo {author} {\bibfnamefont {G.~M.}\ \bibnamefont {Phillips}},\
  }\bibfield  {title} {\bibinfo {title} {The development and application of a
  process for the recovery of over 100 g of protactinium-231 from a uranium
  refinery waste material},\ }\href
  {https://doi.org/10.1016/0022-1902(62)80040-2} {\bibfield  {journal}
  {\bibinfo  {journal} {Journal of Inorganic and Nuclear Chemistry}\ }\textbf
  {\bibinfo {volume} {24}},\ \bibinfo {pages} {441} (\bibinfo {year}
  {1962})}\BibitemShut {NoStop}%
\bibitem [{\citenamefont {Myasoedov}\ \emph {et~al.}(2006)\citenamefont
  {Myasoedov}, \citenamefont {Kirby},\ and\ \citenamefont
  {Tananaev}}]{kirby2006protactinium}%
  \BibitemOpen
  \bibfield  {author} {\bibinfo {author} {\bibfnamefont {B.~F.}\ \bibnamefont
  {Myasoedov}}, \bibinfo {author} {\bibfnamefont {H.~W.}\ \bibnamefont
  {Kirby}},\ and\ \bibinfo {author} {\bibfnamefont {I.~G.}\ \bibnamefont
  {Tananaev}},\ }\bibfield  {title} {\bibinfo {title} {Protactinium},\ }in\
  \href {https://doi.org/10.1007/1-4020-3598-5_4} {\emph {\bibinfo {booktitle}
  {The Chemistry of the Actinide and Transactinide Elements}}},\ \bibinfo
  {editor} {edited by\ \bibinfo {editor} {\bibfnamefont {L.~R.}\ \bibnamefont
  {Morss}}, \bibinfo {editor} {\bibfnamefont {N.~M.}\ \bibnamefont
  {Edelstein}},\ and\ \bibinfo {editor} {\bibfnamefont {J.}~\bibnamefont
  {Fuger}}}\ (\bibinfo  {publisher} {Springer},\ \bibinfo {address}
  {Dordrecht},\ \bibinfo {year} {2006})\ \bibinfo {edition} {3rd}\ ed.,\ pp.\
  \bibinfo {pages} {161--252}\BibitemShut {NoStop}%
\bibitem [{\citenamefont {Romano}\ \emph {et~al.}(2015)\citenamefont {Romano},
  \citenamefont {Horelik}, \citenamefont {Herman}, \citenamefont {Nelson},
  \citenamefont {Forget},\ and\ \citenamefont {Smith}}]{openmc}%
  \BibitemOpen
  \bibfield  {author} {\bibinfo {author} {\bibfnamefont {P.~K.}\ \bibnamefont
  {Romano}}, \bibinfo {author} {\bibfnamefont {N.~E.}\ \bibnamefont {Horelik}},
  \bibinfo {author} {\bibfnamefont {B.~R.}\ \bibnamefont {Herman}}, \bibinfo
  {author} {\bibfnamefont {A.~G.}\ \bibnamefont {Nelson}}, \bibinfo {author}
  {\bibfnamefont {B.}~\bibnamefont {Forget}},\ and\ \bibinfo {author}
  {\bibfnamefont {K.}~\bibnamefont {Smith}},\ }\bibfield  {title} {\bibinfo
  {title} {{OpenMC}: a state-of-the-art {Monte Carlo} code for research and
  development},\ }\href {https://doi.org/10.1016/j.anucene.2014.07.048}
  {\bibfield  {journal} {\bibinfo  {journal} {Ann. Nucl. Energy}\ }\textbf
  {\bibinfo {volume} {82}},\ \bibinfo {pages} {90} (\bibinfo {year}
  {2015})}\BibitemShut {NoStop}%
\bibitem [{\citenamefont {Davis}\ \emph {et~al.}(1986)\citenamefont {Davis}
  \emph {et~al.}}]{davis1986rtns}%
  \BibitemOpen
  \bibfield  {author} {\bibinfo {author} {\bibfnamefont {J.}~\bibnamefont
  {Davis}} \emph {et~al.},\ }\href {https://www.osti.gov/servlets/purl/5717383}
  {\emph {\bibinfo {title} {RTNS-II: experience at 14-MeV source strengths
  between 1 x 1013 and 4 x 1013 n/s}}},\ \bibinfo {type} {Tech. Rep.}\
  (\bibinfo  {institution} {Lawrence Livermore National Lab., CA (USA)},\
  \bibinfo {year} {1986})\BibitemShut {NoStop}%
\bibitem [{\citenamefont {Radchenko}\ \emph {et~al.}(2016)\citenamefont
  {Radchenko}, \citenamefont {Engle}, \citenamefont {Wilson}, \citenamefont
  {Maassen}, \citenamefont {Nortier}, \citenamefont {Birnbaum}, \citenamefont
  {John},\ and\ \citenamefont {Fassbender}}]{Radchenko2016}%
  \BibitemOpen
  \bibfield  {author} {\bibinfo {author} {\bibfnamefont {V.}~\bibnamefont
  {Radchenko}}, \bibinfo {author} {\bibfnamefont {J.~W.}\ \bibnamefont
  {Engle}}, \bibinfo {author} {\bibfnamefont {J.~J.}\ \bibnamefont {Wilson}},
  \bibinfo {author} {\bibfnamefont {J.~R.}\ \bibnamefont {Maassen}}, \bibinfo
  {author} {\bibfnamefont {F.~M.}\ \bibnamefont {Nortier}}, \bibinfo {author}
  {\bibfnamefont {E.~R.}\ \bibnamefont {Birnbaum}}, \bibinfo {author}
  {\bibfnamefont {K.~D.}\ \bibnamefont {John}},\ and\ \bibinfo {author}
  {\bibfnamefont {M.~E.}\ \bibnamefont {Fassbender}},\ }\bibfield  {title}
  {\bibinfo {title} {Formation cross-sections and chromatographic separation of
  protactinium isotopes formed in proton-irradiated thorium metal},\ }\href
  {https://doi.org/10.1515/ract-2015-2486} {\bibfield  {journal} {\bibinfo
  {journal} {Radiochimica Acta}\ }\textbf {\bibinfo {volume} {104}},\ \bibinfo
  {pages} {291} (\bibinfo {year} {2016})}\BibitemShut {NoStop}%
\bibitem [{\citenamefont {Caldwell}\ \emph {et~al.}(1980)\citenamefont
  {Caldwell}, \citenamefont {Dowdy}, \citenamefont {Berman}, \citenamefont
  {Alvarez},\ and\ \citenamefont {Meyer}}]{Caldwell1980}%
  \BibitemOpen
  \bibfield  {author} {\bibinfo {author} {\bibfnamefont {J.~T.}\ \bibnamefont
  {Caldwell}}, \bibinfo {author} {\bibfnamefont {E.~J.}\ \bibnamefont {Dowdy}},
  \bibinfo {author} {\bibfnamefont {B.~L.}\ \bibnamefont {Berman}}, \bibinfo
  {author} {\bibfnamefont {R.~A.}\ \bibnamefont {Alvarez}},\ and\ \bibinfo
  {author} {\bibfnamefont {P.}~\bibnamefont {Meyer}},\ }\bibfield  {title}
  {\bibinfo {title} {Giant resonance for the actinide nuclei: Photoneutron and
  photofission cross sections for $^{235}$u, $^{236}$u, $^{238}$u, and
  $^{232}$th},\ }\href {https://doi.org/10.1103/PhysRevC.21.1215} {\bibfield
  {journal} {\bibinfo  {journal} {Physical Review C}\ }\textbf {\bibinfo
  {volume} {21}},\ \bibinfo {pages} {1215} (\bibinfo {year}
  {1980})}\BibitemShut {NoStop}%
\bibitem [{\citenamefont {Mughabghab}(2018)}]{Mughabghab2018}%
  \BibitemOpen
  \bibfield  {author} {\bibinfo {author} {\bibfnamefont {S.~F.}\ \bibnamefont
  {Mughabghab}},\ }\href
  {https://shop.elsevier.com/books/atlas-of-neutron-resonances/mughabghab/978-0-444-63769-7}
  {\emph {\bibinfo {title} {Atlas of Neutron Resonances}}},\ \bibinfo {edition}
  {6th}\ ed.\ (\bibinfo  {publisher} {Elsevier},\ \bibinfo {address}
  {Amsterdam},\ \bibinfo {year} {2018})\BibitemShut {NoStop}%
\bibitem [{\citenamefont {Aleksandrov}\ \emph {et~al.}(1972)\citenamefont
  {Aleksandrov}, \citenamefont {Bak}, \citenamefont {Krivokhatskii},\ and\
  \citenamefont {Shlyamin}}]{Aleksandrov1972}%
  \BibitemOpen
  \bibfield  {author} {\bibinfo {author} {\bibfnamefont {B.~M.}\ \bibnamefont
  {Aleksandrov}}, \bibinfo {author} {\bibfnamefont {M.~A.}\ \bibnamefont
  {Bak}}, \bibinfo {author} {\bibfnamefont {A.~S.}\ \bibnamefont
  {Krivokhatskii}},\ and\ \bibinfo {author} {\bibfnamefont {{\'E}.~A.}\
  \bibnamefont {Shlyamin}},\ }\bibfield  {title} {\bibinfo {title} {Slow
  neutron capture cross section of {Pa}$^{231}$},\ }\href
  {https://doi.org/10.1007/BF01125115} {\bibfield  {journal} {\bibinfo
  {journal} {Soviet Atomic Energy}\ }\textbf {\bibinfo {volume} {32}},\
  \bibinfo {pages} {203} (\bibinfo {year} {1972})}\BibitemShut {NoStop}%
\bibitem [{\citenamefont {Yurova}\ \emph {et~al.}(1984)\citenamefont {Yurova},
  \citenamefont {Polyakov}, \citenamefont {Rukhlo}, \citenamefont {Titarenko}
  \emph {et~al.}}]{Yurova1984}%
  \BibitemOpen
  \bibfield  {author} {\bibinfo {author} {\bibfnamefont {L.~N.}\ \bibnamefont
  {Yurova}}, \bibinfo {author} {\bibfnamefont {A.~A.}\ \bibnamefont
  {Polyakov}}, \bibinfo {author} {\bibfnamefont {V.~P.}\ \bibnamefont
  {Rukhlo}}, \bibinfo {author} {\bibfnamefont {Y.~E.}\ \bibnamefont
  {Titarenko}}, \emph {et~al.},\ }\bibfield  {title} {\bibinfo {title} {Thermal
  neutron capture cross section of ${}^{231}${Pa}},\ }\href
  {https://www-nds.iaea.org/exfor/servlet/X4sGetSubent?subID=40579003}
  {\bibfield  {journal} {\bibinfo  {journal} {Voprosy Atomnoi Nauki i Tekhniki,
  Ser. Yadernye Konstanty}\ }\textbf {\bibinfo {volume} {1(55)}},\ \bibinfo
  {pages} {3} (\bibinfo {year} {1984})},\ \bibinfo {note} {eXFOR
  40579.003}\BibitemShut {NoStop}%
\bibitem [{\citenamefont {Kang}\ and\ \citenamefont {von
  Hippel}(2001)}]{KangVonHippel2001}%
  \BibitemOpen
  \bibfield  {author} {\bibinfo {author} {\bibfnamefont {J.}~\bibnamefont
  {Kang}}\ and\ \bibinfo {author} {\bibfnamefont {F.~N.}\ \bibnamefont {von
  Hippel}},\ }\bibfield  {title} {\bibinfo {title} {{U}-232 and the
  proliferation-resistance of {U}-233 in spent fuel},\ }\href
  {https://doi.org/10.1080/08929880108426485} {\bibfield  {journal} {\bibinfo
  {journal} {Science \& Global Security}\ }\textbf {\bibinfo {volume} {9}},\
  \bibinfo {pages} {1} (\bibinfo {year} {2001})}\BibitemShut {NoStop}%
\bibitem [{iso(2025)}]{isotek2025}%
  \BibitemOpen
  \href {https://www.ans.org/news/2025-05-07/article-7007/} {\bibinfo {title}
  {Oak {R}idge's {I}sotek dramatically increases world supply of {Th}-229}},\
  \bibinfo {howpublished} {ANS Nuclear Newswire, 7 May 2025} (\bibinfo {year}
  {2025})\BibitemShut {NoStop}%
\bibitem [{\citenamefont {Terrill}\ \emph {et~al.}(1954)\citenamefont
  {Terrill}, \citenamefont {Ingraham},\ and\ \citenamefont
  {Moeller}}]{Terrill1954}%
  \BibitemOpen
  \bibfield  {author} {\bibinfo {author} {\bibfnamefont {J.~G.}\ \bibnamefont
  {Terrill}}, \bibinfo {author} {\bibfnamefont {S.~C.}\ \bibnamefont
  {Ingraham}},\ and\ \bibinfo {author} {\bibfnamefont {D.~W.}\ \bibnamefont
  {Moeller}},\ }\bibfield  {title} {\bibinfo {title} {Radium in the healing
  arts and in industry: radiation exposure in the {United States}},\
  }\href@noop {} {\bibfield  {journal} {\bibinfo  {journal} {Public Health
  Reports}\ }\textbf {\bibinfo {volume} {69}},\ \bibinfo {pages} {255}
  (\bibinfo {year} {1954})}\BibitemShut {NoStop}%
\bibitem [{\citenamefont {Morss}\ \emph {et~al.}(2010)\citenamefont {Morss},
  \citenamefont {Edelstein},\ and\ \citenamefont {Fuger}}]{Morss2010}%
  \BibitemOpen
  \bibfield  {author} {\bibinfo {author} {\bibfnamefont {L.~R.}\ \bibnamefont
  {Morss}}, \bibinfo {author} {\bibfnamefont {N.~M.}\ \bibnamefont
  {Edelstein}},\ and\ \bibinfo {author} {\bibfnamefont {J.}~\bibnamefont
  {Fuger}},\ }\href {https://doi.org/10.1007/978-94-007-0211-0} {\emph
  {\bibinfo {title} {The Chemistry of the Actinide and Transactinide
  Elements}}},\ \bibinfo {edition} {4th}\ ed.\ (\bibinfo  {publisher}
  {Springer},\ \bibinfo {address} {Dordrecht},\ \bibinfo {year}
  {2010})\BibitemShut {NoStop}%
\bibitem [{\citenamefont {Chechev}\ and\ \citenamefont
  {Kuzmenko}(2006)}]{Chechev2006Np236}%
  \BibitemOpen
  \bibfield  {author} {\bibinfo {author} {\bibfnamefont {V.~P.}\ \bibnamefont
  {Chechev}}\ and\ \bibinfo {author} {\bibfnamefont {N.~K.}\ \bibnamefont
  {Kuzmenko}},\ }\href {http://www.lnhb.fr/nuclides/Np-236m_tables.pdf} {\emph
  {\bibinfo {title} {Table de Radionucl\'eides: {Np}-236m}}},\ \bibinfo {type}
  {Tech. Rep.}\ (\bibinfo  {institution} {Laboratoire National Henri Becquerel
  (LNHB), CEA/Saclay},\ \bibinfo {year} {2006})\ \bibinfo {note} {recommended
  decay data; $\beta^-$ branch to $^{236}$Pu 47(1)\% (ENSDF: 48\%)}\BibitemShut
  {NoStop}%
\bibitem [{\citenamefont {Perkin}\ and\ \citenamefont
  {Coleman}(1961)}]{Perkin1961}%
  \BibitemOpen
  \bibfield  {author} {\bibinfo {author} {\bibfnamefont {J.~L.}\ \bibnamefont
  {Perkin}}\ and\ \bibinfo {author} {\bibfnamefont {R.~F.}\ \bibnamefont
  {Coleman}},\ }\bibfield  {title} {\bibinfo {title} {Cross sections for the
  {(n,2n)} reactions of ${}^{232}${Th}, ${}^{238}${U} and ${}^{237}${Np} with
  14-{MeV} neutrons},\ }\href {https://doi.org/10.1016/0368-3230(61)90094-5}
  {\bibfield  {journal} {\bibinfo  {journal} {J. Nucl. Energy A/B}\ }\textbf
  {\bibinfo {volume} {14}},\ \bibinfo {pages} {69} (\bibinfo {year}
  {1961})}\BibitemShut {NoStop}%
\bibitem [{\citenamefont {Landrum}\ \emph {et~al.}(1973)\citenamefont
  {Landrum}, \citenamefont {Nagle},\ and\ \citenamefont
  {Lindner}}]{Landrum1973}%
  \BibitemOpen
  \bibfield  {author} {\bibinfo {author} {\bibfnamefont {J.~H.}\ \bibnamefont
  {Landrum}}, \bibinfo {author} {\bibfnamefont {R.~J.}\ \bibnamefont {Nagle}},\
  and\ \bibinfo {author} {\bibfnamefont {M.}~\bibnamefont {Lindner}},\
  }\bibfield  {title} {\bibinfo {title} {{(n,2n)} cross sections for
  ${}^{238}${U} and ${}^{237}${Np} in the region of 14 {MeV}},\ }\href
  {https://doi.org/10.1103/PhysRevC.8.1938} {\bibfield  {journal} {\bibinfo
  {journal} {Phys. Rev. C}\ }\textbf {\bibinfo {volume} {8}},\ \bibinfo {pages}
  {1938} (\bibinfo {year} {1973})}\BibitemShut {NoStop}%
\bibitem [{\citenamefont {Lindeke}\ \emph {et~al.}(1975)\citenamefont
  {Lindeke}, \citenamefont {Specht},\ and\ \citenamefont {Born}}]{Lindeke1975}%
  \BibitemOpen
  \bibfield  {author} {\bibinfo {author} {\bibfnamefont {K.}~\bibnamefont
  {Lindeke}}, \bibinfo {author} {\bibfnamefont {S.}~\bibnamefont {Specht}},\
  and\ \bibinfo {author} {\bibfnamefont {H.-J.}\ \bibnamefont {Born}},\
  }\bibfield  {title} {\bibinfo {title} {Determination of the
  ${}^{237}${Np}{(n,2n)}${}^{236}${Np} cross section at 15 {MeV} neutron
  energy},\ }\href {https://doi.org/10.1103/PhysRevC.12.1507} {\bibfield
  {journal} {\bibinfo  {journal} {Phys. Rev. C}\ }\textbf {\bibinfo {volume}
  {12}},\ \bibinfo {pages} {1507} (\bibinfo {year} {1975})}\BibitemShut
  {NoStop}%
\bibitem [{\citenamefont {Veyssi{\`e}re}\ \emph {et~al.}(1973)\citenamefont
  {Veyssi{\`e}re}, \citenamefont {Beil}, \citenamefont {Berg{\`e}re},
  \citenamefont {Carlos}, \citenamefont {Lepr{\^e}tre},\ and\ \citenamefont
  {Kernbach}}]{Veyssiere1973}%
  \BibitemOpen
  \bibfield  {author} {\bibinfo {author} {\bibfnamefont {A.}~\bibnamefont
  {Veyssi{\`e}re}}, \bibinfo {author} {\bibfnamefont {H.}~\bibnamefont {Beil}},
  \bibinfo {author} {\bibfnamefont {R.}~\bibnamefont {Berg{\`e}re}}, \bibinfo
  {author} {\bibfnamefont {P.}~\bibnamefont {Carlos}}, \bibinfo {author}
  {\bibfnamefont {A.}~\bibnamefont {Lepr{\^e}tre}},\ and\ \bibinfo {author}
  {\bibfnamefont {K.}~\bibnamefont {Kernbach}},\ }\bibfield  {title} {\bibinfo
  {title} {A study of the photofission and photoneutron processes in the giant
  dipole resonance of ${}^{232}${Th}, ${}^{238}${U} and ${}^{237}${Np}},\
  }\href {https://doi.org/10.1016/0375-9474(73)90333-3} {\bibfield  {journal}
  {\bibinfo  {journal} {Nucl. Phys. A}\ }\textbf {\bibinfo {volume} {199}},\
  \bibinfo {pages} {45} (\bibinfo {year} {1973})}\BibitemShut {NoStop}%
\bibitem [{\citenamefont {Berman}\ \emph {et~al.}(1986)\citenamefont {Berman},
  \citenamefont {Caldwell}, \citenamefont {Dowdy}, \citenamefont {Dietrich},
  \citenamefont {Meyer},\ and\ \citenamefont {Alvarez}}]{Berman1986}%
  \BibitemOpen
  \bibfield  {author} {\bibinfo {author} {\bibfnamefont {B.~L.}\ \bibnamefont
  {Berman}}, \bibinfo {author} {\bibfnamefont {J.~T.}\ \bibnamefont
  {Caldwell}}, \bibinfo {author} {\bibfnamefont {E.~J.}\ \bibnamefont {Dowdy}},
  \bibinfo {author} {\bibfnamefont {S.~S.}\ \bibnamefont {Dietrich}}, \bibinfo
  {author} {\bibfnamefont {P.}~\bibnamefont {Meyer}},\ and\ \bibinfo {author}
  {\bibfnamefont {R.~A.}\ \bibnamefont {Alvarez}},\ }\bibfield  {title}
  {\bibinfo {title} {Photofission and photoneutron cross sections and
  photofission neutron multiplicities for ${}^{233}${U}, ${}^{234}${U},
  ${}^{237}${Np}, and ${}^{239}${Pu}},\ }\href
  {https://doi.org/10.1103/PhysRevC.34.2201} {\bibfield  {journal} {\bibinfo
  {journal} {Phys. Rev. C}\ }\textbf {\bibinfo {volume} {34}},\ \bibinfo
  {pages} {2201} (\bibinfo {year} {1986})}\BibitemShut {NoStop}%
\bibitem [{\citenamefont {Gardner}\ and\ \citenamefont
  {Gardner}(1989)}]{Gardner1989}%
  \BibitemOpen
  \bibfield  {author} {\bibinfo {author} {\bibfnamefont {D.~G.}\ \bibnamefont
  {Gardner}}\ and\ \bibinfo {author} {\bibfnamefont {M.~A.}\ \bibnamefont
  {Gardner}},\ }\href {https://www.osti.gov/servlets/purl/5665453} {\emph
  {\bibinfo {title} {Some neutron and photon reactions on the ground and
  isomeric states of ${}^{236,237}${Np}}}},\ \bibinfo {type} {Tech. Rep.}\
  \bibinfo {number} {UCRL-101613}\ (\bibinfo  {institution} {Lawrence Livermore
  National Laboratory},\ \bibinfo {year} {1989})\ \bibinfo {note} {preprint
  UCRL-101613; Int. Conf. on Nuclear Data for Science and Technology,
  M{\"u}nchen-Garching}\BibitemShut {NoStop}%
\bibitem [{\citenamefont {Bateman}(1910)}]{Bateman1910}%
  \BibitemOpen
  \bibfield  {author} {\bibinfo {author} {\bibfnamefont {H.}~\bibnamefont
  {Bateman}},\ }\bibfield  {title} {\bibinfo {title} {Solution of a system of
  differential equations occurring in the theory of radio-active
  transformations},\ }\href
  {https://archive.org/details/cbarchive_122715_solutionofasystemofdifferentia1843}
  {\bibfield  {journal} {\bibinfo  {journal} {Proc. Cambridge Philos. Soc.}\
  }\textbf {\bibinfo {volume} {15}},\ \bibinfo {pages} {423} (\bibinfo {year}
  {1910})}\BibitemShut {NoStop}%
\end{thebibliography}%

\end{document}